\pdfoutput=1
\documentclass[journal]{new-aiaa} 
\usepackage[utf8]{inputenc}
\usepackage{graphicx}
\usepackage{amsmath}
\usepackage{longtable,tabularx}
\usepackage{multirow}
\usepackage{comment}
\usepackage{mdframed}
\usepackage[norelsize]{algorithm2e}
\usepackage{xcolor}

\newcommand{\mbf}{\mathbf}
\newcommand{\mbb}{\mathbb}
\newcommand{\mc}{\mathcal}
\renewcommand{\f}{\frac}
\newcommand{\bs}{\boldsymbol}
\newtheorem{theorem}{Theorem}[section]

\title{Aerodynamic Data Fusion Towards the Digital Twin Paradigm \footnote{Submitted to the editors}}

\author{S. Ashwin Renganathan \footnote{Postdoctoral Appointee, Division of Mathematics \& Computer Science, 9700 S Cass Ave. Member}}
\affil{Argonne National Laboratory \\
 Lemont, IL 60439}
\author{Kohei Harada \footnote{Ph.D. Candidate, School of Aerospace Engineering, 270 Ferst Dr NW. Student Member}}
\author{Dimitri N. Mavris \footnote{Regents Professor, School of Aerospace Engineering, 270 Ferst Dr NW. Fellow}}
\affil{Georgia Institute of Technology \\
Atlanta, GA 30332}

\begin{document}

\maketitle

\begin{abstract}
We consider the fusion of two aerodynamic data sets originating from differing fidelity physical or computer experiments. We specifically address the fusion of: 1) noisy and in-complete fields from wind tunnel measurements and 2) deterministic but biased fields from numerical simulations. These two data sources are fused in order to estimate the \emph{true} field that best matches measured quantities that serves as the ground truth. For example, two sources of pressure fields about an aircraft are fused based on measured forces and moments from a wind-tunnel experiment. A fundamental challenge in this problem is that the true field is unknown and can not be estimated with 100\% certainty. We employ a Bayesian framework to infer the true fields conditioned on measured quantities of interest; essentially we perform a \emph{statistical correction} to the data. The fused data may then be used to construct more accurate surrogate models suitable for early stages of aerospace design. We also introduce an extension of the Proper Orthogonal Decomposition with constraints to solve the same problem. Both methods are demonstrated on fusing the pressure distributions for flow past the RAE2822 airfoil and the Common Research Model wing at transonic conditions. Comparison of both methods reveal that the Bayesian method is more robust when data is scarce while capable of also accounting for uncertainties in the data. Furthermore, given adequate data, the POD based and Bayesian approaches lead to \emph{similar} results.

\end{abstract}

\section*{Nomenclature}

{\renewcommand\arraystretch{1.0}
\noindent\begin{longtable*}{@{}l @{\quad=\quad} l@{}}
$\mc{N}$  & normal distribution \\
$\bs{\mu}$ & mean of multivariate normal distribution\\
$\sigma, \tau$ & standard deviation \\
$\pi$ & probability density \\
$z$ & measurements \\
$n$ & dataset size \\
$\mbf{I}_n$ & identity matrix of size $n\times n$ \\
$\mbf{H}$ & output operator \\
$\mbb{E}$ & expectation of a random variable \\
$\Sigma, \Gamma$ & variance-covariance matrix \\
$\mbf{x}$ & spatial-coordinates \\
$f(\cdot)$ & forward model \\
$\ell$ & correlation length scale \\
$MAP$ & maximum a-posteriori estimate \\
$M/Re/ \alpha$ & Mach/Reynolds number/ angle of attack \\
$C_P$ & coefficient of pressure \\
$\theta$ & fusion parameter \\
$C_m/C_M, C_l/C_L$ & coefficient of moment, lift (2D/3D)
\end{longtable*}}

\section{Introduction}
\label{s:Intro}


The Digital Twin (DT) concept in aerospace systems design is a vision that aims to achieve paradigm shift in flight certification, fleet management and sustainment~\cite{glaessgen2012digital}. The basic idea is to develop a simulator integrating high-fidelity physics models, historical flight-test data and sensor updates amongst others, to mirror the operation of the corresponding flying twin. This would in turn enable real-time health monitoring of those vehicles during their operation that could potentially result in reliable and efficient designs that meet the design requirements of the future. The trend in aircraft design over the past couple decades have resulted in an abundance of data being generated from three main sources namely, numerical simulations, wind-tunnel measurements and flight testing; thereby laying the foundation towards realizing an aerodynamic digital twin. Particularly, the steady growth in computational capabilities in the past few decades have assisted aircraft manufacturers to numerically simulate the aerodynamics of an aircraft under real flight conditions. This comes at a significantly cheaper cost than wind-tunnel or flight tests and thereby replacing or supplementing them to a good extent ~\cite{BoeingCFD}.  On the other hand, flight test and wind tunnel data, though not abundant are available in non-negligible amounts. Numerical simulations at the practical scale are seldom exact and suffer from \emph{model bias}, whereas measurements are \emph{noisy}, incomplete and could be contaminated by errors introduced due to model scale and instrumentation among other factors. Therefore, as a means for realizing the DT vision, there is a need to take advantage of available data by combining them in a fashion that is likely to make them more accurate than their individual selves. We refer to this problem as multifidelity data fusion and within the context of this work, we consider aerodynamic pressure fields arriving from computational fluid dynamics (CFD) simulations and wind-tunnel measurements. We distinguish the present work from some of the past work where the term \emph{data fusion} has been used in the context of using variable fidelity simulation codes to accelerate design optimization; for instance see Keane (2003)~\cite{keane2003wing}. On the contrary, the present work is purely data-driven and uses only \emph{domain knowledge} about the actual models involved as will be demonstrated in later sections.

An associated challenge with the data fusion problem is to also quantify the confidence in the fused data. As mentioned earlier, neither the CFD predictions nor the wind-tunnel field measurements are devoid of uncertainties and therefore their fusion is expected to inherit those uncertainties. Uncertainty in CFD results could originate from inadequate physical models (such as turbulence closures ~\cite{Edeling2014, Emory2013}), geometry discrepancy, and discretization errors that lead to numerical diffusion~\cite{mathelin2005stochastic}, just to name a few. The wind-tunnel measurements are affected by model installation effects (such as due to wall and sting), simulation of proper boundary-layer development, and flow angle correction that could lead to errors in directional static stability. Quantifying these uncertainties is a challenge in itself; however we assume they are known and show how to properly account for them in the proposed approach. The goal of data fusion therefore, is to synergistically combine datasets from multiple fidelity sources, to improve the overall prediction in addition to quantifying uncertainty in the overall prediction. 

Combining aero/fluid dynamic data from computer and physical experiments is commonly done with the Proper Orthogonal Decomposition (POD) originally introduced by Lumley~\cite{Lumley1998} and Sirovich~\cite{Sirovich1987} in the context of turbulent fluid flow, where the combined flow snapshots are stacked in a matrix and a set of orthonormal modes are extracted via the singular value decomposition (SVD); the state of the flow is then assumed to be a linear combination of these modes. In a way, the POD leads to a dimensionality reduction from the original state-space dimension to the number of POD modes which are typically much fewer. Taking advantage of this fact, the \emph{gappy} POD introduced by Sirovich~\cite{everson1995karhunen} imputes missing data in fluid dynamic datasets where the POD modes are first extracted from the complete datasets. Then the missing data from the incomplete sets are estimated as the best linear combination of the POD modes nearest to the incomplete dataset in the least-squares sense. See \cite{bui2004aerodynamic} for further details. Ruscher et al \cite{Ruscher2016} supplement the gappy POD with wavelet methods to ensure continuity in the reconstructed data, while Wen et al \cite{Wen2019} apply it when there are two sets of missing data but in a pattern that complement each other.  Another technique that has a very similar goal as data fusion is the so-called Data Assimilation (DA), which refers to combining physical measurements with mathematical models particularly in the context of dynamical systems. DA originated in the geosciences field for weather prediction~\cite{dee2011era, kalnay2003atmospheric}. The methodology proposed in this work differs from DA by the fact that it is purely data-driven. Although domain knowledge about the models that generated the data is leveraged, the models themselves don't feature in the proposed method, as in DA. Overall, our method is more generic, robust to \emph{small} data and applies to any situation where there are multiple sources of data and some measurement of the ground truth. 

\begin{figure}
    \centering
    \includegraphics[width = 5in]{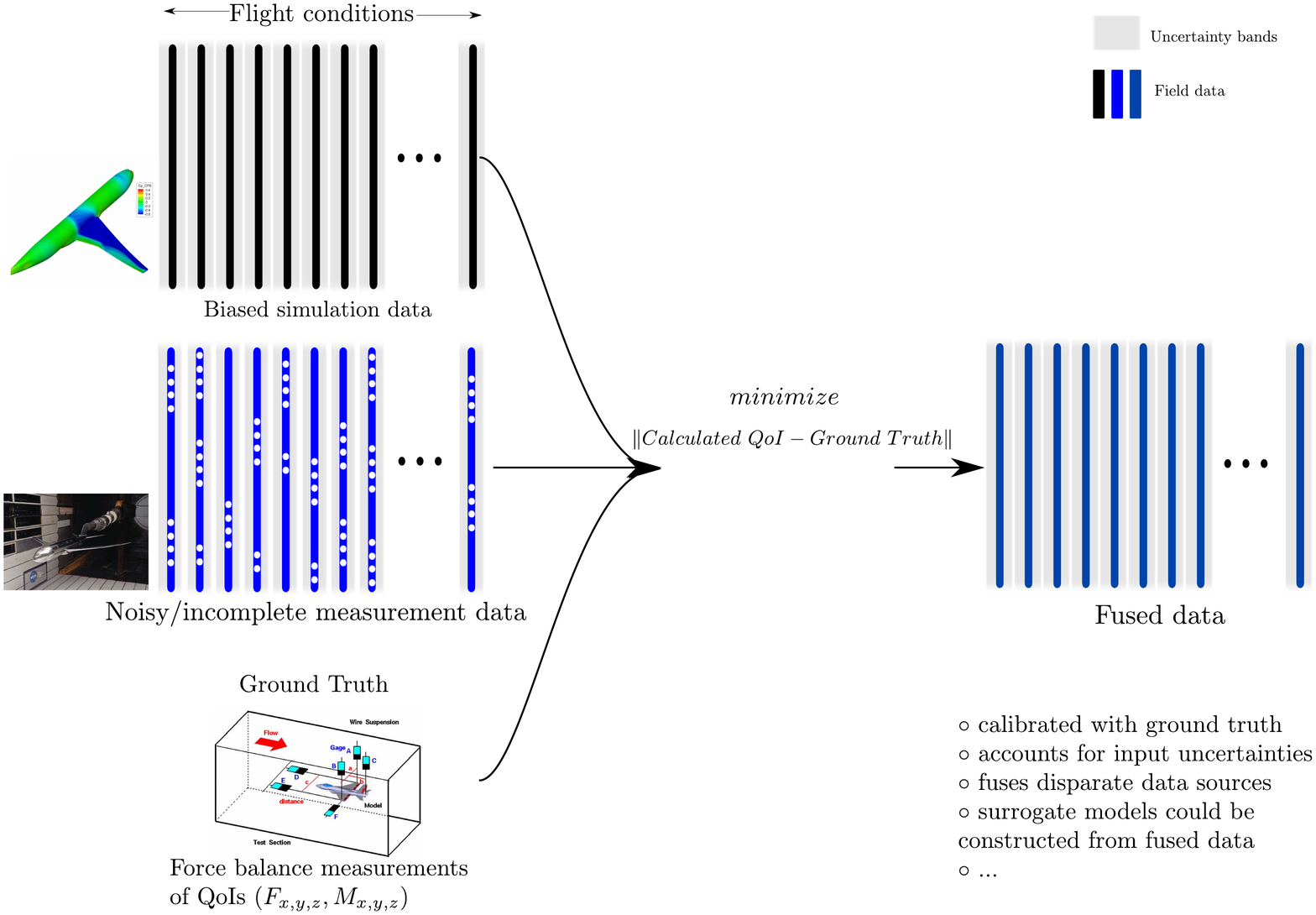}
    \caption{Schematic of the overall methodology}
    \label{f:overall_method}
\end{figure}

As a first step, this work proposes a methodology to fuse two datasets that could be corrupted by noise,  bias and/or incompleteness. As mentioned before, we consider field data that represent the state of the system; such as the distribution of pressure on the surface of an aircraft wing at a specific flight condition. Using this information and additional measurements of quantities of interest (QoIs) such as forces and moments, that are considered the ground truth, the inverse problem of estimating the true field by fusing the available datasets that best matches the measurements is posed and solved. The measurements of QoIs, although corrupted by noise and wind-tunnel scaling effects like the field measurements are extracted from more mature and reliable instrumentation such as force balance. Furthermore, these measurements are independent of the field measurements, i.e. they are not calculated as a function of the fields but are directly measured. For these reasons, the QoI measurements are treated as the ground truth in this work. An schematic of the overall methodology is provided in Figure~\ref{f:overall_method}.

The inversion of the pressure field from the measured forces and moments is an \emph{ill-posed} problem which is easy to see. For instance an arbitrary shift in the pressure distribution over a closed aerodynamic surface does not change the the value of its integral; i.e. $\int_S p ds = \int_S (p-p_{\infty}) ds$. In this regard, we incorporate a Bayesian framework such that regularizing priors can be specified to tackle the ill-posedness of the inverse problem, in addition to accounting for uncertainties associated with the datasets; this method was originally introduced in \cite{renganathan2019multifidelity}. This way, the method performs a statistical correction on the datasets and infers the probability distribution of the fused data. The output of the proposed method may then be used to construct more accurate surrogate models of the fields which can be queried cheaply to solve problems in aerospace design such as optimization and uncertainty quantification. Additionally, we extend the work in \cite{renganathan2019multifidelity} and introduce a POD-based method to solve the same problem, namely the POD with constraints (CPOD). In contrast to the Bayesian approach, this method searches the POD subspace generated from the data to find the fused data corresponding to a given flight condition. However, the CPOD, similar to the Bayesian method also treats the measured forces and moments as the ground truth.

This work makes the following assumptions. Firstly, we consider only two levels of fidelity for the sake of simplicity although it extends to a hierarchy of fidelities without any modification except for the notations. Secondly, we assumes identical geometric fidelity for all the sources. For instance, it does not fuse wing and airfoil pressure distributions. Thirdly, we assume that a pair of fidelity data are available for all flight conditions. In practice, the case might be that the wind-tunnel and CFD date are do not correspond to identical flight conditions. To meet this requirement, the CFD data used in this work is generated to match the flight conditions of a pre-existing wind-tunnel dataset. Fourthly, we assume that the uncertainty in the available data sources are known. Whereas in the wind-tunnel measurements an estimate of these uncertainties are available from replications and sensor calibration tests, quantifying all forms of uncertainty in CFD data is not trivial. While this assumption makes the proposed method subjective, it also adds the flexibility of leveraging expert domain knowledge when available. Finally, viscous effects are not accounted for in the methodology because field measurements of shear stress are typically hard to obtain. For this reason, the QoIs considered in this work are restricted to the lift and moment coefficients where the contributions from viscosity are relatively marginal.

The rest of the article is organized as follows. The Bayesian methodology is outlined in the next section. The two test cases and their associated detail are discussed in section~\ref{s:Test_Cases} . Demonstration of the proposed Bayesian method is shown in section~\ref{s:Method_Demo} and the CPOD method and comparison of both methods are discussed in section~\ref{s:CPOD}. Finally the paper concludes with a summary of main outcomes of the study and some directions for future work.

\section{Data Fusion via Bayesian Inference}
\label{s:Methodology}
In section \ref{s:Method_Demo} we demonstrate the methodology on the coefficient of pressure ($C_P$) distribution on an aircraft wing-section as well as the entire aircraft. However, here we keep the exposition general and consider two independent fidelity sources of data $\mbf{y}_1 \in \mbb{R}^n$ and $\mbf{y}_2 \in \mbb{R}^n$ represented by random variable $Y$ with densities

\begin{equation}
\begin{split}
\pi_{Y} (\mbf{y}_1) = \mathcal{N}(\bs{\mu}_1, \sigma_1^2 \mbf{I}_n)\\
\pi_{Y} (\mbf{y}_2) = \mathcal{N}(\bs{\mu}_2, \sigma_2^2 \mbf{I}_n)
\end{split}
\label{e:y1y2}
\end{equation}

\noindent
where $\sigma_1^2$ and $\sigma_2^2$ are the variance in the datasets and are assumed known. Furthermore, $\bs{\mu}_1$ and $\bs{\mu}_2$ are the expected values of the datasets which in the case of computer experiments are the direct predictions and in the case of physical experiments could be the ensemble averaged measurements of the fields.
Since the source of $\bs{\mu}_1$ and $\bs{\mu}_2$ could be vectors of disparate lengths, we interpolate them onto a common grid of length $n$. Let $\mbf{z} = [z_1,\hdots,z_m]^\top, ~z_i \in \mbb{R}$ be $m$ QoI's which are a function of $\mbf{y}$ and whose direct measurements are also available. Furthermore, we assume that the forward problem $\mbf{z} = f(\mbf{y})$ is known and we are interested in solving the \emph{inverse} problem $\mbf{y} = f^{-1}(\mbf{z})$ given $\mbf{z}$. Finally we restrict the current scope of the problem to be \emph{linear} and hence $\mbf{z} = f(\mbf{y}) = \mbf{H}^\top \mbf{y}$, where $\mbf{H}$ is a linear operator $\in \mbb{R}^{n\times m}$ that maps the field to the QoI. In the field of applied aerodynamics such a linear assumption is valid since quantities such as the lift, drag and moment coefficients are linear functions of the pressure distribution \footnote{note that we do not account for viscous effects in this context since they are typically unavailable from physical experiments}. The $\mbf{z}$ are assumed to be noisy and are related to the forward model via the following relationship
\begin{equation}
    \mbf{z} = \mbf{H}^\top \mbf{y} + \mbf{\epsilon}
    \label{e:fwd_model}
\end{equation}
where $\mbf{\epsilon}$ represents additive Gaussian white noise with probability density $\pi_E$ given by $\mathcal{N}(\mbf{0}, \tau^2 \mbf{I}_m)$, where $\tau^2$ is again assumed known. Our goal is now to estimate the probability distribution of the unknown $\mbf{y}$ given the measurements $\mbf{z}$ and this is done via Bayes' rule explained as follows.

\subsection{Bayes' Rule}
The cornerstone of the present methodology is the Bayes' rule, stated in \eqref{eqn:bayes}, whose highlight is that it operates on the probability densities of the variables rather than the variables themselves, thereby incorporating the uncertainty in the data naturally. The main idea is that, to infer the distribution of some unknown parameter conditioned on available data, $\pi(\mbf{y}|\mbf{z})$ \footnote{Note that we remove the subscripts in the density for convenience of notation}, the likelihood of observing the data given the model and its inputs, $\pi(\mbf{z}|\mbf{y})$ can be consolidated with whatever prior belief is available about the unknown parameter itself, denoted by $\pi(\mbf{y})$ and are related to each other as follows
\begin{equation}
    \pi(\mbf{y}|\mbf{z})=\frac{\pi(\mbf{z}|\mbf{y})\pi(\mbf{y})}{\pi(\mbf{z})}
    \label{eqn:bayes}
\end{equation}

\noindent 
In the equation above, $\pi(\mbf{y}|\mbf{z})$ is called the posterior density and $\pi(\mbf{z})$ is the marginal density of $\mbf{z}$ which can be evaluated by integrating out $\mbf{y}$ from the joint density $\pi(\mbf{z},\mbf{y})$. For practical purposes we are interested in some moment of the posterior density such as its mode or expected value. Since $\pi(\mbf{z})$ evaluates to a constant, its evaluation can be ignored in such situations. In this work we shall work with the value of the parameter estimate that maximizes the posterior probability which is called the maximum \emph{aposteriori} (MAP) estimate given by

\begin{equation}
    \mbf{y}^* = \underset{\mbf{y}}{arg \mathtt{max}~} \pi(\mbf{z}|\mbf{y})\pi(\mbf{y})
    \label{e:MAP}
\end{equation}

\noindent
which can be evaluated via gradient-based non-linear optimization methods. We now proceed to give specific details on the prior and likelihood models. 

\subsection{Prior Distribution}
A prior distribution on the unknown $\mbf{y}$ can be thought of as a regularizer that restricts the posterior distribution to physically valid solutions. This is where we take advantage of the available information in the form of $\mbf{y}_1$ and $\mbf{y}_2$ and combine them to define the prior. In this work, we introduce a fusion parameter $\theta \in \mbb{R} \subseteq [0,1]$ which combines $\mbf{y}_1$ and $\mbf{y}_2$ as

\begin{equation}
\begin{split}
    \tilde{\bs{\mu}}  &= \theta \times \mbb{E}(\mbf{y}_1) + (1-\theta) \times \mbb{E}(\mbf{y}_2) \\
    &= \theta \times \bs{\mu}_1 + (1-\theta) \times \bs{\mu}_2
\end{split}
    \label{e:prior_mean}
\end{equation}

where $\theta$ is chosen as the solution of
\begin{equation}
    \theta = \underset{\theta}{arg\text{min}}~ \|\mbf{H}^\top \tilde{\bs{\mu}} - \mbf{z}\|_2^2
    \label{e:theta_est}
\end{equation}

The the prior distribution on $\mbf{y}$ is then set as a multivariate Gaussian distribution centered on $\tilde{\bs{\mu}}$ given by

\begin{equation}
    \pi(\mbf{y}) = \f{1}{(2\pi)^{n/2} |\Sigma|^{1/2}} \text{exp} \left\lbrace-\f{1}{2}(\mbf{y}-\tilde{\bs{\mu}})^\top \Sigma^{-1}(\mbf{y}-\tilde{\bs{\mu}}) \right\rbrace
    \label{e:prior}
\end{equation}

In other words, we first chose the linear combination of the two sources of information that best matches the observed data (in the least squares sense) and define a Gaussian distribution centered around this estimate. This is built on the belief that the true value of $\mbf{y}$ lies somewhere in the neighborhood of $\mbf{y}_1$ and $\mbf{y}_2$ which is approximated as a linear weighted combination. Although the current way of estimating $\tilde{\bs{\mu}}$ is not unique, it provides a very simple and intuitive way of specifying the prior; i.e. it is a linearly weighted average of the available data. Furthermore, such prior specification are called \emph{sample based priors}~\cite{calvetti2018inverse} where prior belief is expressed via a combination of sample solutions of the unknown. Note that alternatively, one can treat $\theta$ as a random variable and use a hierarchical Bayes~\cite{santner2003design} framework and instead infer the posterior $\pi(\theta|\mbf{z})$ with some prior on $\pi(\theta)$. In that case, with a uniform prior on theta and a Gaussian likelihood, using the MAP estimate of $\pi(\theta|\mbf{z})$ is equivalent to the present approach. However we do not take that route in the present work to favor simplicity of exposition. In \eqref{e:prior} we treat $\mbf{y}$ to be spatially correlated and hence define the covariance matrix $\Sigma$ as
\begin{equation}
    \Sigma_{ij} = Cov(\mbf{y}(\mbf{x}_i) , \mbf{y}(\mbf{x}_j)) = \sigma^2 \text{exp} \left(-\f{\|\mbf{x}_i - \mbf{x}_j\|_2^2}{2\ell^2} \right)
    \label{e:prior_covariance}
\end{equation}
where, $\mbf{x}$ denote the spatial coordinates with $\|\cdot\|_2$ denoting the Euclidean distance, the parameter $\ell$ represents the length scale that is assumed to be known and (due to independence of $\mbf{y}_1$ and $\mbf{y}_2$) $\sigma^2 = \theta^2 \sigma_1^2 + (1-\theta)^2 \sigma_2^2$. The choice of the \emph{squared-exponential} kernel in \eqref{e:prior_covariance} is to specify smoothness in the prior realizations although other choice of kernels may be considered depending on the problem. See Ch.4 of \cite{williams2006gaussian} for a compendium of covariance kernels. 

\subsection{Likelihood Model}
Assuming that $\mbf{\epsilon}$ and $\mbf{y}$ are mutually independent, the probability density of $\mbf{z}$, conditional on $Y=\mbf{y}$ is obtained by shifting the density $\pi_E$ around $\mbf{H}^\top \mbf{y}$ leading to the likelihood density

\begin{equation}
\begin{split}
    \pi(\mbf{z}|\mbf{y}) &\sim \mathcal{N}(\mbf{H}^\top \mbf{y}, \tau^2 \mbf{I}_m) \\
    &=   \f{1}{\sqrt{2\pi}\tau} \text{exp} \left\lbrace-\f{1}{2\tau^2}(\mbf{z}-\mbf{H}^\top \mbf{y})^\top (\mbf{z}-\mbf{H}^\top \mbf{y}) \right\rbrace
\end{split}
    \label{e:likelihood}
\end{equation}

where again, $\tau^2$ is the measurement noise. 

\subsection{Maximum a-Posteriori (MAP) Estimation}
We are interested in solving for the inverse problem of estimating the true field given the measurements in terms of its probability density function, $\pi(\mbf{y}|\mbf{z})$, which is given by 
 
 \begin{equation}
    \begin{split}
        \pi(\mbf{y}|\mbf{z}) &\propto \pi(\mbf{z}|\mbf{y}) \times \pi(\mbf{y}) \\
        \pi(\mbf{y}|\mbf{z}) &\propto \text{exp} \left \lbrace -\frac{1}{2\tau^2} \|\mbf{z} - \mathbf{H}^\top \mbf{y} \|^2 \right \rbrace \times \text{exp} \left \lbrace -\frac{1}{2} (\mbf{y} - \tilde{\bs{\mu}})^\top \Sigma^{-1} (\mbf{y} - \tilde{\bs{\mu}}) \right \rbrace
     \end{split}
     \label{e:posterior}
 \end{equation}
 
 By \eqref{e:posterior} what we mean is that we estimate the true $\mbf{y}$ distribution that best fits the measured value of the quantity of interest while also being similar to the prior elicited for $\mbf{y}$ via \eqref{e:prior}. The \emph{mode} of the resulting posterior probability distribution is the MAP estimate we are interested in. It should be noted that since the posterior distribution in the present context is Gaussian, the mean, median and mode are identical and hence the choice does not matter. Therefore we are interested in solving the following optimization problem
 
 \begin{equation}
     \mbf{y}_{MAP} = ~\underset{\mbf{y}}{arg\text{max}}~ \text{exp} \left \lbrace -\frac{1}{2\tau^2} \|\mbf{z} - \mathbf{H}^\top \mbf{y} \|^2 \right \rbrace \times \text{exp} \left \lbrace -\frac{1}{2} (\mbf{y} - \tilde{\bs{\mu}})^\top \Sigma^{-1} (\mbf{y} - \tilde{\bs{\mu}}) \right \rbrace
     \label{e:MAP_problem}
 \end{equation}
 
 which is equivalent to solving
  \begin{equation}
  \begin{split}
     \mbf{y}_{MAP} &= ~\underset{\mbf{y}}{arg\text{min}}~ \frac{1}{2\tau^2} \|\mbf{z} - \mbf{H}^\top \mbf{y} \|^2 + \frac{1}{2} (\mbf{y} - \tilde{\bs{\mu}})^\top \Sigma^{-1} (\mbf{y} - \tilde{\bs{\mu}}) \\
      &= ~\underset{\mbf{y}}{arg\text{min}}~ J(\mbf{y})~\text{(say)} 
       \end{split}
     \label{e:MAP_problem2}
 \end{equation}
 
where $J$ denotes the entire cost function. Since \eqref{e:MAP_problem2} is differentiable everywhere we evaluate its gradient and set it to zero. Furthermore, $J(\mbf{y})$ is a symmetric positive-definite quadratic form which has a unique minimizer. Therefore we write the gradient as
 
 \begin{equation}
     \frac{\partial J}{\partial \mbf{y}} = \frac{1}{\tau^2} \left[- (\mbf{z} - \mathbf{H}^\top \mbf{y})^\top \mathbf{H}^\top + (\mbf{y} - \tilde{\bs{\mu}})^\top \Sigma^{-1} \right] = 0
 \end{equation}
 
 Rearranging above equation gives
 \[\mbf{y}_{MAP}^\top \left[ \frac{1}{\tau^2}\mathbf{H} \mathbf{H}^\top + \tilde{}^{-1} \right] = \frac{1}{\tau^2}\mbf{z}^\top \mathbf{H}^\top + \tilde{\bs{\mu}}^\top \Sigma^{-1} \]

which gives
\begin{equation}
   \mbf{y}_{MAP}^\top =  \left[ \frac{1}{\tau^2} \mbf{z}^\top \mbf{H}^\top + \tilde{\bs{\mu}}^\top \Sigma^{-1}\right] \left[\frac{1}{\tau^2}\mbf{H} \mbf{H}^\top + \Sigma^{-1} \right]^{-1}
   \label{e:yMAP}
\end{equation}  
It can then be shown that the posterior distribution is given by
\begin{equation}
    \begin{split}
        \pi(\mbf{y}|\mbf{z}) &\sim \mathcal{N}(\mbf{y}_{MAP}^\top, \boldsymbol{\Gamma}) \\
        \text{where, } 
        \boldsymbol{\Gamma} &= \left[\frac{1}{\tau^2}\mathbf{H} \mathbf{H}^\top + \Sigma^{-1} \right]^{-1}
        \label{e:posterior2}
    \end{split}
\end{equation}
 
The diagonal elements of $\boldsymbol{\Gamma}$ contain the pointwise variance of $\mbf{y}|\mbf{z}$ which may be used to construct confidence intervals on the predictions. The hyperparameters of the methodology are $\bs{\gamma} = \lbrace\sigma_1^2, \sigma_2^2, \tau^2, \ell \rbrace$. Among them, $\sigma_1^2, \sigma_2^2, \tau^2$ represent the uncertainty in the available dataset including the measured QoIs. Quantifying these uncertainties in the data is a very elaborate task which is not undertaken in this work. Instead we present a method that fuse the datasets by accounting for these uncertainties. As for the length-scale parameter $\ell$, although it can be estimated from data, in the present work we fix its value for each of the test cases listed in section~\ref{s:Test_Cases}. It is chosen from trial-and-error such that the realizations from the prior look physically reasonable. Note that one of the primary advantages of the Bayesian framework is the ability to specify subjective priors that leverage domain knowledge. The method is summarized in Algorithm~\ref{a:Alg_1}.

\begin{algorithm}[H]
\SetAlgoLined
\KwResult{$\mbf{y}_{MAP}$, $\mbf{\Gamma}$, confidence intervals}
\KwData{fields: $\bs{\mu}_1,\bs{\mu}_2$, measurements: $\mbf{z}$, parameters: $\bs{\gamma}$} \; 
 \begin{enumerate}
     \item Estimate $\theta$ from \eqref{e:theta_est}
     \item Compute $\mbf{y}_{MAP}$ from \eqref{e:yMAP}
     \item Compute posterior covariance matrix from \eqref{e:posterior2}
     \item Extract standard deviation to construct confidence intervals; $\pm \sqrt{\text{diag}(\mbf{\Gamma})}$
 \end{enumerate}
 \caption{Bayesian Data Fusion}
 \label{a:Alg_1}
\end{algorithm}

\section{Test Cases and Data}
\label{s:Test_Cases}
The two main test cases for the methods used in this work are the viscous transonic flow past the RAE2822 airfoil and the NASA Common Research Model (CRM). The corresponding wind-tunnel data are extracted from \cite{AGARD} and \cite{vassberg2008development} respectively. Specifically, for the CRM test case, the pressure distributions from pressure-sensitive paint (PSP)~\cite{mclachlan1995pressure, morris1993aerodynamic} measurements are obtained from the NASA Ames 11ft Transonic Wind Tunnel provided in ~\cite{rivers2014experimental}. All the CFD simulations were performed using the commercial, finite-volume based unstructured code STAR-CCM+~\cite{STARCCM+}. 

\subsection*{RAE-2822}
\begin{figure}[htb!]
\centering
\includegraphics[scale=0.2]{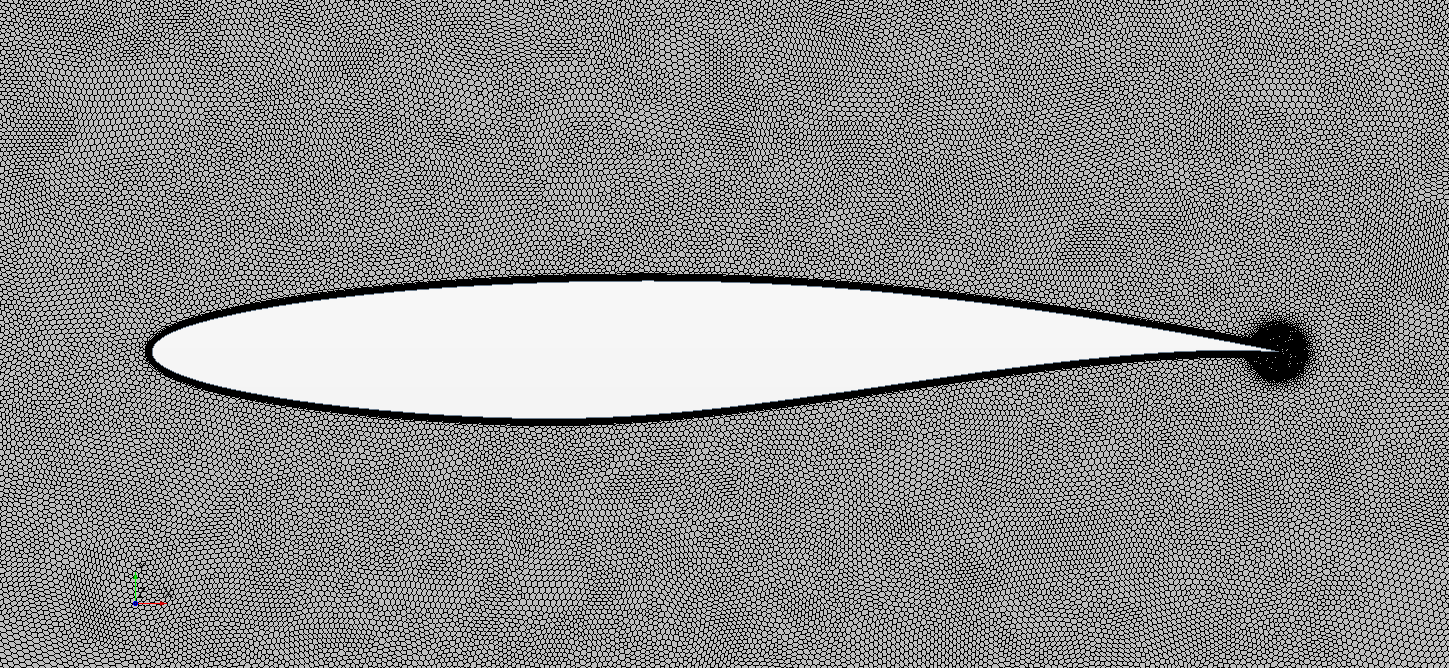}
\caption{The RAE2822 airfoil shape and near-field mesh with polyhedral elements.}
\label{f:RAE}
\end{figure}

The RAE2822 is a commonly used benchmark test case for transonic flight conditions. The airfoil shape and the mesh for the RANS analysis are shown in Figure~\ref{f:RAE}. A mesh with approximately 122,500 polyhedral mesh elements is generated with 41 prism layers to capture the boundary layer. The first layer of the prism layer is placed approximately $1.6\times 10^{-6}$ m away from the wall such that the wall $y^+~\approx 1$ for the Reynolds number ranges considered in this work. The AGARD~\cite{AGARD} wind-tunnel measurements are available for both the pressure distribution as well as the lift and moment coefficients and the operating conditions are summarized in Table~\ref{t:RAE_cases}. The airfoil surface is discretized into an $n=128$ size equally-spaced grid on which both the CFD and wind-tunnel data are interpolated before performing the fusion.

\begin{table}[htb!]
    \centering
    \caption{RAE2822: summary of test cases}
    \begin{tabular}{cccc}
    \hline
    Case & Mach & Reynolds number (millions) & Angle of attack (degrees) \\
    \hline
    1    & 0.676  & 5.7            &  2.40\\
    2    & 0.676  & 5.7            & -2.18\\
    3    & 0.600  & 6.3            & 2.57 \\
    4    & 0.725  & 6.5            & 2.92 \\
    5    & 0.725  & 6.5            & 2.55 \\
    6    & 0.728  & 6.5            & 3.22 \\
    7    & 0.730  & 6.5            & 3.19 \\
    8    & 0.750  & 6.2            & 3.19 \\    
    9    & 0.730  & 2.7            & 3.19 \\
   10    & 0.745  & 2.7            & 3.19 \\
   11    & 0.740  & 2.7            & 3.19 \\
    \hline
    \end{tabular}
    \label{t:RAE_cases}
\end{table}

\newpage
\subsection*{Common Research Model}
\begin{figure}
    \centering
    \includegraphics[scale = 0.4]{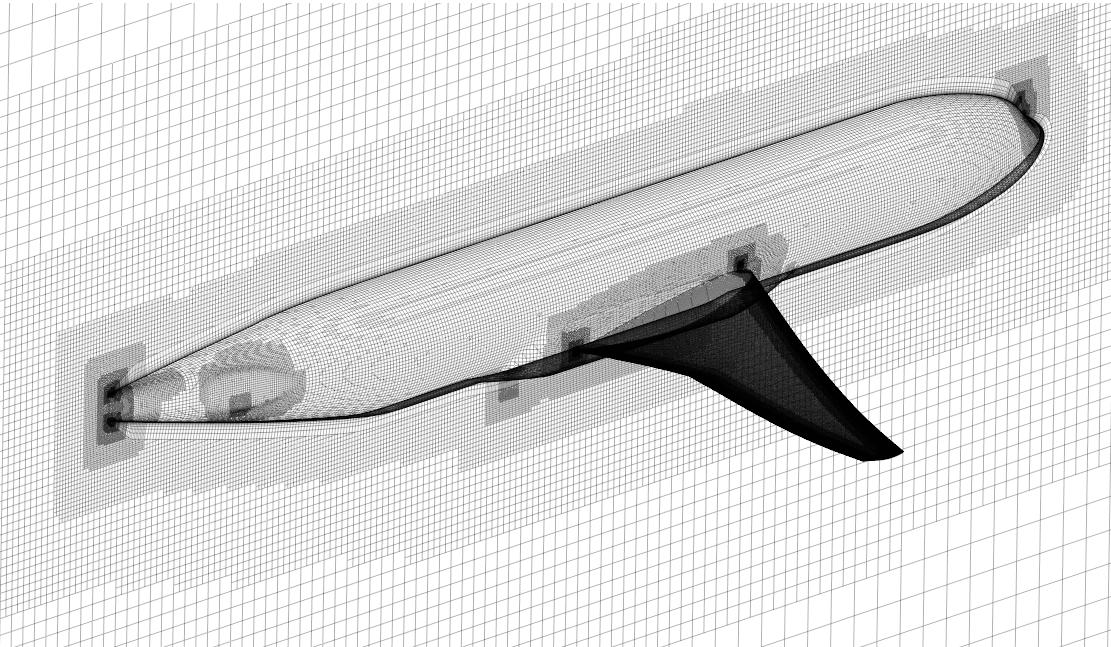}
    \caption{Trimmed hexahedral mesh for the CRM wing-body geometry containing 24M elements}
    \label{f:CRM_Mesh}
\end{figure}

The clean wing-body configuration of the CRM without tail-planes is used in this work. A mesh with approximately 24 million trimmed hexahedral elements were generated as shown in Figure~\ref{f:CRM_Mesh}, where the boundary layer is resolved with 41 prism layers and the wall $y^+ \approx 1$. In order to keep the computational costs of the method tractable, the CFD and PSP data are interpolated onto a coarse grid consisting of $n = 8688$ cells (further details are provided in the Appendix~\ref{app:pre-process}). The viscous contributions to the lift coefficient is in $\mathcal{O}(10^{-5})$ and the moment coefficient is in $\mathcal{O}(10^{-3})$ across all the test cases; while this is negligible for the lift, it is not so for the moments. In this work, the viscous contributions are ignored to keep the exposition simple although they can be easily included by adding an offset paramter in \eqref{e:theta_est} and \eqref{e:likelihood} of the form $\|\mbf{z} - \mbf{H}^\top \mbf{y} - \delta \|$ where $\delta > 0$ is a \emph{discrepancy} parameter and is in the same order as the viscous contributions. The test data for the CRM are set at a Reynolds number of 5 million and a full-factorial design of Mach = $\lbrace 0.70, 0.85, 0.87 \rbrace$, angle of attack = $\lbrace 0,0.5, 1.0, 1.5, 2.0, 3.0, 4.0 \rbrace$. The results are demonstrated in the next section for a selected combination of operating conditions.

\section{Method Demonstration}
\label{s:Method_Demo}
Here we demonstrate the methodology on inferring the true $C_P$ distributions given measurements. Here $z_1 = C_L/C_l$, $z_2 = C_M/C_m$, $\bs{\mu}_1 = C_{P, PSP/WT}$, $\bs{\mu}_2 = C_{P, CFD}$ and $\tilde{\bs{\mu}} = \tilde{C}_P$. The matrix $\mathbf{H} \in \mbb{R}^{n\times 2}$ where the two columns contain information about the local surface normals, cell areas and free-stream flow necessary to numerically integrate the pressure distributions to compute $C_L/C_l$ and $C_M/C_m$ respectively. 

\subsection{RAE2822}
\begin{figure}[htb!]
    \centering
    \begin{subfigure}{1\textwidth}
        \centering
        \includegraphics[width = 3.5in]{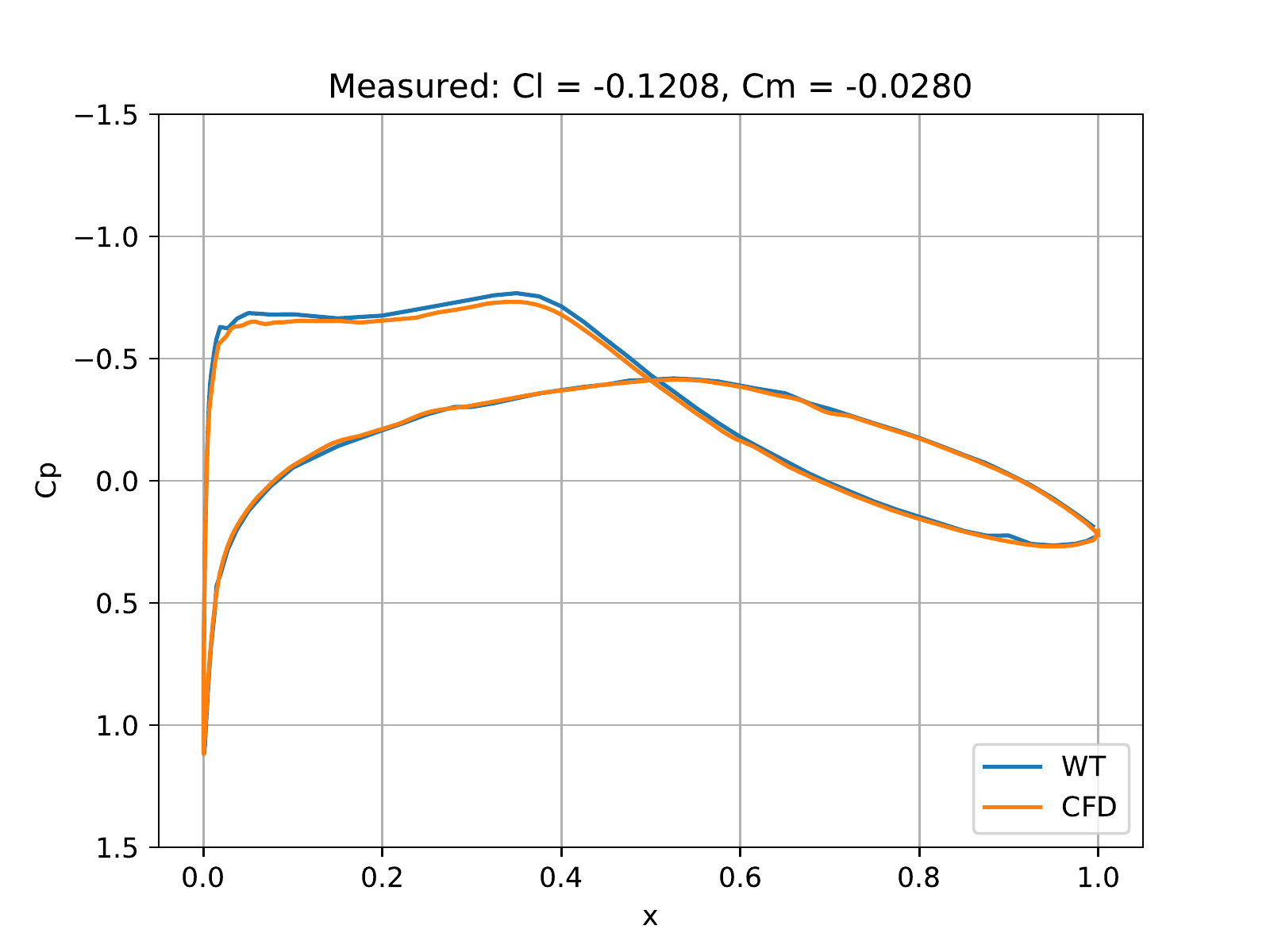}
        \caption{Original $C_P$ distributions}
        \label{f:RAE_Case2}
    \end{subfigure}\\
        \begin{subfigure}{0.5\textwidth}
        \centering
        \includegraphics[width = 3.5in]{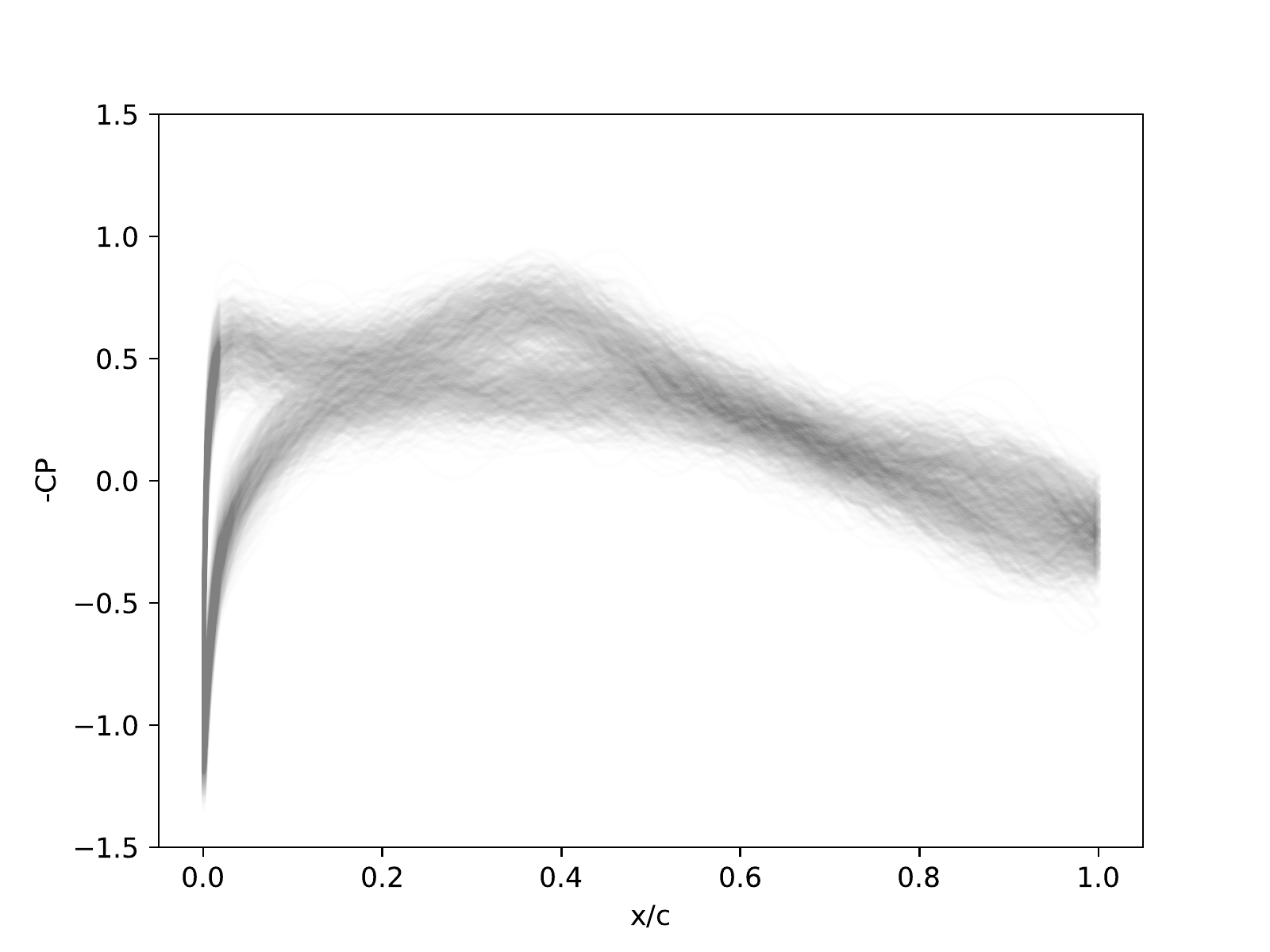}
        \caption{500 posterior draws}
        \label{f:RAE_Case2_post}
    \end{subfigure}~
        \begin{subfigure}{0.5\textwidth}
        \centering
        \includegraphics[width = 3.5in]{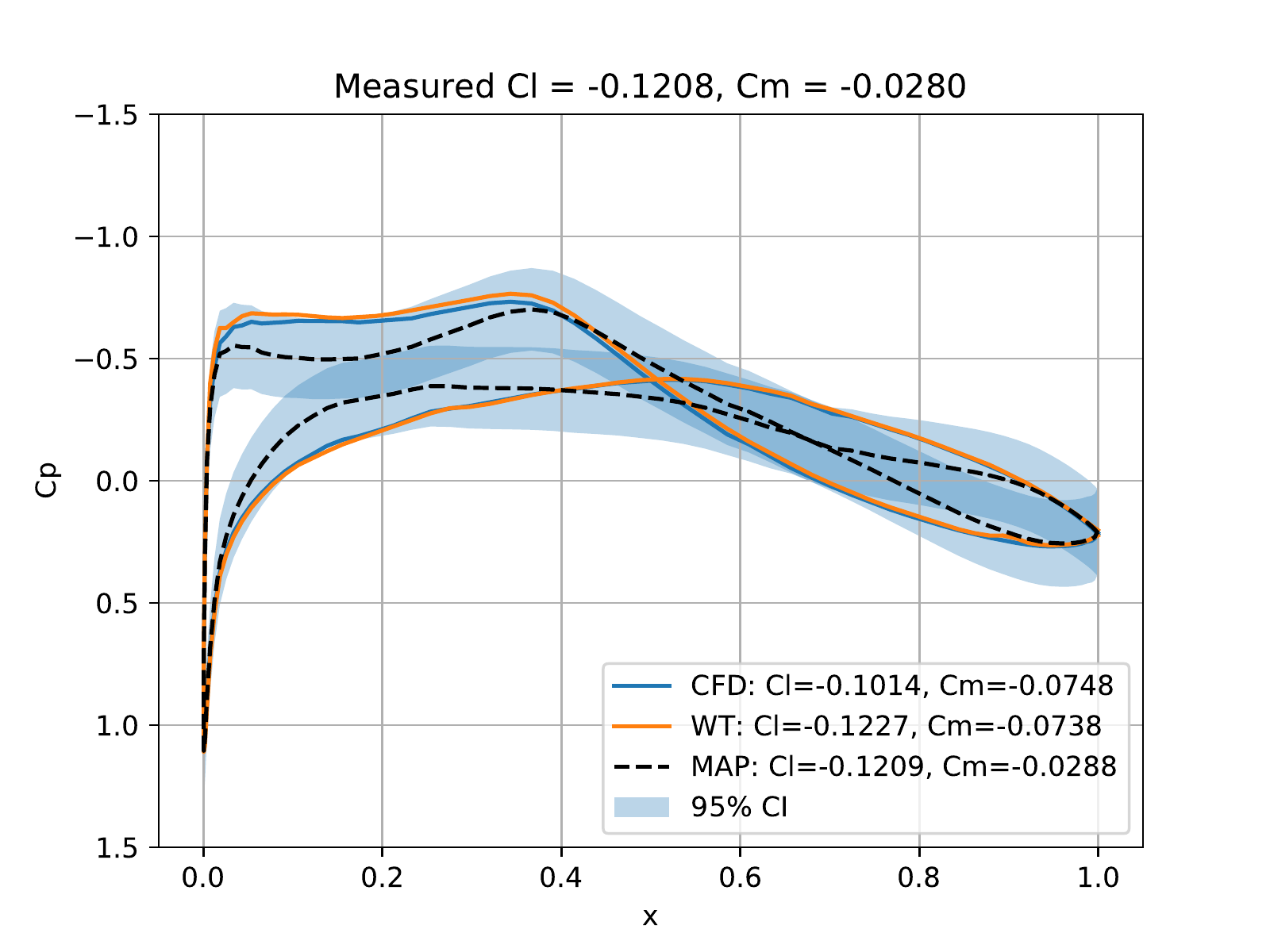}
        \caption{MAP with confidence bands}
        \label{f:RAE_Case2_MAP}
    \end{subfigure}
    \caption{$C_P$ distribution corresponding to Case-2 ($Mach=0.676$, $Re=5.7M$, $\alpha = -2.18~deg.$)}
    \label{fig:RAE_Res}
\end{figure}

Consider the Figure \ref{f:RAE_Case2} where the CFD prediction and the wind-tunnel measurement of the $C_P$ distribution (for flight condition $Mach=0.676$, $Re=5.7M$, $\alpha = -2.18~deg.$) match quite closely. However, the QoI's $(C_l, C_m)$ obtained by \emph{integrating} the $C_P$ curves (evaluating the forward model) from the wind-tunnel and CFD $C_P$'s are $(-0.122, -0.074)$ and $(-0.101, -0.075)$ respectively. Both QoI's are off from the \emph{measured} QoI's $(-0.121, -0.028)$ although, the discrepancy is more pronounced for $C_m$. In this case, no matter what value of $\theta$ is chosen, the resulting $\tilde{C}_P$ is not expected to produce outputs that match the measurements. Therefore, we are interested in adjusting the curves in Fig.~\ref{f:RAE_Case2} such that the QoI's derived from the adjusted curve matches the measured values in the least-squares sense. We set $\tau^2 = 10^{-6},~\sigma_1^2 = 10^{-2},~\sigma_2^2 = 10^{-2}$ based on the belief that the the pressure distributions from CFD predictions and wind-tunnel measurements have greater uncertainty than the force and moment measurements; while $\ell$ is set to $10^{-4}$. The statistically adjusted solution is then given by the posterior distribution~\eqref{e:posterior2} and 500 draws from this distribution are shown in Fig.\ref{f:RAE_Case2_post}. The MAP estimate is the expectation of the posterior distribution and is shown in Fig.\ref{f:RAE_Case2_MAP} overlaid with plots for the CFD and wind-tunnel distributions. Notice that the current approach shrinks the best combination of the two distributions in order to minimize the misfit between the derived QoIs and the measurements. It should be noted that the MAP estimate should not be treated as the \emph{true} $C_P$ distribution that uniquely defines the state of the system at the given operating conditions. This is because the MAP estimate depends on the chosen values of the hyperparameters. Additionally, it tries to minimize the misfit between noisy data and a deterministic model while not accounting for viscous effects. Therefore the fused $C_P$ should be analyzed in terms of its probability densities which account for the uncertainty rather than treating it as a deterministic estimate. Having said that, the MAP estimate provides an interpretable visualization of the posterior distribution and is used while comparing various curves.

\begin{figure}[htb!]
    \centering
        \begin{subfigure}{0.5\textwidth}
        \centering
        \includegraphics[width = 3.5in]{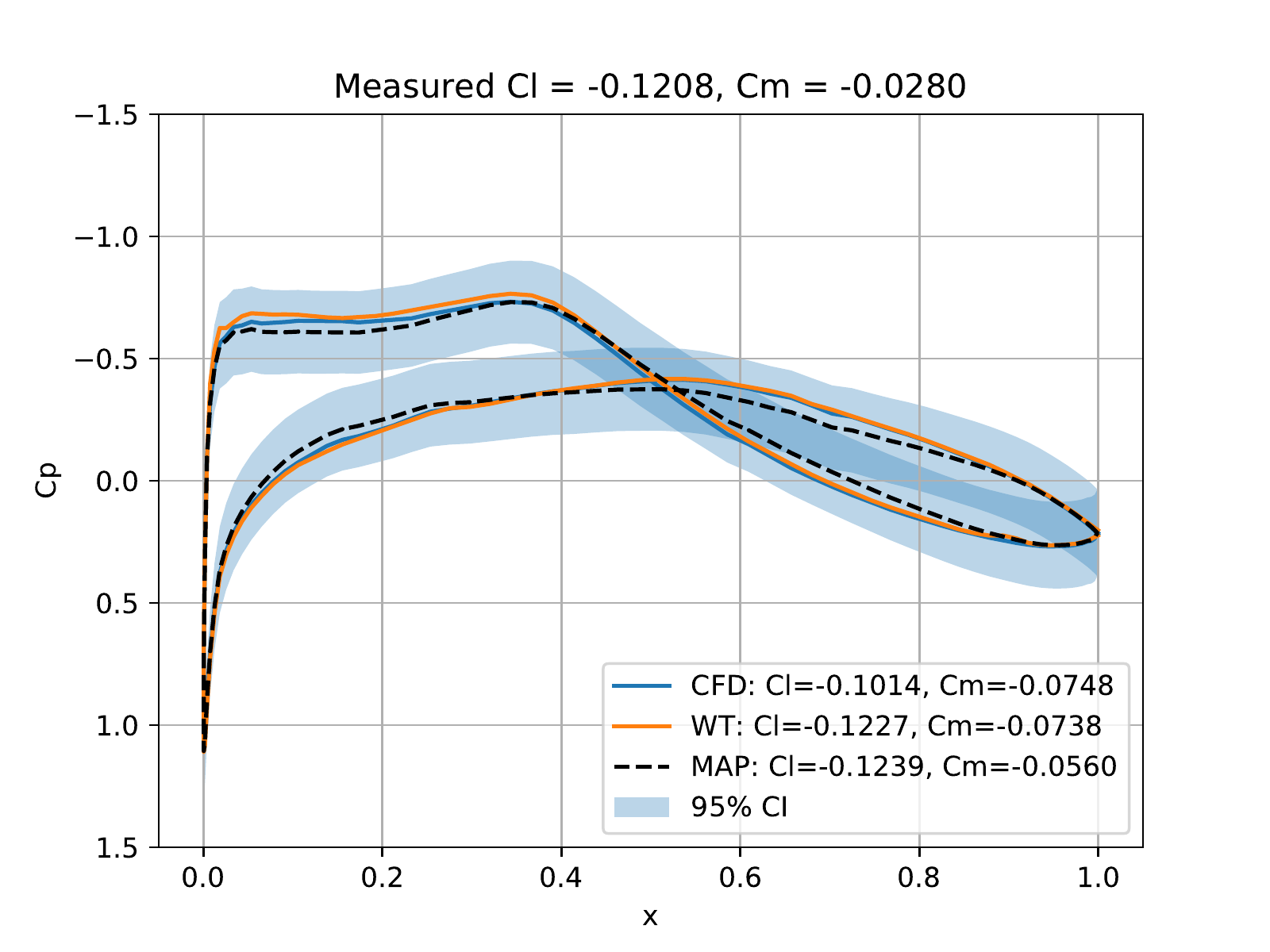}
        \caption{$\tau^2 = 10^{-4}$}
        \label{f:tau2=1e-4}
    \end{subfigure}~
        \begin{subfigure}{0.5\textwidth}
        \centering
        \includegraphics[width = 3.5in]{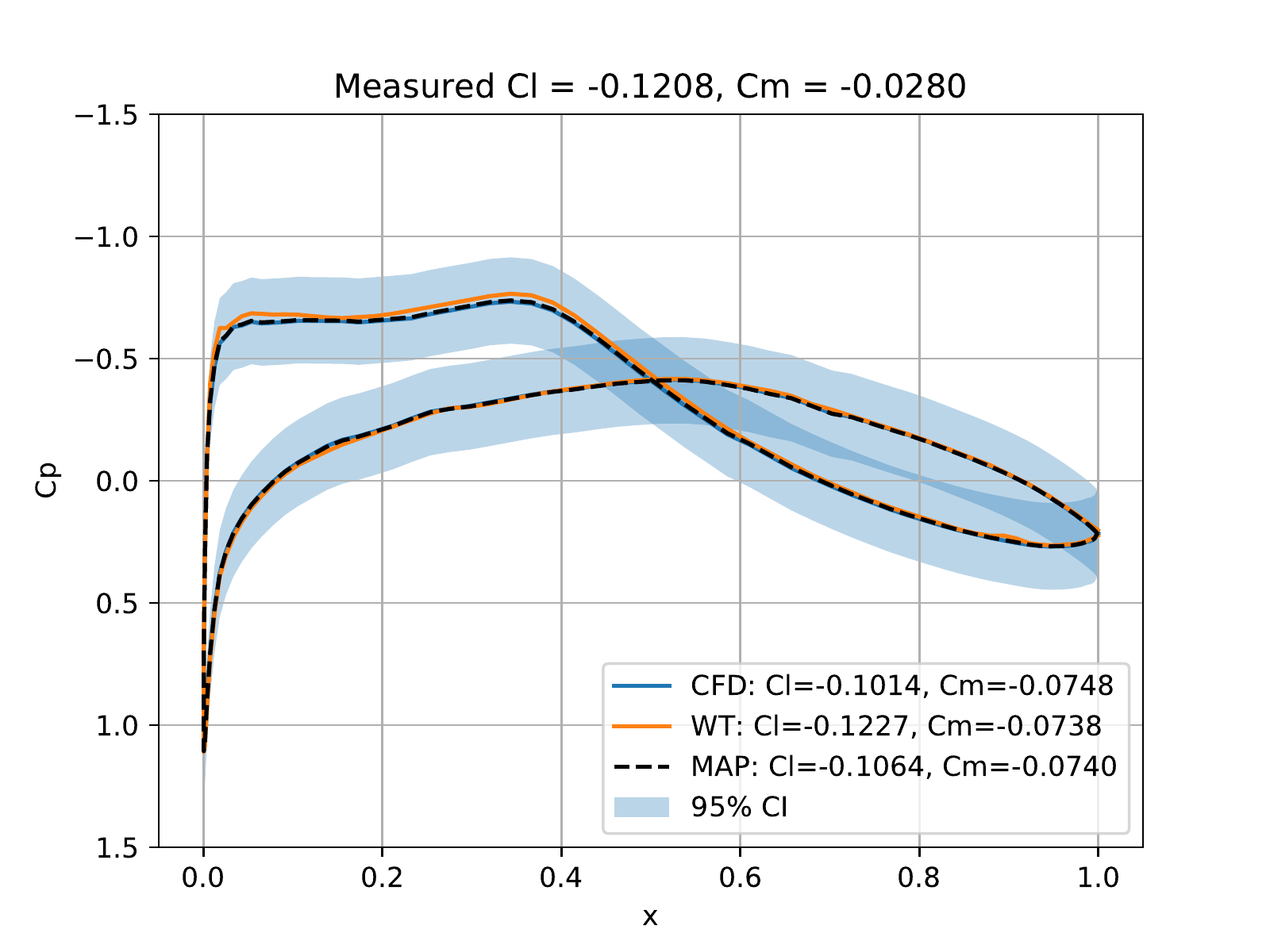}
        \caption{$\tau^2 = 10^{-2}$}
        \label{f:tau2=1e-2}
    \end{subfigure}
    \caption{Effect of the measurement noise parameter $\tau^2$. Smaller value forces method to adjust the $C_P$ curves more.}
    \label{fig:effect_of_tau}
\end{figure}

A point worth mentioning is the impact on the results due to the parameters. In Fig.\ref{fig:RAE_Res}, the parameter $\tau^2$ was set to $10^{-6}$ which implies that we trust the measured QoIs to posses very high signal-noise ratio. As a result the proposed approach tries to match them as closely as possible leading to relatively more adjustment on the original $C_P$ curves. On the other hand if we admit our ignorance about the actual value of the measurements and set a higher $\tau^2$, then the method leads to relatively less adjustment. This is demonstrated in figures~\ref{f:tau2=1e-4} and \ref{f:tau2=1e-2} where the $\tau^2$ is set to $10^{-4}$ and $10^{-2}$ respectively. Notice that when $\tau^2 = 10^{-2}$ the uncertainty in the QoIs (particularly $C_m$) is very high that the misfit term in \eqref{e:MAP_problem2} becomes less important and the prior dominates; as a result $C_{P,MAP} \approx \tilde{C}_P$. Larger the $\tau^2$ parameter, greater the misfit between the measured QoIs and those obtained by integrating the MAP estimate of $C_P$ (see the legend entries of Figures~\ref{fig:effect_of_tau}).

\begin{figure}[htb!]
    \centering
    \begin{subfigure}{1\textwidth}
        \centering
        \includegraphics[width = 3.5in]{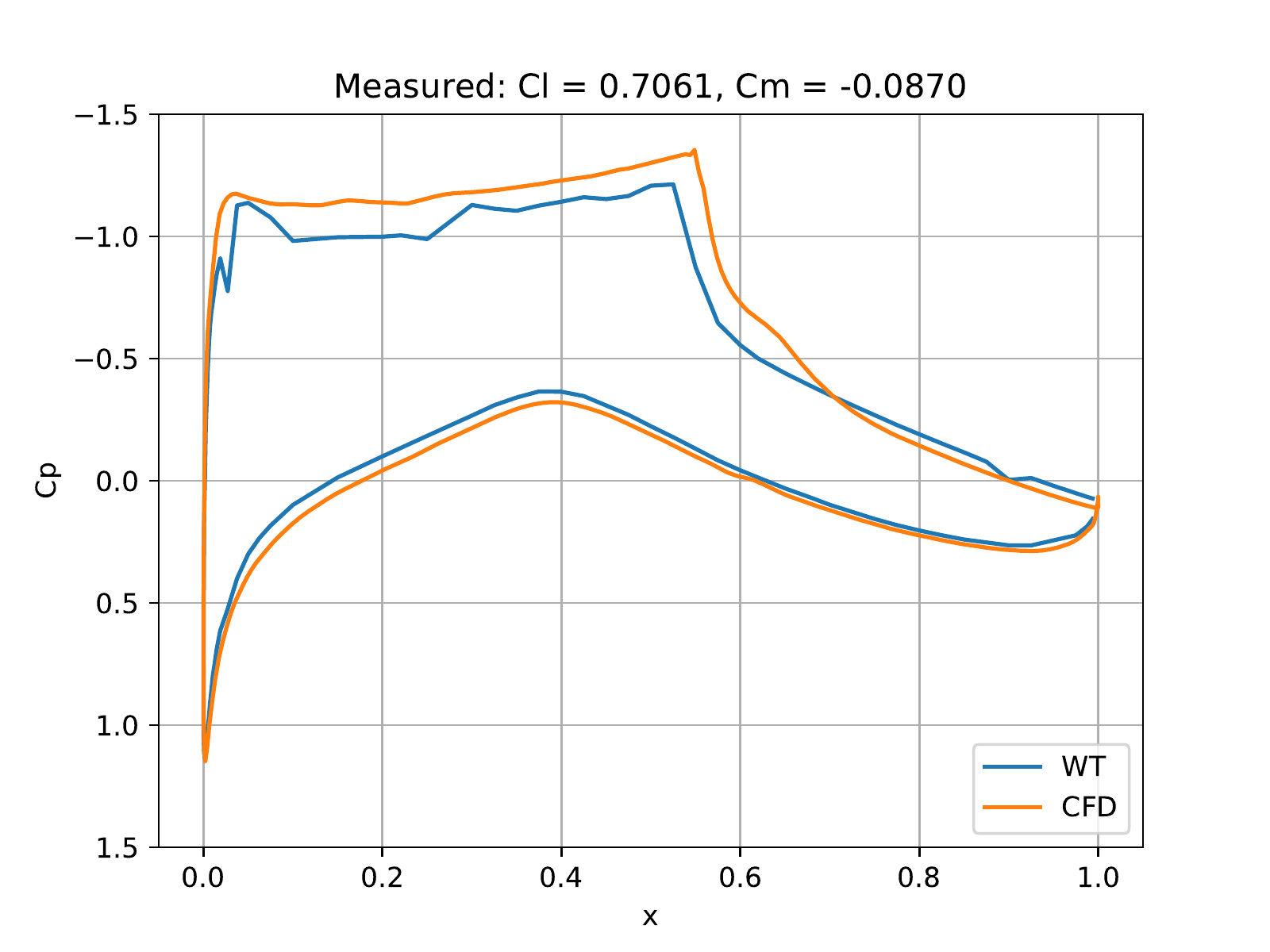}
        \caption{Original $C_P$ distributions}
        \label{f:RAE_Case13}
    \end{subfigure}\\
        \begin{subfigure}{0.5\textwidth}
        \centering
        \includegraphics[width = 3.5in]{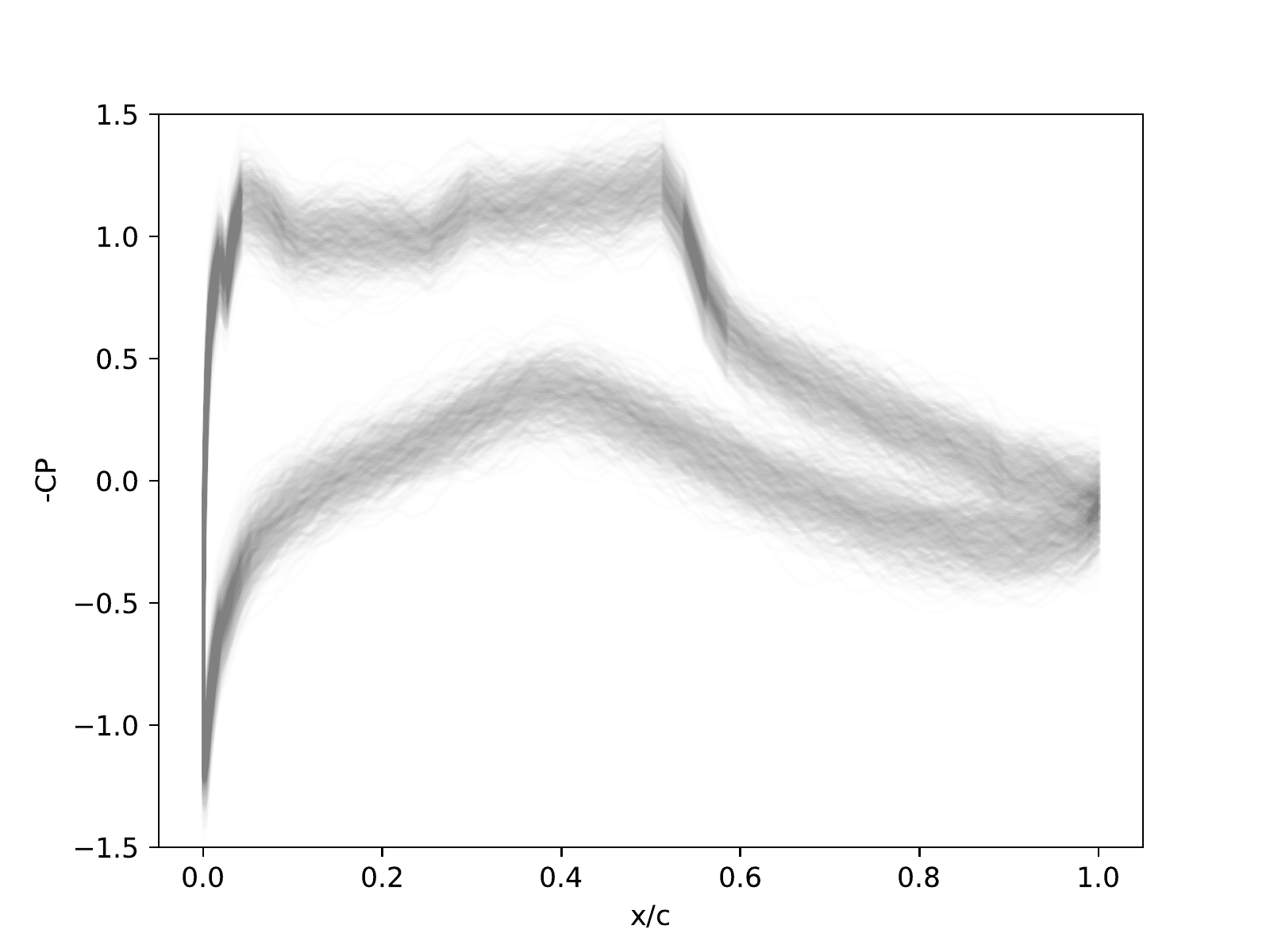}
        \caption{500 posterior draws}
        \label{f:RAE_Case13_post}
    \end{subfigure}~
        \begin{subfigure}{0.5\textwidth}
        \centering
        \includegraphics[width = 3.5in]{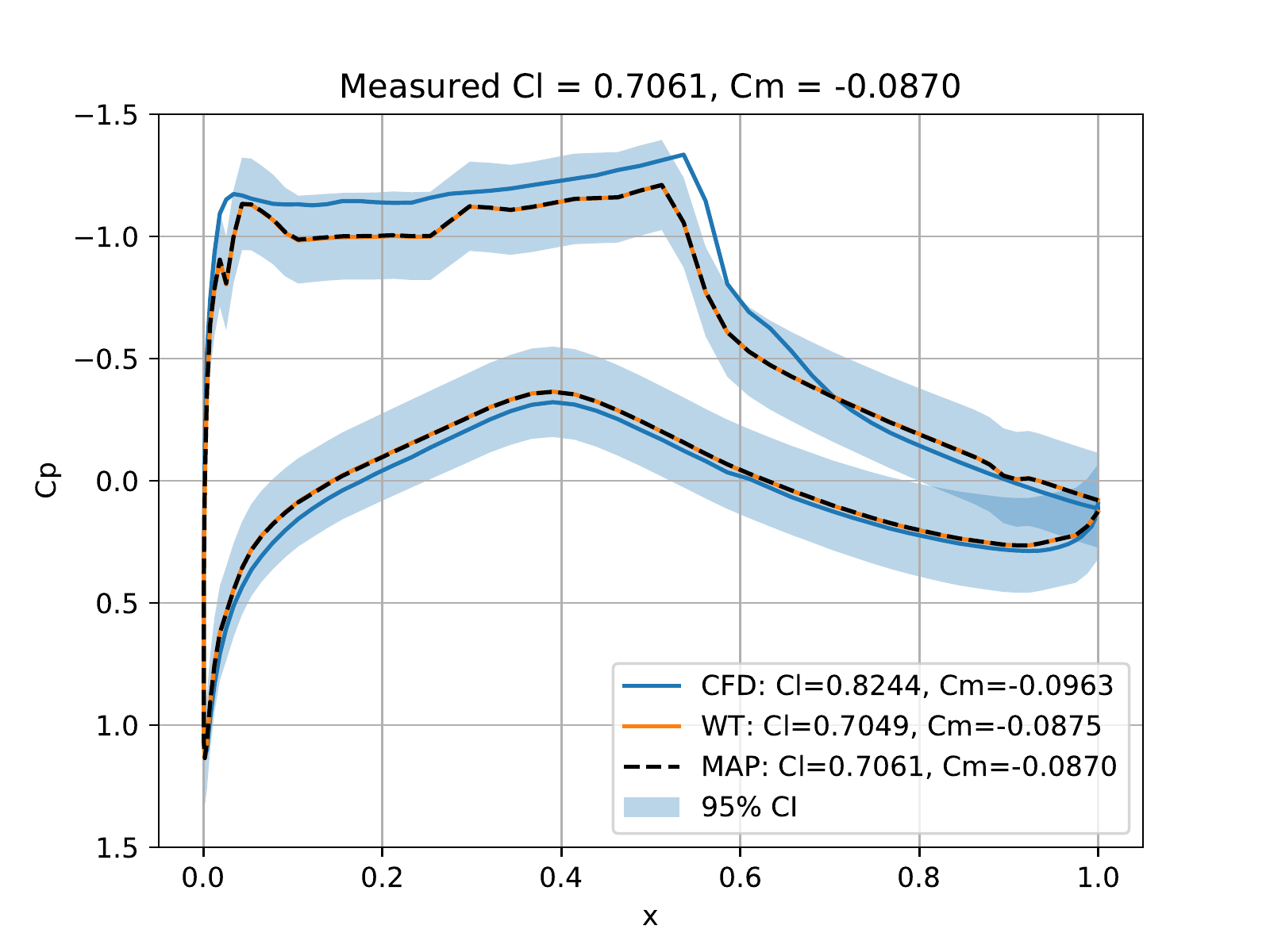}
        \caption{MAP with confidence bands}
        \label{f:RAE_Case13_MAP}
    \end{subfigure}
    \caption{$C_P$ distribution corresponding to Case-11 ($Mach=0.740$, $Re=2.7M$, $\alpha = 3.19~deg.$)}
    \label{fig:RAE_Case13_Res}
\end{figure}

As a counter example, consider the case shown in Figure~\ref{fig:RAE_Case13_Res} which corresponds to the flight conditions ($Mach=0.740$, $Re=2.7M$, $\alpha = 3.19~deg.$). For this case the QoI's integrated from the wind-tunnel curve (0.7049, -0.0875) match very closely with the measurements (0.7061, -0.087) and as a result the predicted MAP estimate for $C_P$ almost overlaps with the wind-tunnel curve (Fig.~\ref{f:RAE_Case13_MAP}). Therefore in this case, the choice of the parameters play a relatively minor role. 

\subsection{Common Research Model}
\begin{figure}[htb!]
    \centering
        \begin{subfigure}{0.33\linewidth}
    \centering
        \includegraphics[width = 2.0in]{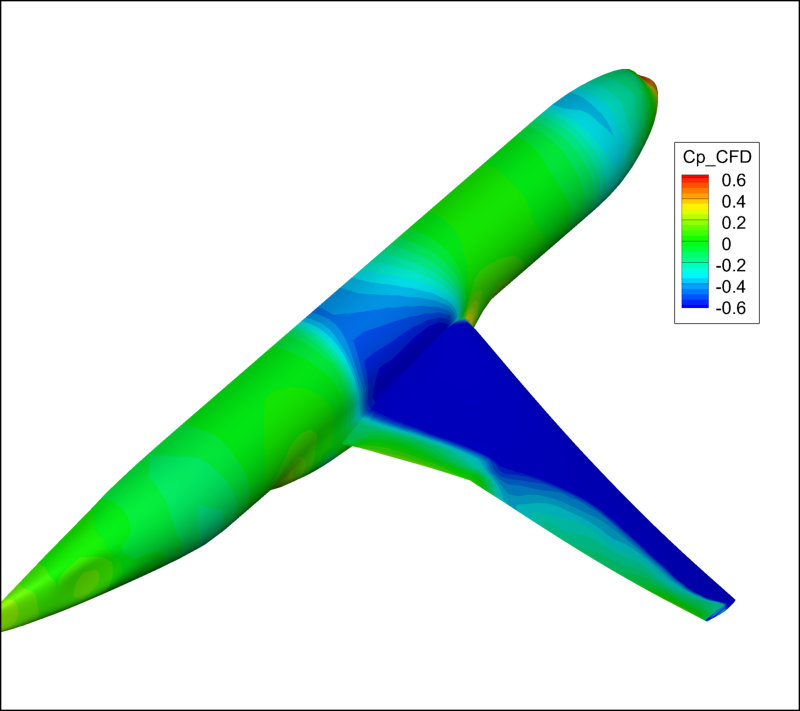}
        \caption{CFD}
        \label{f:3609_CFD}
    \end{subfigure}~
    \begin{subfigure}{0.33\linewidth}
        \centering
        \includegraphics[width = 2.0in]{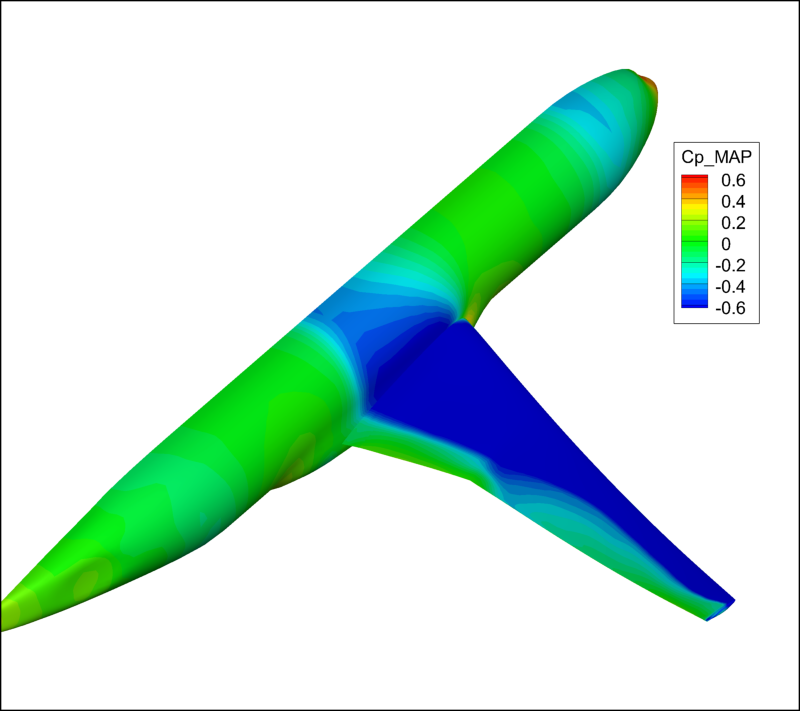}
        \caption{MAP}
        \label{f:3609_MAP}
    \end{subfigure}~
        \begin{subfigure}{0.33\linewidth}
    \centering
        \includegraphics[width = 2.0in]{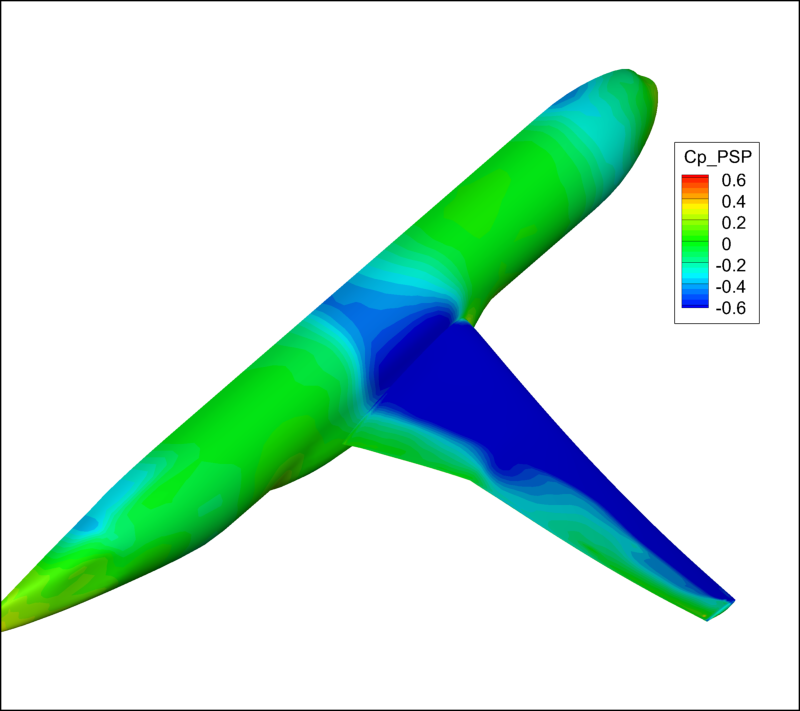}
        \caption{PSP}
        \label{f:3609_PSP}
    \end{subfigure}\\
    \begin{subfigure}{0.33\linewidth}
    \centering
        \includegraphics[width = 2.3in]{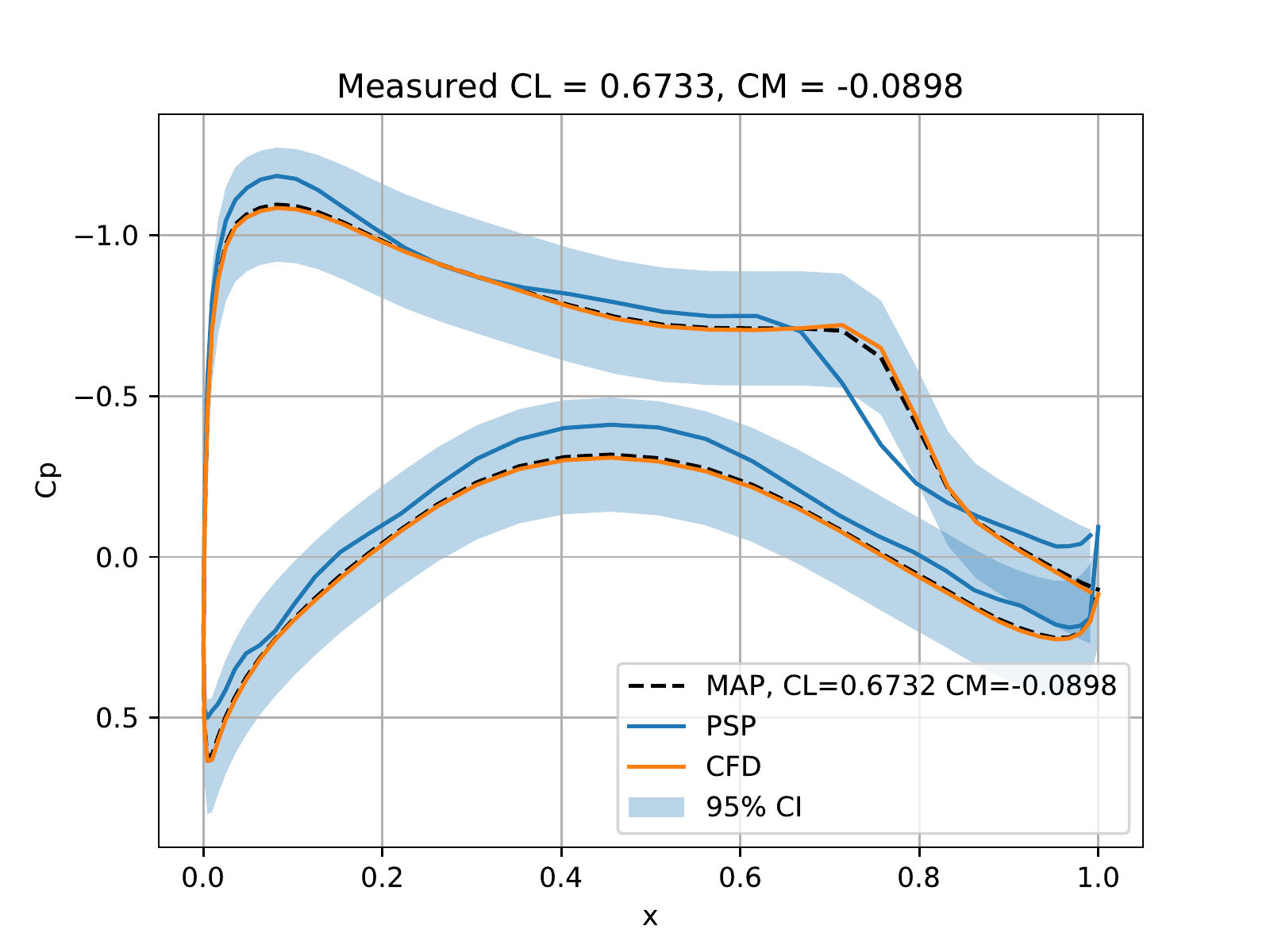}
        \caption{20\% Span}
        \label{f:3609_CFD}
    \end{subfigure}~
    \begin{subfigure}{0.33\linewidth}
        \centering
        \includegraphics[width = 2.3in]{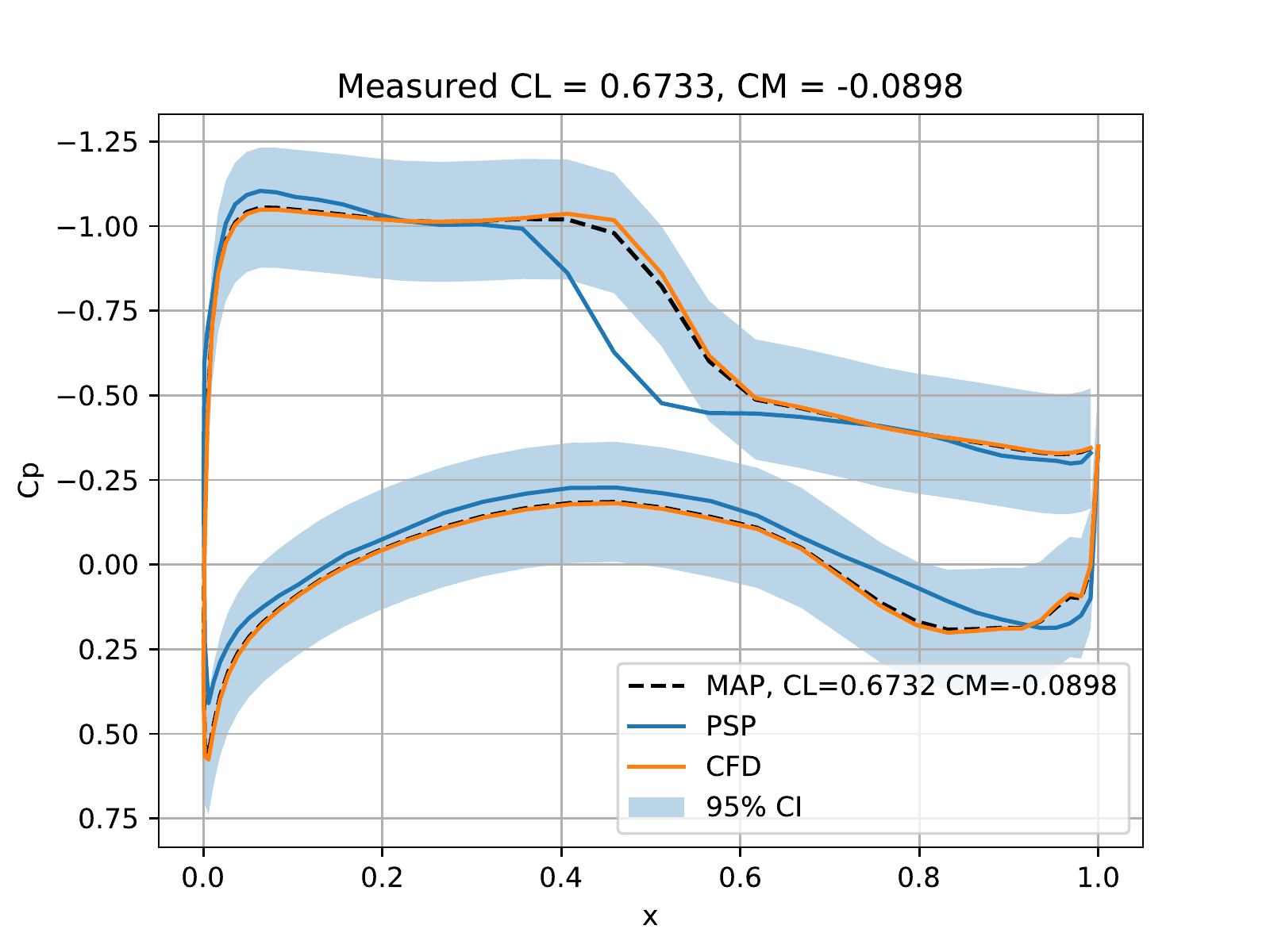}
        \caption{50\% Span}
        \label{f:3609_MAP}
    \end{subfigure}~
        \begin{subfigure}{0.33\linewidth}
    \centering
        \includegraphics[width = 2.3in]{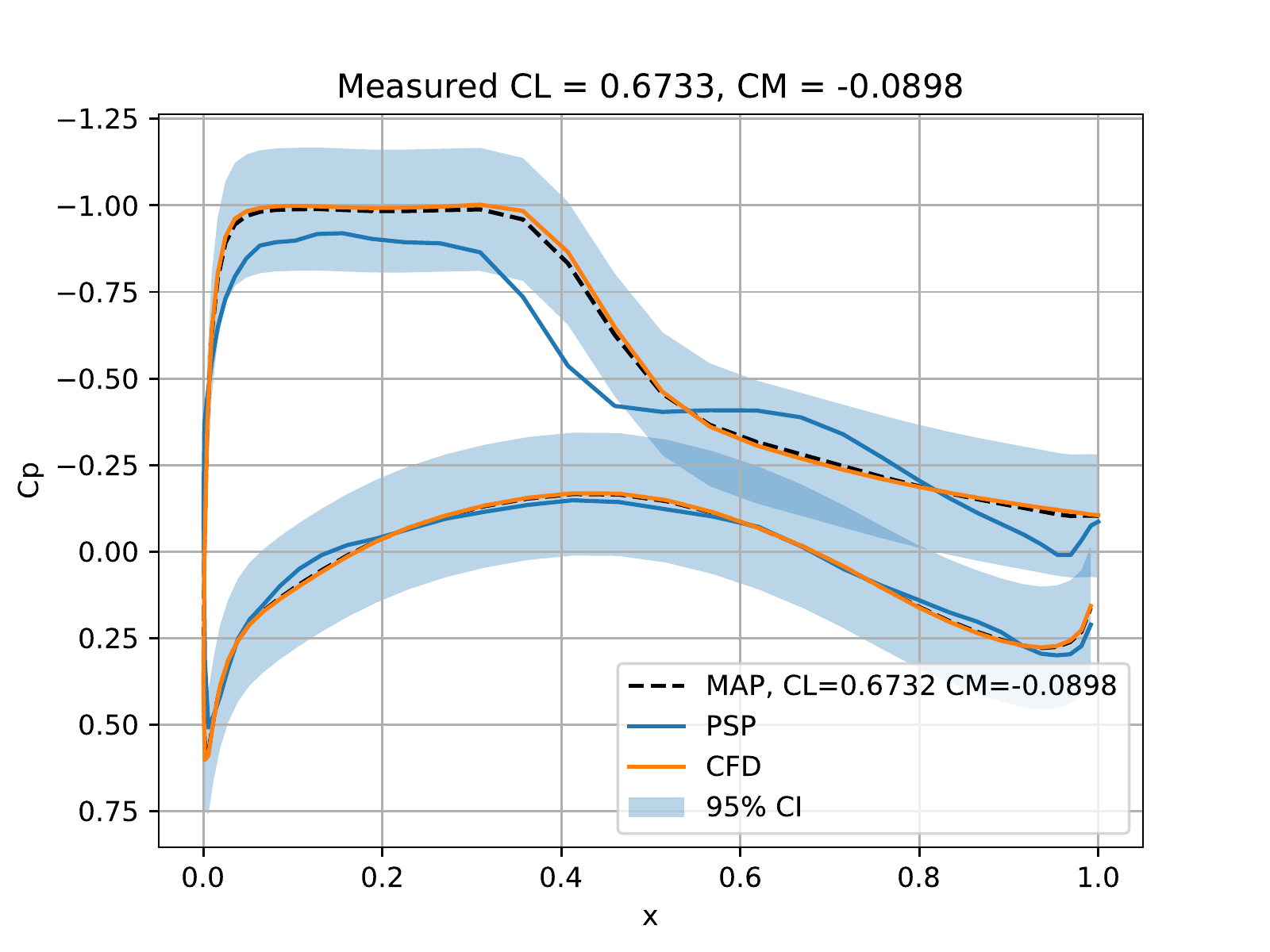}
        \caption{80\% Span}
        \label{f:3609_PSP}
    \end{subfigure}
    \caption{$Mach=0.87$, $Re=5M$, $\alpha = 4.0~deg.$}
    \label{f:CRM_Case3609}
\end{figure}

\begin{figure}[htb!]
    \centering
        \begin{subfigure}{0.33\linewidth}
    \centering
        \includegraphics[width = 2.0in]{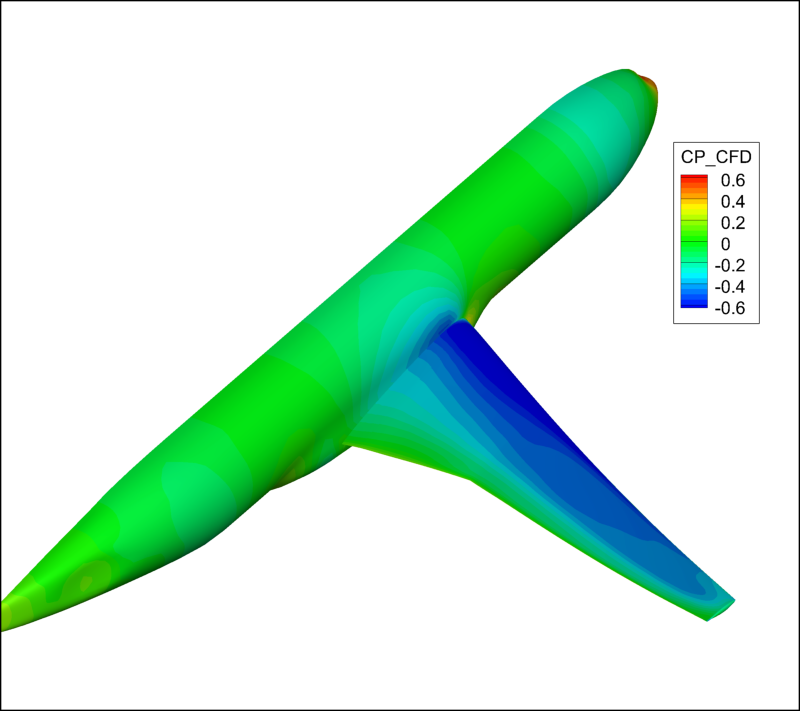}
        \caption{CFD}
        \label{f:3404_CFD}
    \end{subfigure}~
    \begin{subfigure}{0.33\linewidth}
        \centering
        \includegraphics[width = 2.0in]{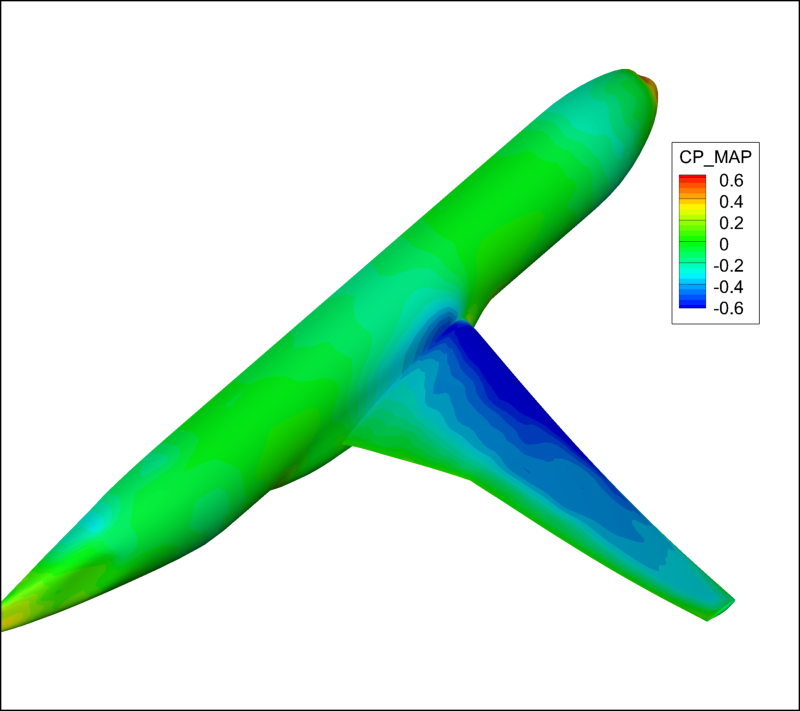}
        \caption{MAP}
        \label{f:3404_MAP}
    \end{subfigure}~
        \begin{subfigure}{0.33\linewidth}
    \centering
        \includegraphics[width = 2.0in]{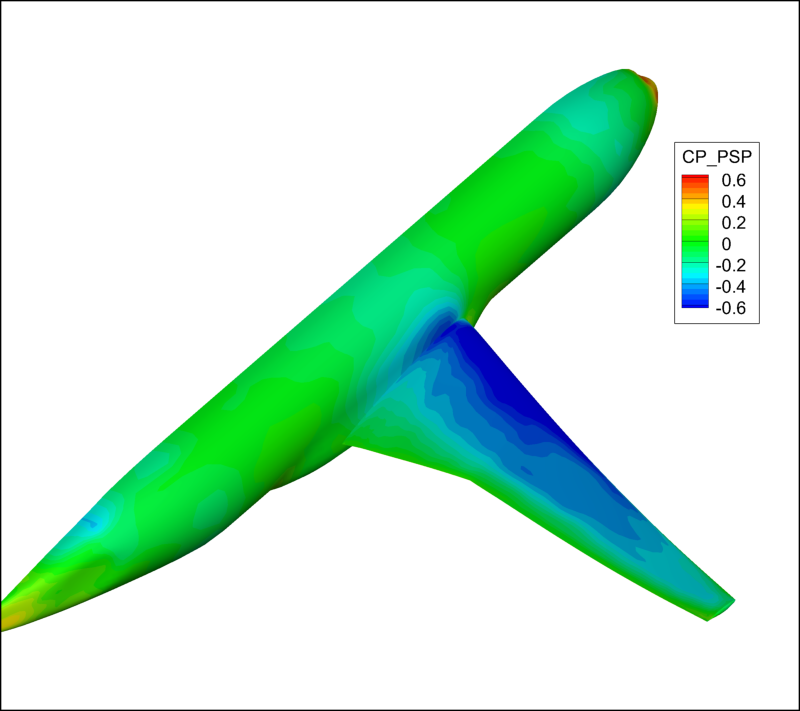}
        \caption{PSP}
        \label{f:3404_PSP}
    \end{subfigure}\\
    \begin{subfigure}{0.33\linewidth}
    \centering
        \includegraphics[width = 2.3in]{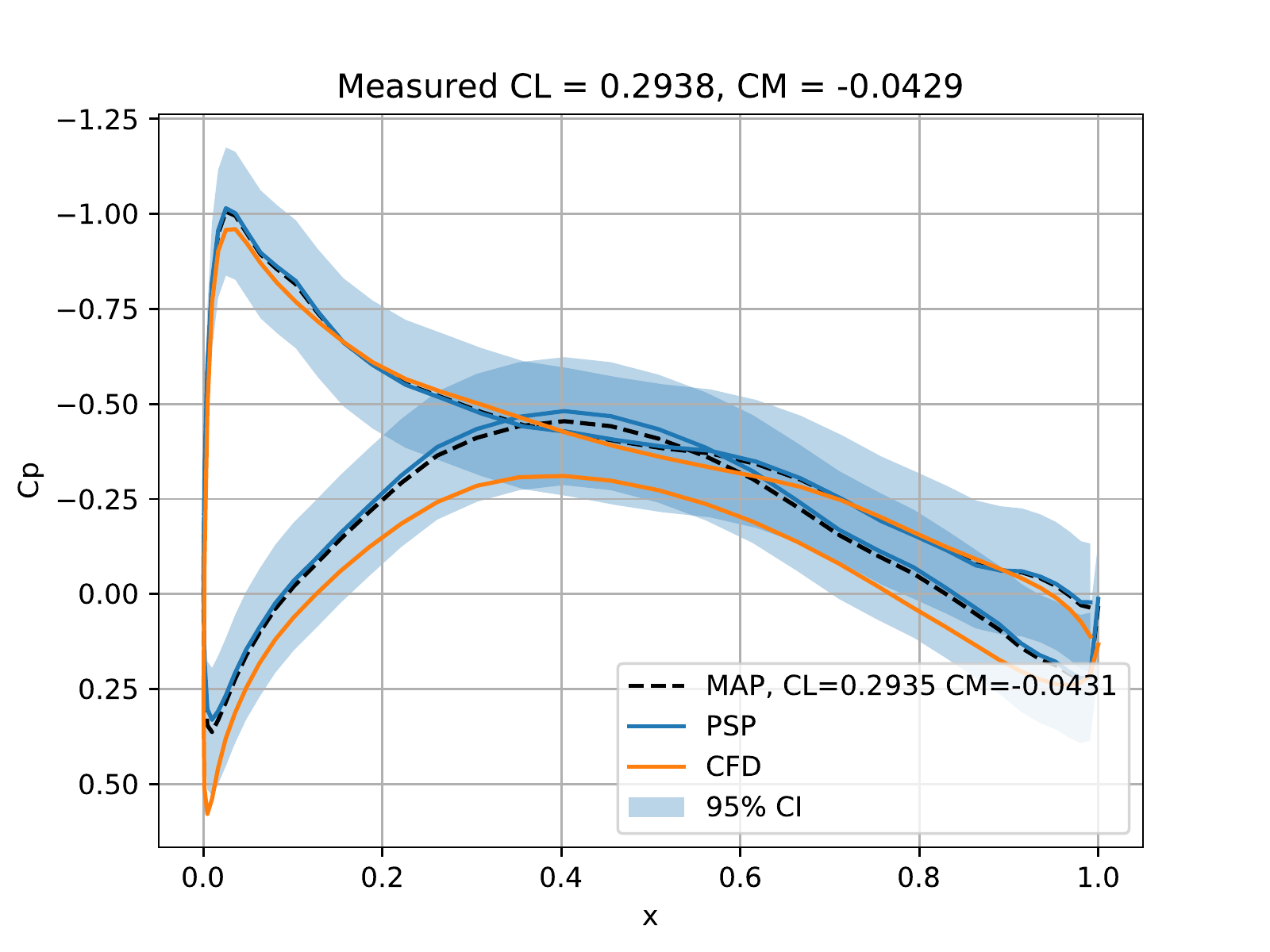}
        \caption{20\% Span}
        \label{f:3404_CFD}
    \end{subfigure}~
    \begin{subfigure}{0.33\linewidth}
        \centering
        \includegraphics[width = 2.3in]{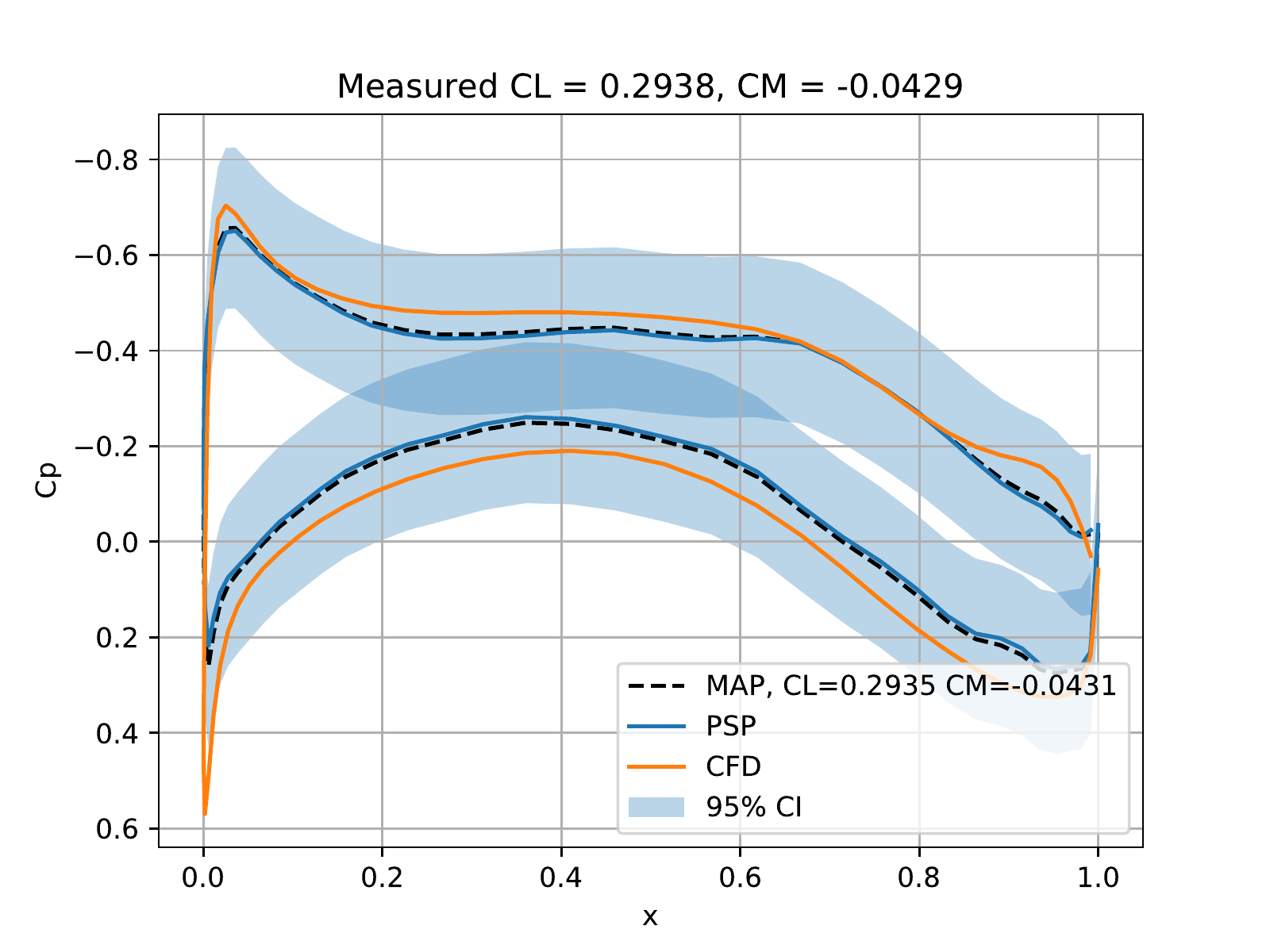}
        \caption{50\% Span}
        \label{f:3404_MAP}
    \end{subfigure}~
        \begin{subfigure}{0.33\linewidth}
    \centering
        \includegraphics[width = 2.3in]{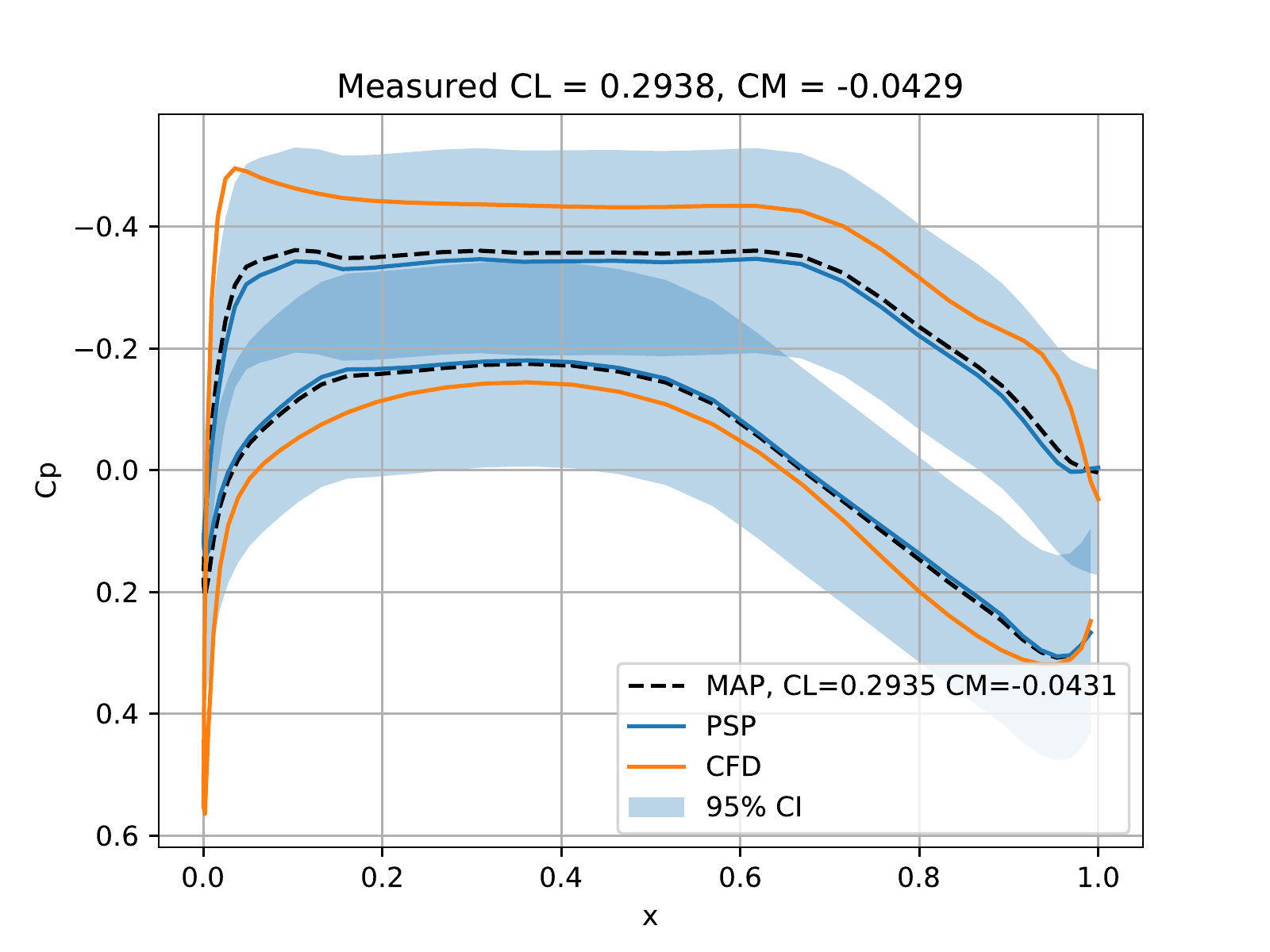}
        \caption{80\% Span}
        \label{f:3404_PSP}
    \end{subfigure}
    \caption{$M=0.7$, $Re=5M$, $\alpha = 1.5$}
    \label{f:CRM_Case4}
\end{figure}

\begin{figure}[htb!]
    \centering
        \begin{subfigure}{0.33\linewidth}
    \centering
        \includegraphics[width = 2.0in]{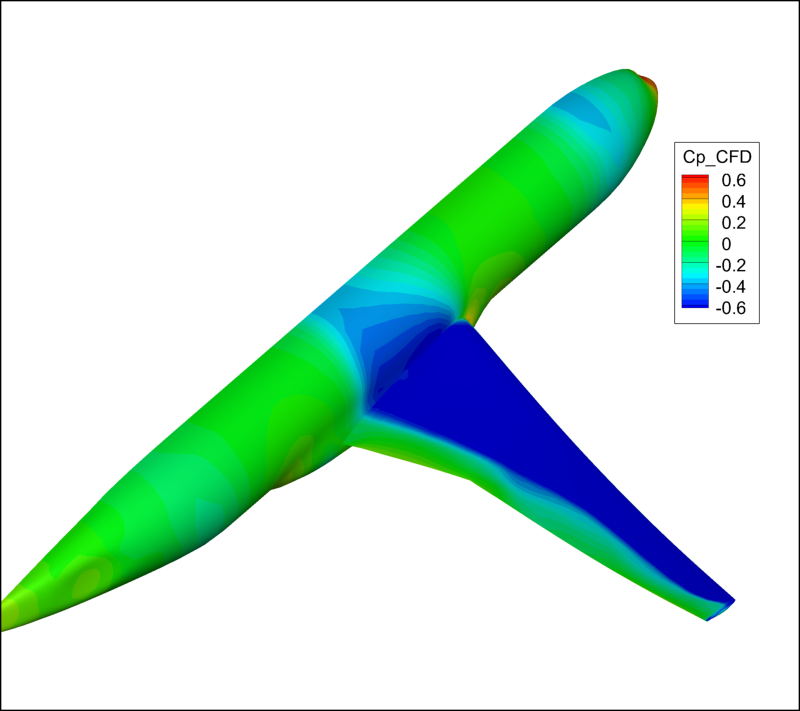}
        \caption{CFD}
        \label{f:3607_CFD}
    \end{subfigure}~
    \begin{subfigure}{0.33\linewidth}
        \centering
        \includegraphics[width = 2.0in]{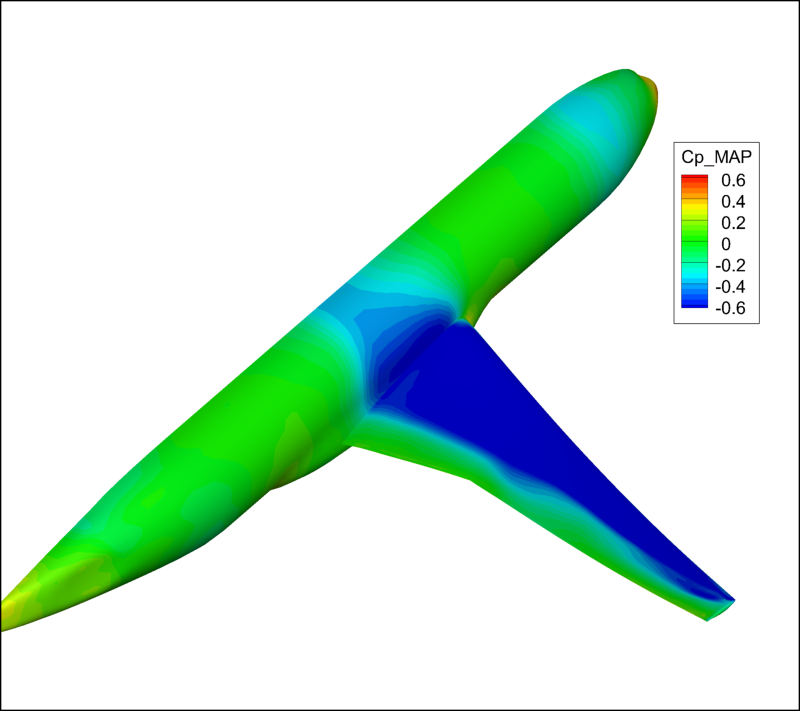}
        \caption{MAP}
        \label{f:3607_MAP}
    \end{subfigure}~
        \begin{subfigure}{0.33\linewidth}
    \centering
        \includegraphics[width = 2.0in]{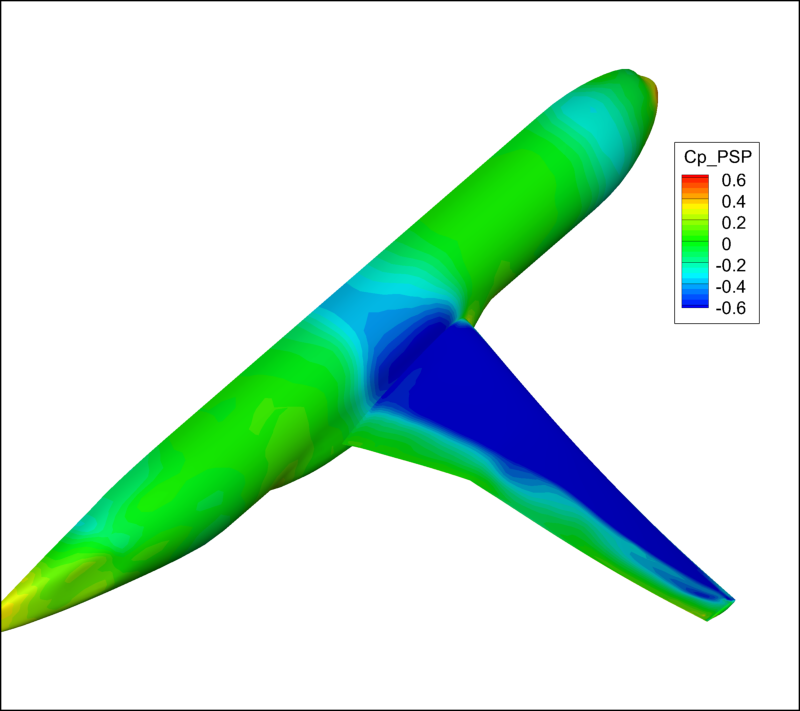}
        \caption{PSP}
        \label{f:3607_PSP}
    \end{subfigure}\\
    \begin{subfigure}{0.33\linewidth}
    \centering
        \includegraphics[width = 2.3in]{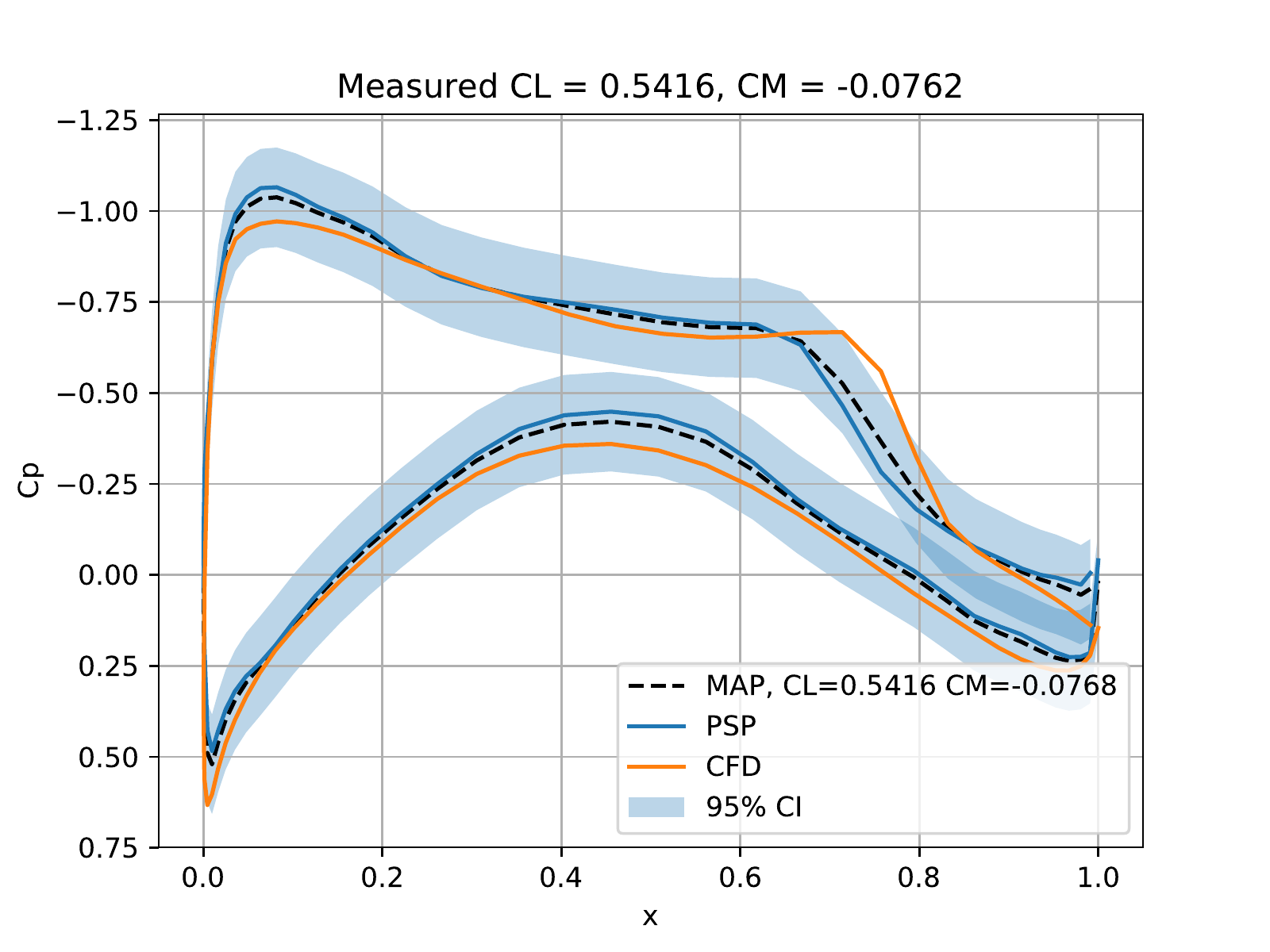}
        \caption{20\% Span}
        \label{f:3607_CFD}
    \end{subfigure}~
    \begin{subfigure}{0.33\linewidth}
        \centering
        \includegraphics[width = 2.3in]{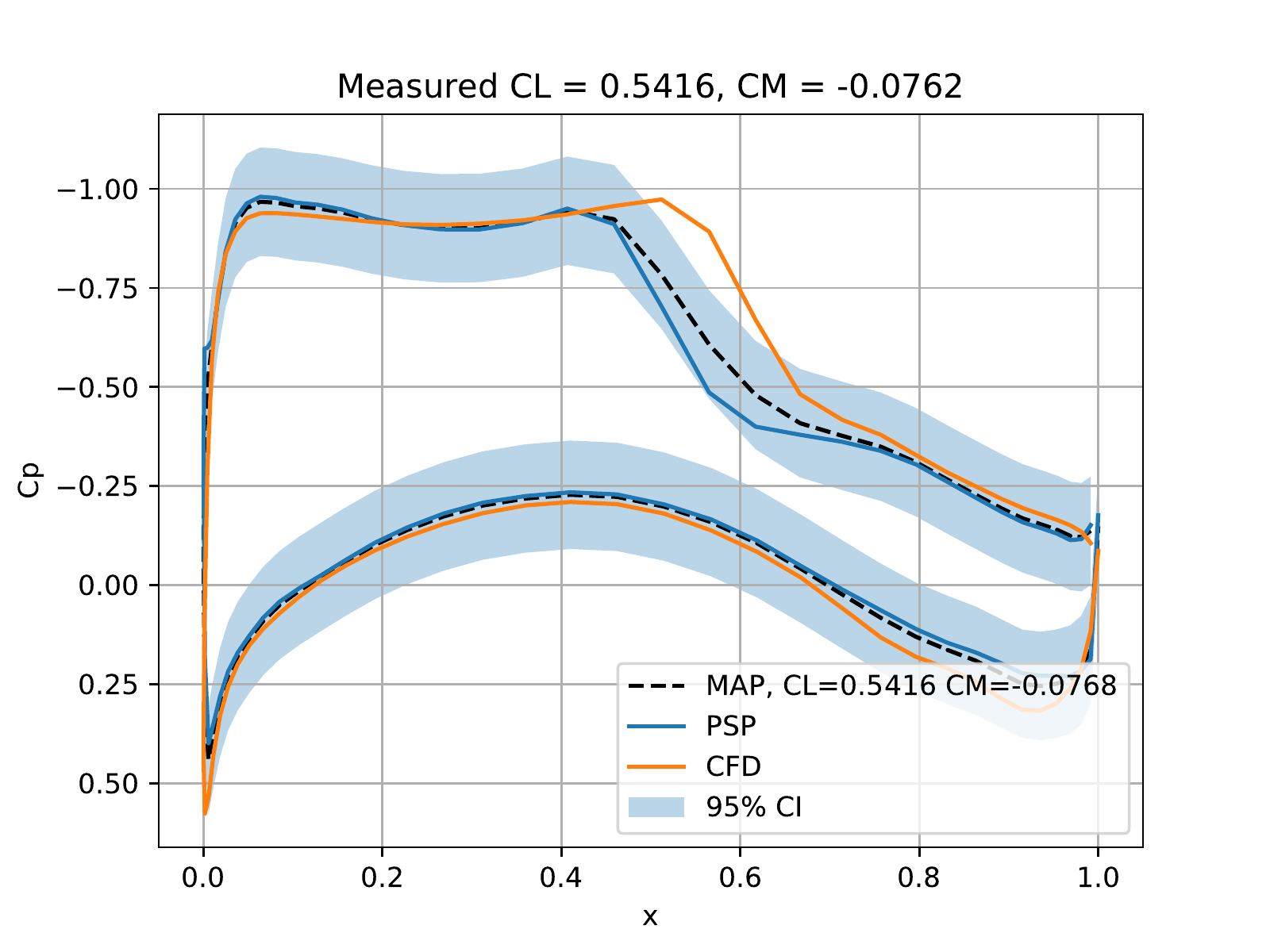}
        \caption{50\% Span}
        \label{f:3607_MAP}
    \end{subfigure}~
        \begin{subfigure}{0.33\linewidth}
    \centering
        \includegraphics[width = 2.3in]{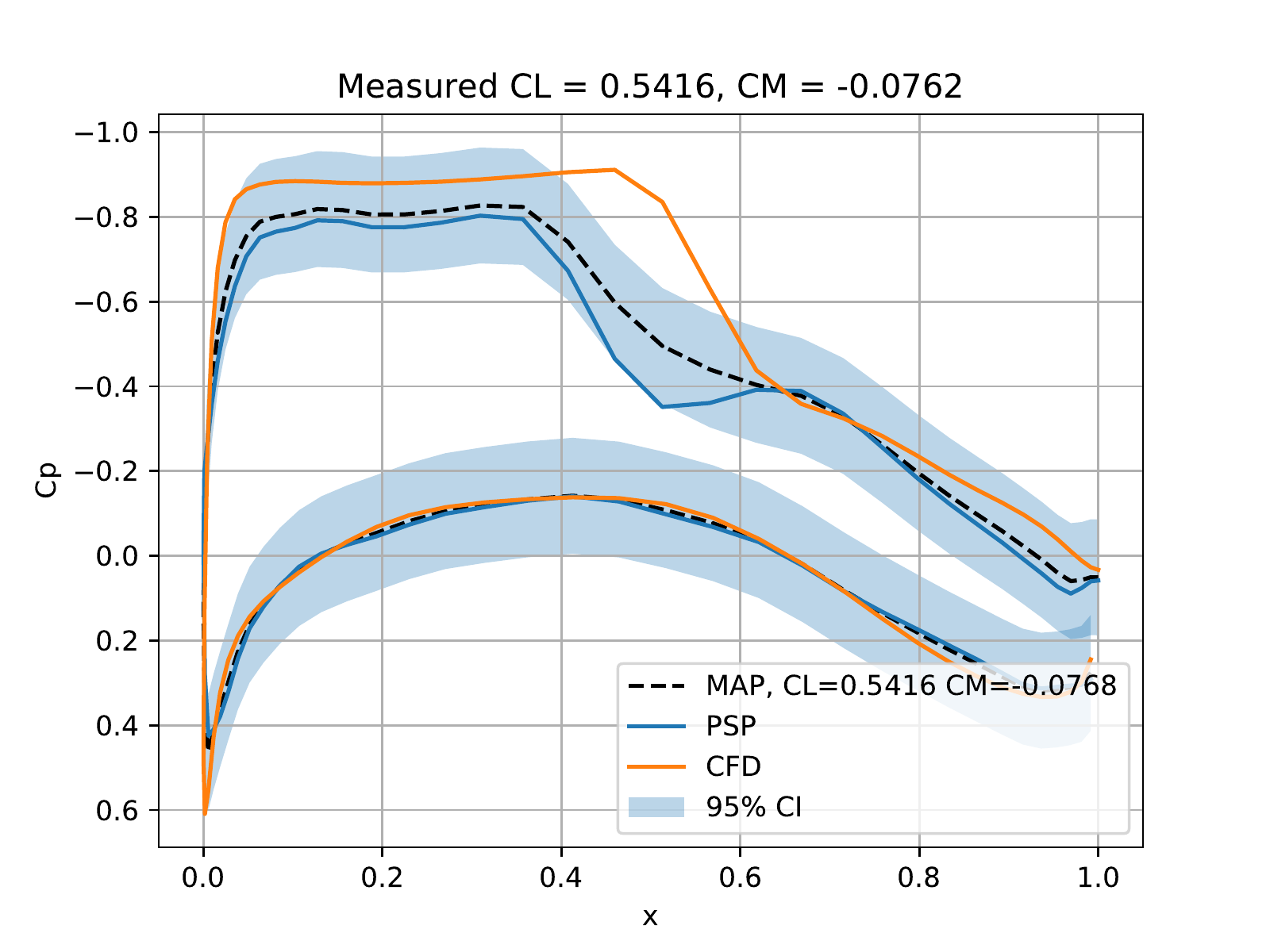}
        \caption{80\% Span}
        \label{f:3607_PSP}
    \end{subfigure}
    \caption{$Mach=0.87$, $Re=5M$, $\alpha = 3.0~deg.$}
    \label{f:CRM_Case3607}
\end{figure}

For the CRM test case, the approach is repeated and results are presented for a select few cases as follows. The hyperparameters are set to $\tau^2 = 10^{-6},~\sigma_1^2 = 10^{-2},~\sigma_2^2 = 10^{-2}$ and $\ell = 0.01$. Figure~\ref{f:CRM_Case3609} shows results for $Mach=0.87$ and $\alpha = 4.0~deg.$ where the CFD predicts a relatively stronger shock compared to PSP measurements. Incidentally, the MAP estimate matches more closely with the CFD results compared to PSP. Furthermore, the PSP curve falls outside the 95\% confidence region of the MAP in the vicinity of the shock, suggesting that the PSP measurements in this region are less reliable. While a different choice of parameter values will change the size of the confidence bands, the results clearly demonstrate that the CFD results agree more closely with the measurements than PSP for this specific case. A contrasting behavior is observed in Figure~\ref{f:CRM_Case4} where the MAP estimate matches more closely with the PSP results compared to CFD. Finally, in the case of Figure~\ref{f:CRM_Case3607}, the MAP estimate lies between the CFD and PSP estimates where the CFD predictions fall outside of the 95\% confidence bounds around the MAP estimate, in the vicinity of the shock. These results demonstrate that the proposed approach allows one to quantify the spatially distributed confidence in the fused data by accounting for the uncertainties in the original datasets. We claim that this makes the fused data more useful that the original datasets and one may then use the inferred posterior distributions to construct surrogate models of the fields using model reduction methods~\cite{renganathan2018koopman, renganathan2018methodology} for instance.

\begin{figure}[htb!]
    \centering
        \begin{subfigure}{0.33\linewidth}
    \centering
        \includegraphics[width = 2.0in]{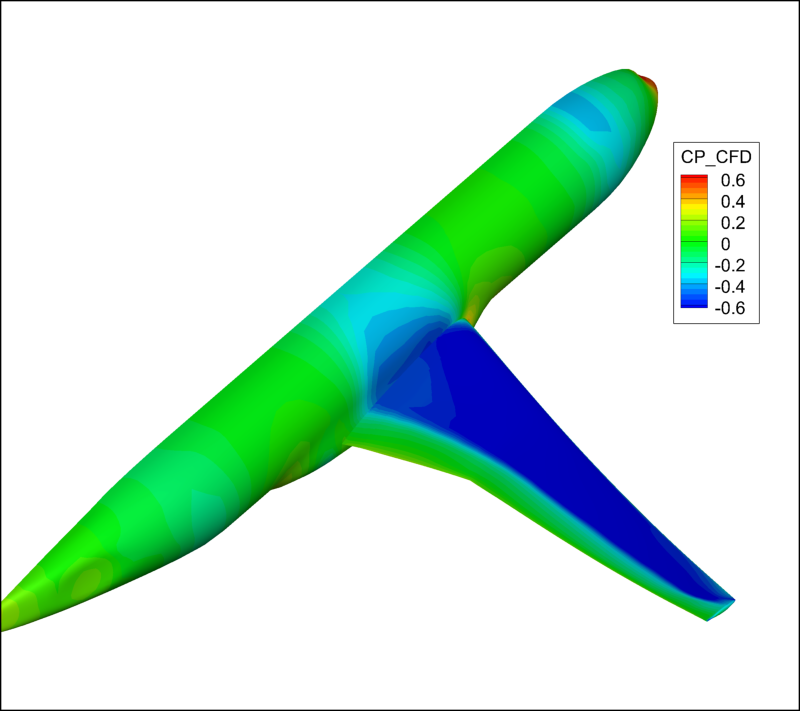}
        \caption{CFD}
        \label{f:3604_CFD}
    \end{subfigure}~
    \begin{subfigure}{0.33\linewidth}
        \centering
        \includegraphics[width = 2.0in]{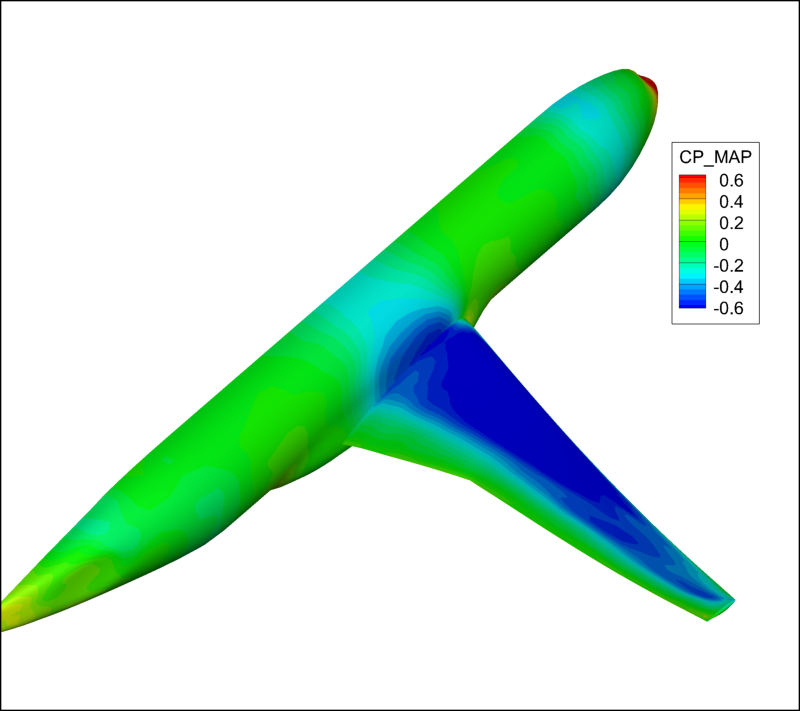}
        \caption{MAP}
        \label{f:3604_MAP}
    \end{subfigure}~
        \begin{subfigure}{0.33\linewidth}
    \centering
        \includegraphics[width = 2.0in]{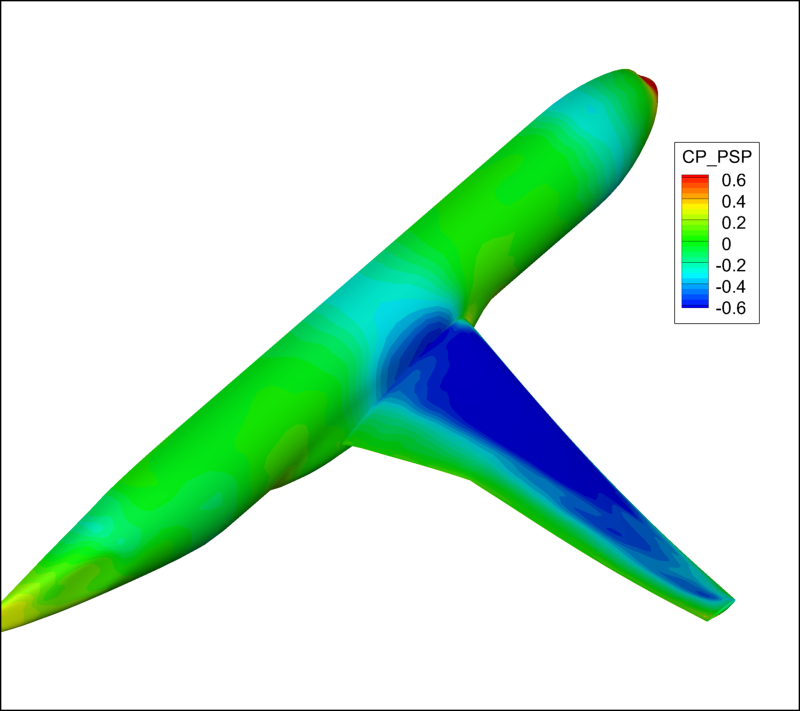}
        \caption{PSP}
        \label{f:3604_PSP}
    \end{subfigure}\\
    \begin{subfigure}{0.33\linewidth}
    \centering
        \includegraphics[width = 2.3in]{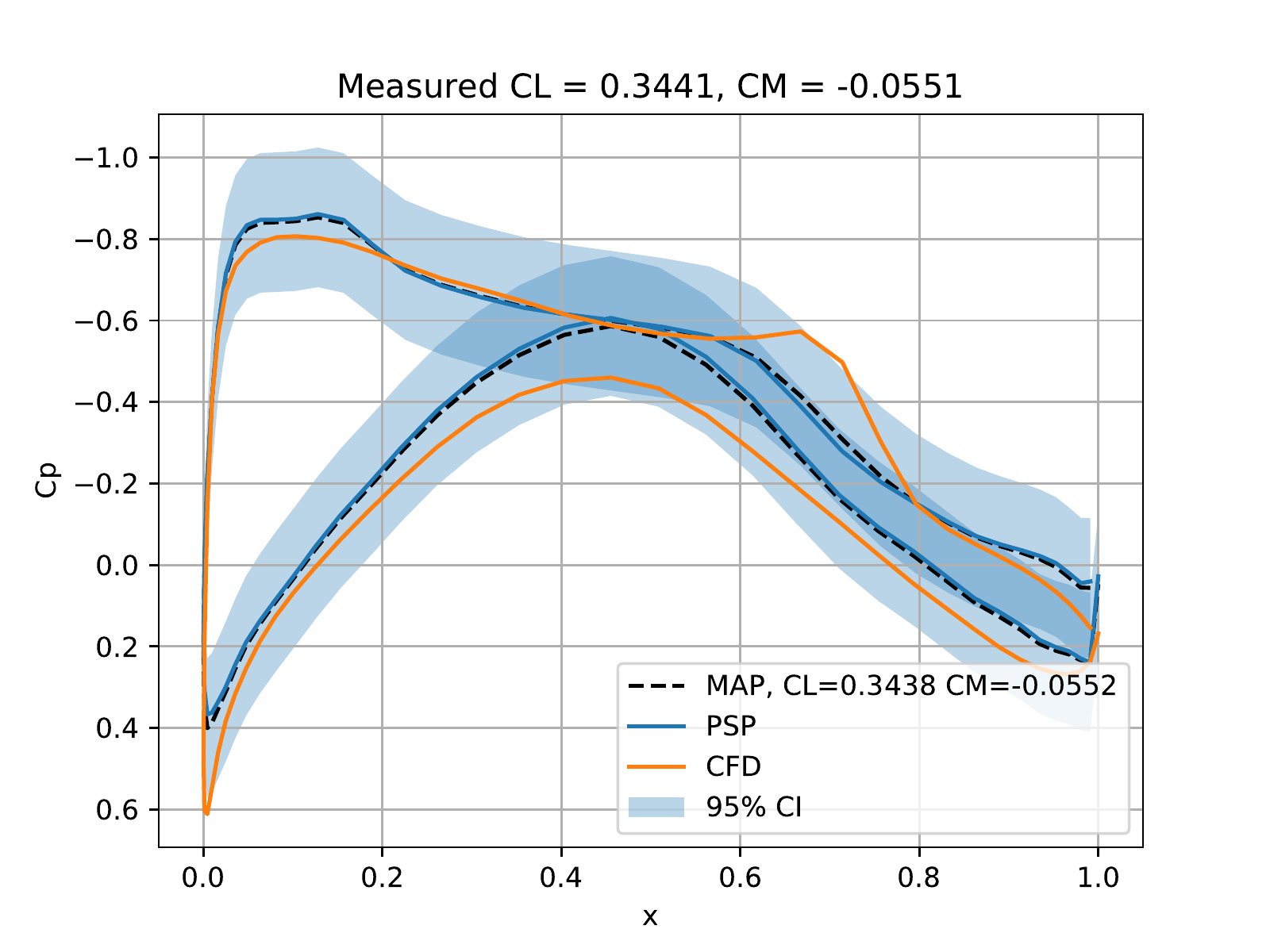}
        \caption{20\% Span}
        \label{f:3604_CFD}
    \end{subfigure}~
    \begin{subfigure}{0.33\linewidth}
        \centering
        \includegraphics[width = 2.3in]{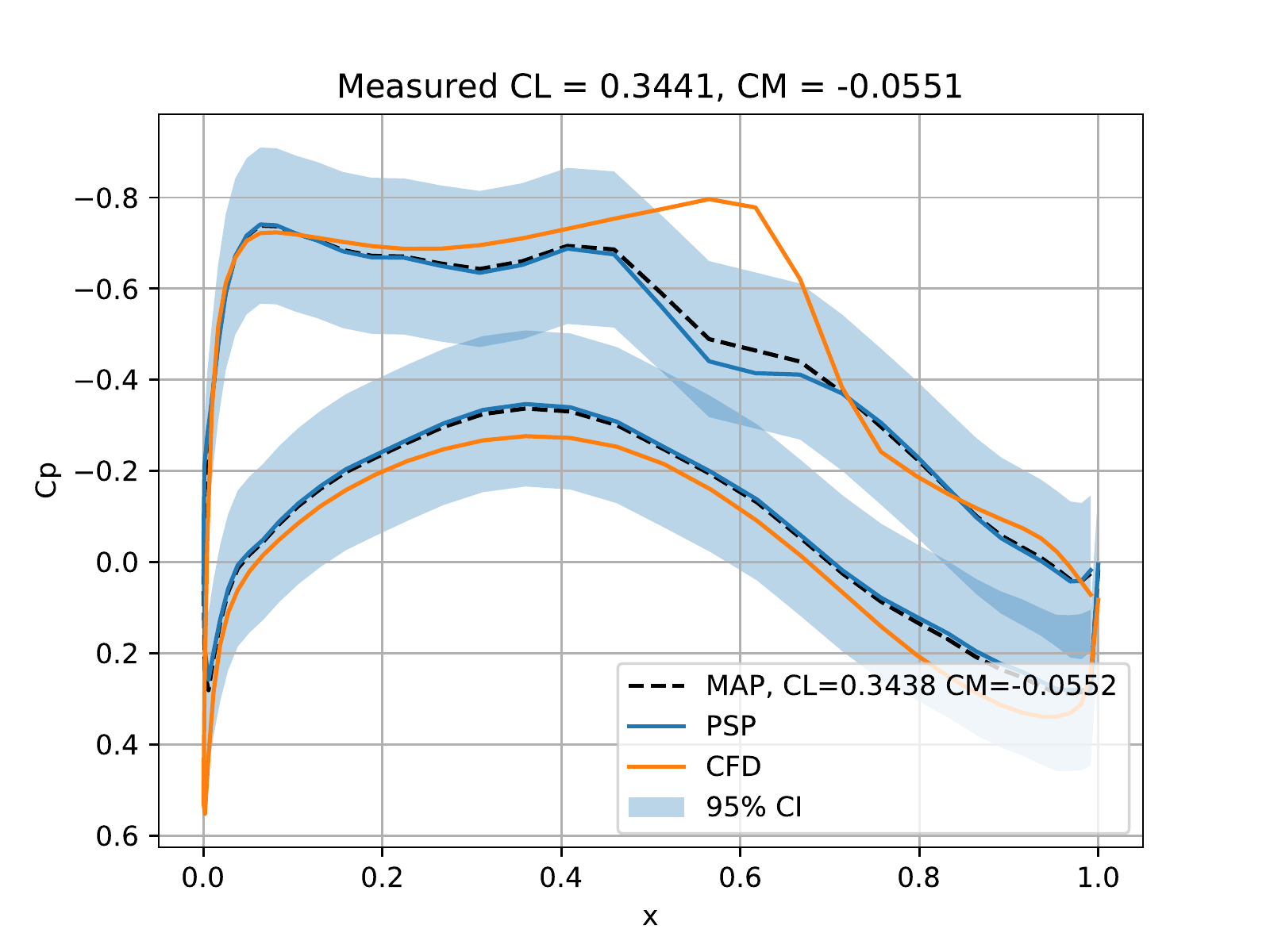}
        \caption{50\% Span}
        \label{f:3604_MAP}
    \end{subfigure}~
        \begin{subfigure}{0.33\linewidth}
    \centering
        \includegraphics[width = 2.3in]{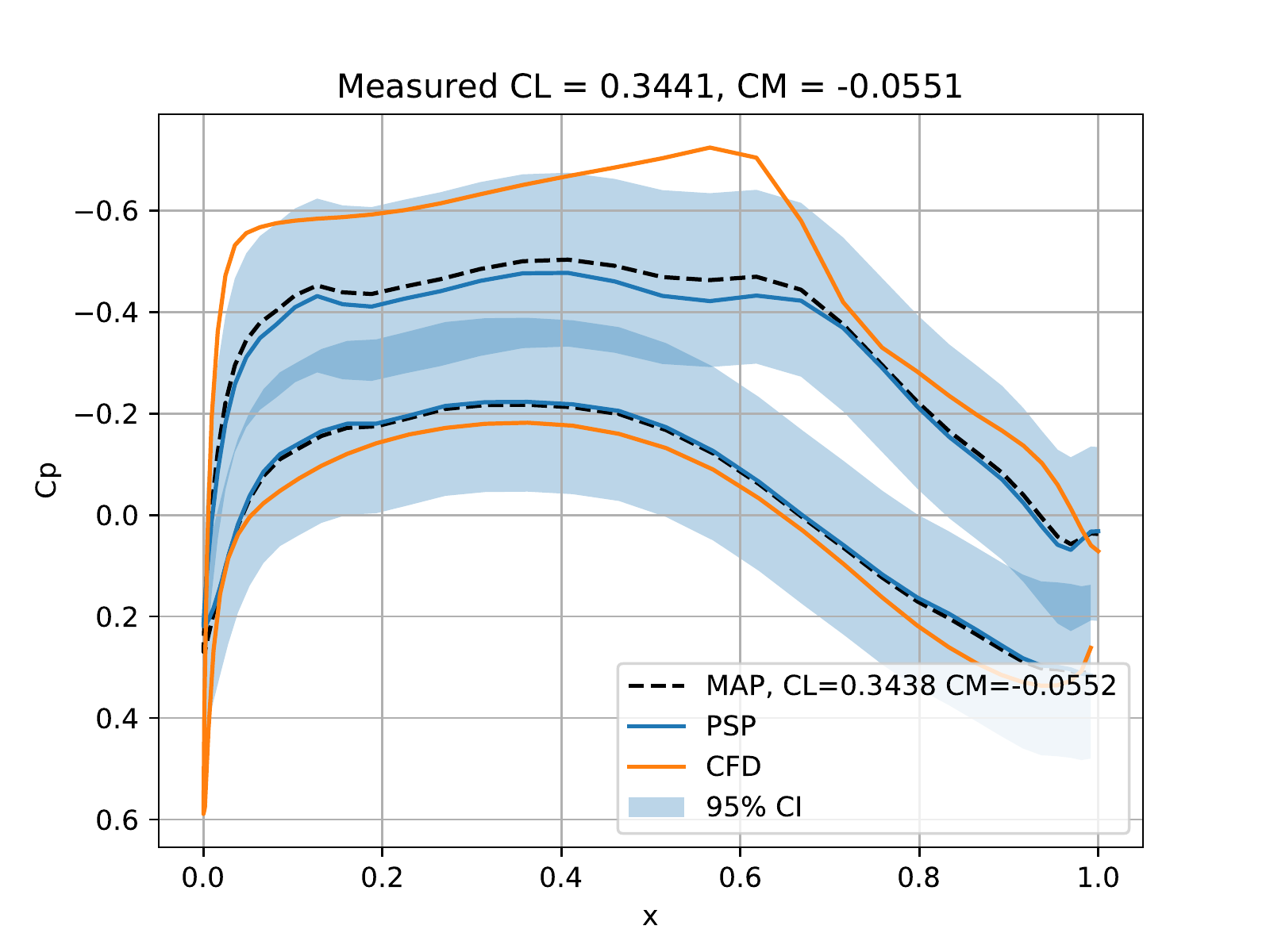}
        \caption{80\% Span}
        \label{f:3604_PSP}
    \end{subfigure}
    \caption{$M=0.87$, $Re=5M$, $\alpha = 1.5$. $\theta$ estimated from \eqref{e:theta_est}}
    \label{f:CRM_Case12}
\end{figure}

\begin{figure}[htb!]
    \centering
        \begin{subfigure}{0.33\linewidth}
    \centering
        \includegraphics[width = 2.0in]{figures/CRM/3604/Iso_CFD.png}
        \caption{CFD}
        \label{f:3604_CFD}
    \end{subfigure}~
    \begin{subfigure}{0.33\linewidth}
        \centering
        \includegraphics[width = 2.0in]{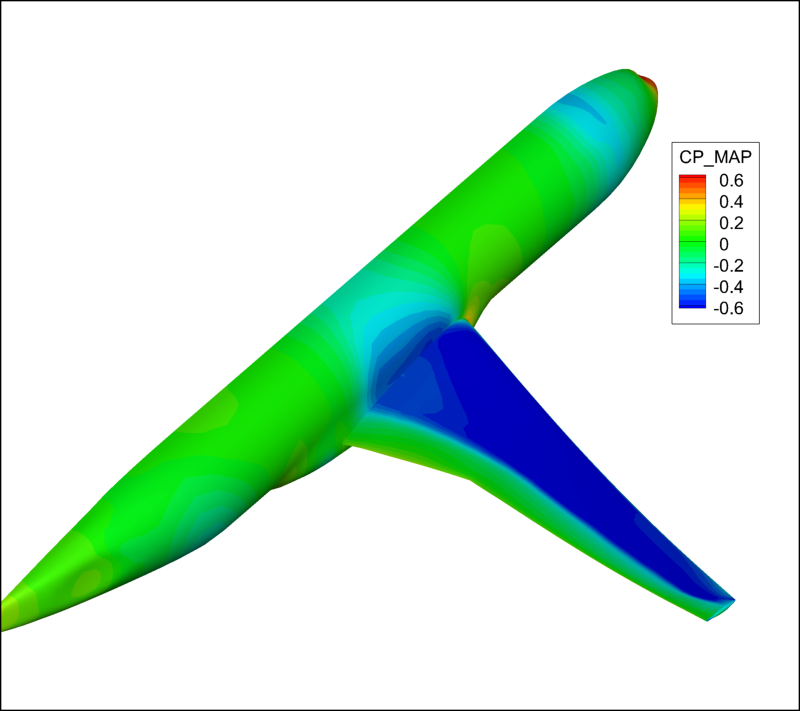}
        \caption{MAP}
        \label{f:3604_theta=0_MAP}
    \end{subfigure}~
        \begin{subfigure}{0.33\linewidth}
    \centering
        \includegraphics[width = 2.0in]{figures/CRM/3604/Iso_WTT.png}
        \caption{PSP}
        \label{f:3604_PSP}
    \end{subfigure}\\
    \begin{subfigure}{0.33\linewidth}
    \centering
        \includegraphics[width = 2.3in]{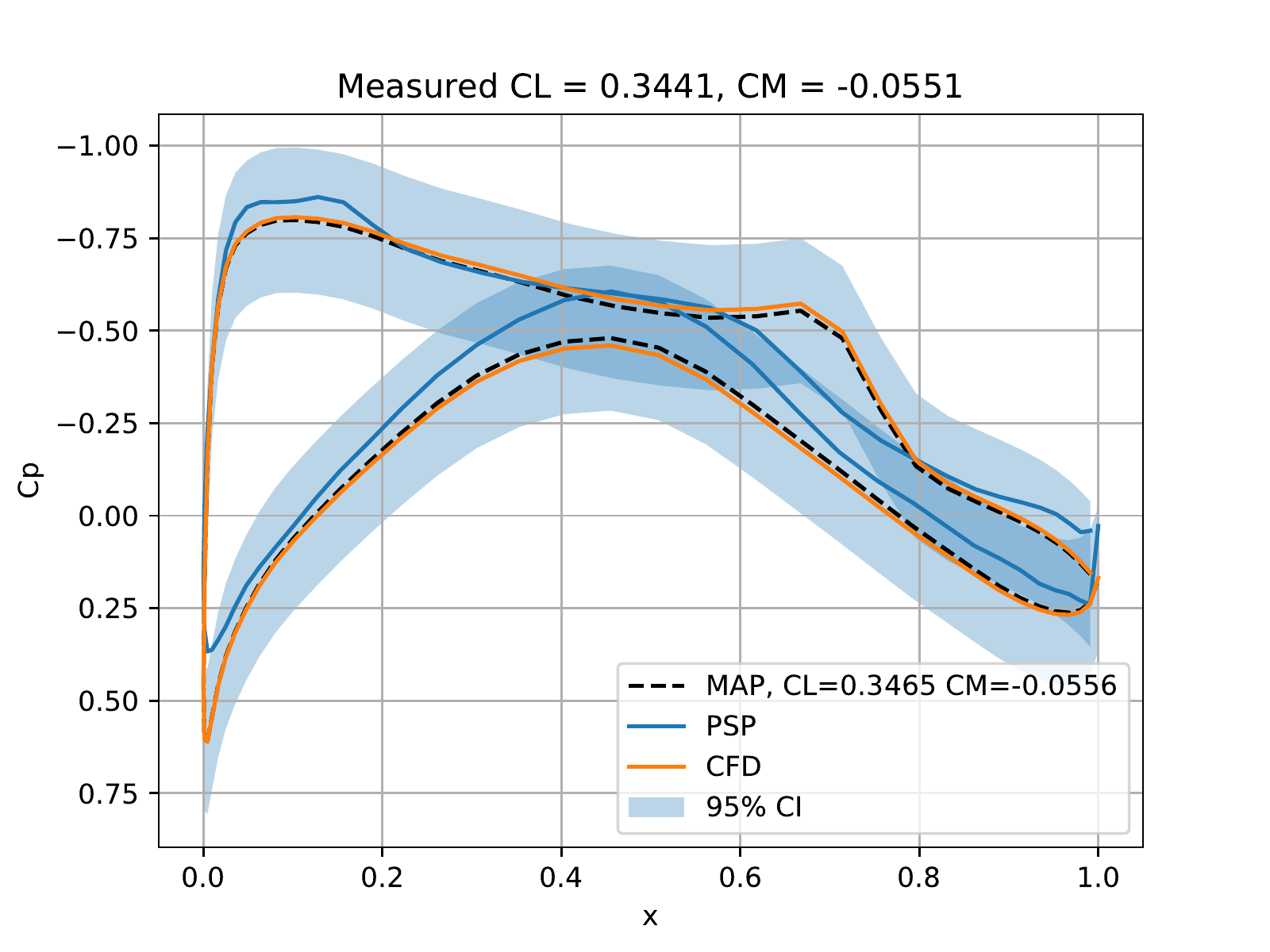}
        \caption{20\% Span}
        \label{f:3604_line_cp1}
    \end{subfigure}~
    \begin{subfigure}{0.33\linewidth}
        \centering
        \includegraphics[width = 2.3in]{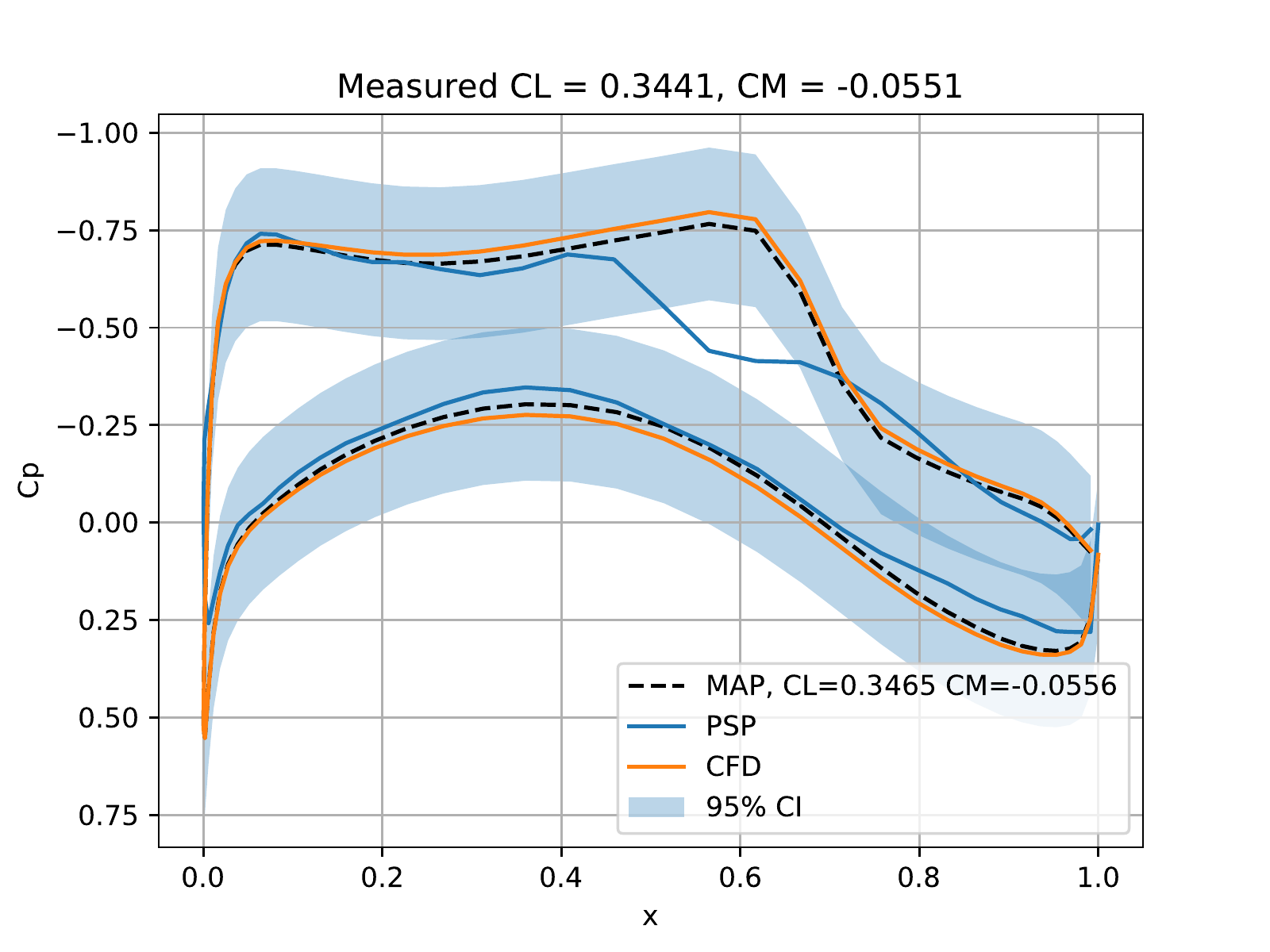}
        \caption{50\% Span}
        \label{f:3604_line_cp2}
    \end{subfigure}~
        \begin{subfigure}{0.33\linewidth}
    \centering
        \includegraphics[width = 2.3in]{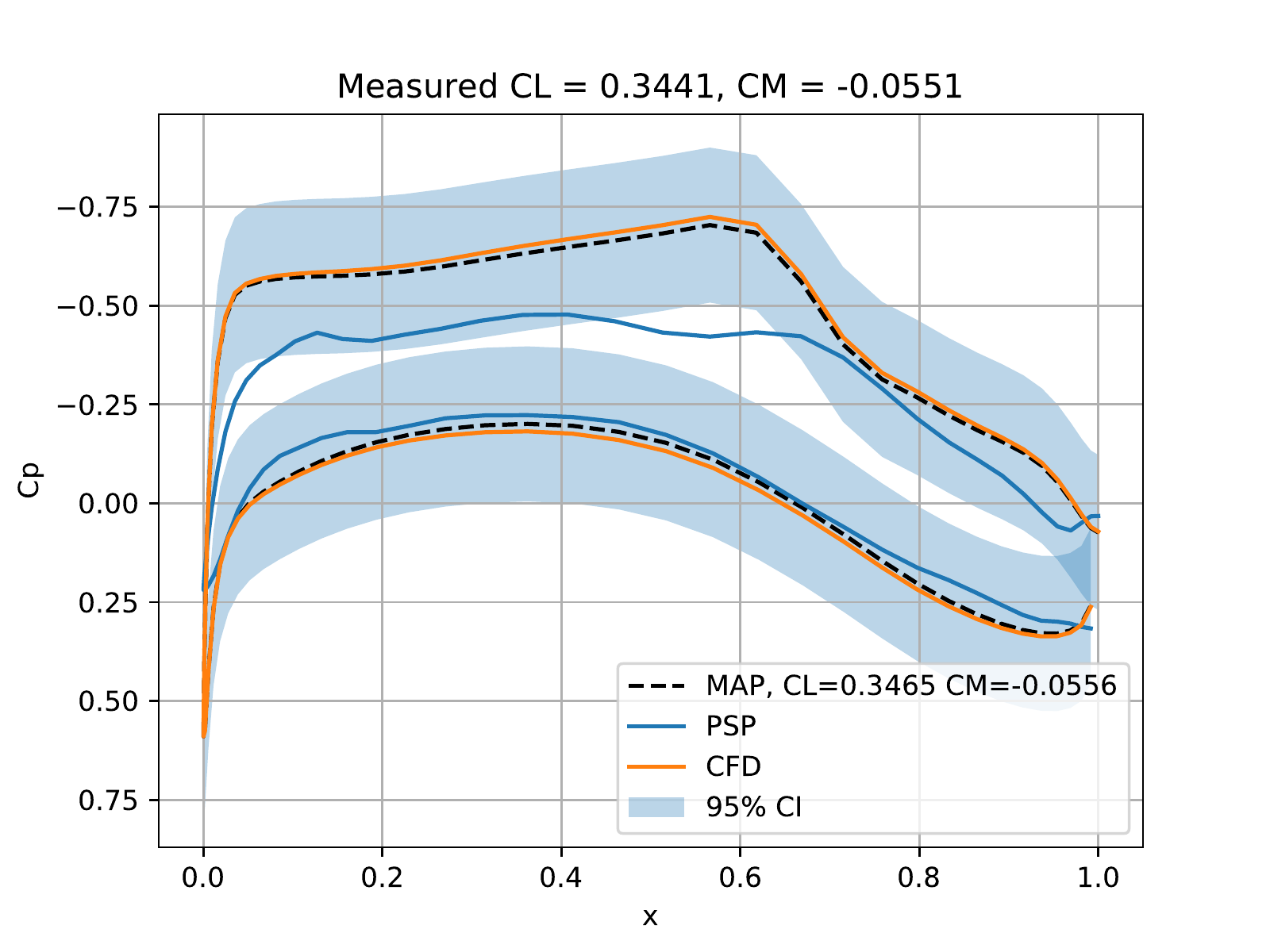}
        \caption{80\% Span}
        \label{f:3604_line_cp3}
    \end{subfigure}
    \caption{$Mach=0.87$, $Re=5M$, $\alpha = 1.5~deg.$. $\theta$ set to $0$ to express greater prior belief on the CFD predictions}
    \label{f:CRM_Case12_theta=0}
\end{figure}

An alternative way to use the proposed method is to invoke domain knowledge in the prior specification. For instance, if domain knowledge recommends more trust in one amongst the PSP or CFD data, then the $\theta$ parameter can be treated as a tuning parameter whose value can be directly set as opposed to estimating it. The plots in Figure~\ref{f:CRM_Case12} represent an operating condition ($Mach=0.87,~\alpha=1.5~deg.$) where a strong shock wave is expected on the wing. While this is predicted by the CFD simulations, the PSP measurements show rather shallow gradients in  the coefficient of pressure (Figs.~\ref{f:3604_line_cp1} through \ref{f:3604_line_cp3}). Therefore the user might choose to put more prior belief in the CFD predictions by setting $\theta = 0$ in \eqref{e:prior_mean}. In this case, the method tries to adjust the CFD predictions so as match the measured QoI which results in a MAP estimate that looks relatively similar to the actual CFD prediction. This is demonstrated in Figure~\ref{f:CRM_Case12_theta=0}. 
It should be noted that the MAP estimate is itself non-unique - i.e. there can be more than one MAP estimate that agrees with the measurements $\mbf{z}$ equally well. The estimate depends on the prior specification as illustrated through Figures~\ref{f:CRM_Case12} and \ref{f:CRM_Case12_theta=0}; re-iterating the ill-posedness issue of the inverse problem mentioned in section~\ref{s:Intro}. Therefore, it is critical to specify the prior more judiciously. 

The Table~\ref{tab:CRM_Results} summarizes the integrated QoIs (by evaluating forward model) from the MAP, CFD and PSP $C_P$ distributions and compares them against the measured values. It can be seen that the MAP estimates match the measured values very closely. Therefore, the methodology overall finds the best estimate that agrees with measurements given the multiple sources of data, regularized by the specified prior belief on the estimate. The methodology is computationally efficient since the posterior distributions are analytically derived as opposed to making numerical approximations and requires only approximately 1 minute of wall-clock time to fuse one pair of dataset ($n\approx 10000$) on an Intel 4-core i7 processor with 16gb RAM. This establishes the scalability of the proposed approach for large datasets.

One of the key characteristics of the proposed Bayesian approach is that it assumes that more than one source of the dataset is available for a given operating condition and operates only on this dataset. However, typically windtunnel and CFD campaigns are conducted in batches and the available datasets span a range of operating conditions. In the following section, we propose another solution for the same problem that relies on learning flow features from the entire dataset, unlike the proposed Bayesian approach following which, we show comparisons of both methods.

\begin{table}[htb!]
    \centering
        \caption{Summary of the CRM results in terms of the QoIs ($C_L$ and $C_M$). The data under columns labeled MAP, PSP and CFD contain the QoI's from evaluating the forward model at the $C_P$ distributions. Bold-face entries indicate best match to measurements }
    \begin{tabular}{c|cc|cc|cc|cc}
        \hline
         & \multicolumn{2}{c}{Measurements} & \multicolumn{2}{c}{MAP} & \multicolumn{2}{c}{PSP} & \multicolumn{2}{c}{CFD}\\
         \hline
            $Mach/Re/\alpha$  & $C_L$ & $C_M$ & $C_L$ & $C_M$ & $C_L$ & $C_M$ & $C_L$ & $C_M$ \\
        \hline
         0.87/5m/4.0  & 0.6733 & -0.0898 & \textbf{0.6732} & \textbf{-0.0898} & 0.6024 & -0.1111 & 0.6773 & -0.0932\\
         0.87/5m/3.0  & 0.5416 & -0.0762 & \textbf{0.5416} & \textbf{-0.0767} & 0.5105 & -0.1159 & 0.2593 & -0.0870\\
         0.7/5m/1.5   & 0.2938 & -0.0429 & \textbf{0.2935} & \textbf{-0.0431}  & 0.2703 & -0.0654 & 0.3646 & -0.0959\\
         0.85/5m/1.5  & 0.3383 & -0.0486 & \textbf{0.3378} & \textbf{-0.0489}  & 0.2758 & -0.0482 & 0.4206 & -0.1206\\
         \hline
    \end{tabular}

    \label{tab:CRM_Results}
\end{table}

\clearpage

\section{Data Fusion via Proper Orthogonal Decomposition}
\label{s:CPOD}
\begin{figure}
    \centering
    \includegraphics[width = 5in]{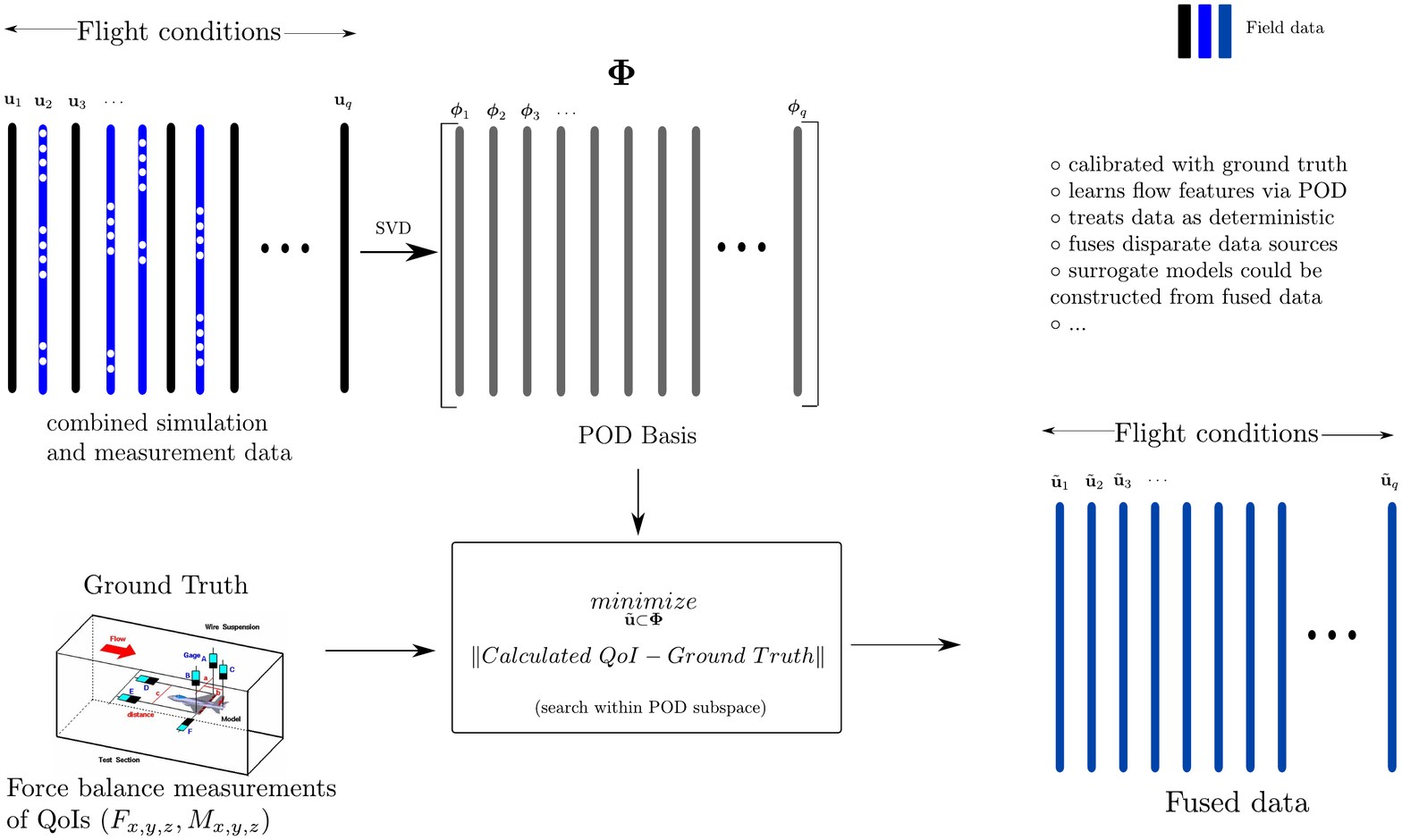}
    \caption{Schematic of the CPOD methodology}
    \label{f:cpod_method}
\end{figure}

We now present the proper orthogonal decomposition with constraints (CPOD), which uses the POD~\cite{Lumley1998, Sirovich1987} to first construct an orthogonal sub-space (also known as 'POD modes') from data (available $C_P$ distributions). Then the unknown fused $C_P$ is approximated as a linear expansion of the POD modes, whose parameters (coordinates) can be estimated via linear least-squares methodology. A schematic of the method is provided in Figure~\ref{f:cpod_method}. We begin with a brief review of POD before proceeding to outline the proposed CPOD method.


\subsection{Proper Orthogonal Decomposition}
POD was originally introduced in the context of turbulent flow modeling by Holmes et al~\cite{Lumley1998}, where it was used to characterize the coherent structures in the flow from wind tunnel measurements. POD has a special characteristic of \textit{optimality} in that it provides the most efficient means to capture the dominant components of a process~\cite{HolmesPhilip.LumleyJohnL.BerkoozGahlandRowley1998}. Given a state variable $\mbf{u} \in \mbb{R}^n$ which may be the numerical solution of a PDE on a computational mesh of size $n$ or measurements from a physical experiment (such as Particle Image Velocimetry), the POD expresses $\mbf{u}$ as the linear expansion on a finite number of $k$ orthonormal basis vectors $\mbf{\phi}_i \in \mbb{R}^n$. That is,
\begin{equation}
\mbf{u} \approx \sum_{i=1}^{k} a_i \bs{\phi}_i
\label{e:POD}
\end{equation}

where, $a_i$ is the $i$th component of $\mbf{a} \in \mbb{R}^k$ and are the coefficients of the basis expansion. It can be shown that \cite{HolmesPhilip.LumleyJohnL.BerkoozGahlandRowley1998, Chatterjee2000} the POD modes in the above equation are the same as the left singular vectors of the snapshot matrix (obtained by stacking $q$ snapshots of $\mbf{u}$), $\mbf{U} = [\mbf{u}_1, \hdots, \mbf{u}_{q}]$. That is,

\begin{equation}
\mbf{U}  \underset{\text{thin-svd}}{=} \mbf{\Phi} \mbf{D} \mbf{\Psi}^\top
\label{e:svd}
\end{equation}

then $\mbf{\Phi}_k$ represents the first $k$ columns of $\mbf{\Phi} \in \mbb{R}^{n \times q}$, after truncating the last $q-k$ columns based on the relative magnitudes of the cumulative sum of their singular values. The $L_2$ error in approximation of the state variables due to the POD basis expansion is then given as

\begin{equation}
\sum_{j=1}^q \left\Vert \mbf{u}^j - (\mbf{\Phi}_k \mbf{\Phi}_k^\top) \mbf{u}^j \right \Vert ^2_2 = \sum_{i=k+1}^q d_i^2
\end{equation}

where $d_i$ is the singular value corresponding to the $i$th column of $\mbf{\Phi}$ and is also the $i$th diagonal element of $\mbf{D}$. We choose $k$ such that $\sum_{i=1}^k d_i/\sum_{i=1}^q d_i \approx 0.99$, which essentially means we retain the modes that explain approximately 99\% of the variability in the dataset.

Let the (unknown) fused $C_P$ be denoted $\tilde{\mbf{u}}$. Firstly, we make the assumption that $\tilde{\mbf{u}}$ is missing from the dataset but is in the subspace spanned by $\mbf{\Phi}_k$ and hence $\tilde{\mbf{u}} \approx \mbf{\Phi}_k \mbf{a}$, where $\mbf{a}$ is to be estimated. Secondly, we know that $\tilde{\mbf{u}}$ has to match the measurements $\mbf{z}$ and hence one may consider estimating $\mbf{a}$ by minimizing $\| \mbf{H}^\top \mbf{\Phi}_k \mbf{a} - \mbf{z}\|_2 ^2$, however this is an underdetermined problem unless $k\leq m$, which is practically an unlikely scenario. To overcome this we combine it with the minimization of $\|\mbf{\Phi}_k \mbf{a} - \tilde{\mbf{u}}\|_2 ^2$, however $\tilde{\mbf{u}}$ is unknown. Therefore, we introduce an \emph{iterative} method that implements the CPOD which is described as follows. 

\subsection{CPOD}
The unknown $\tilde{\mbf{u}}$ is first provided an initial guess following which $\mbf{a}$ is estimated via solving the minimization problem

\begin{equation}
\begin{split}
    \mbf{a}^* = \underset{\mbf{a}}{arg\text{min}} \quad &\f{1}{2} \|\mbf{\Phi}_k \mbf{a} - \tilde{\mbf{u}} \|_2^2 \\
    \text{subject to} \quad & \mbf{H}^\top \mbf{\Phi}_k \mbf{a} = \mbf{z}
\end{split}
\label{e:CPOD_min}
\end{equation}

Then, the guess for $\tilde{\mbf{u}}$ is updated as $\tilde{\mbf{u}} \leftarrow \mbf{\Phi}_k \mbf{a}^*$. The updated $\tilde{\mbf{u}}$ is then used to enrich the POD modes, i.e. $\mbf{\Phi}_k \leftarrow \text{thin-svd}\left( [\mbf{U}, \tilde{\mbf{u}}]\right)$ and Eq.\eqref{e:CPOD_min} is solved again. The procedure is repeated until the cost function ($J$, to be defined in \eqref{e:CPOD_cost}) reaches a steady value, which in this work is measured based on the standard deviation ($\Delta_c$) of $J$ over the previous $c$ iterations falling within some specified threshold $\epsilon_c$, i.e. $\Delta_c = \f{1}{c}\sum_{i=1}^c J_i^2 - \left(\f{1}{c}\sum_{i=1}^c J_i\right)^2 \leq \epsilon_c$, where $J_i$ is the value of the cost function at the $i$th iteration. This stopping criterion is used based on the assumption that we are interested in the objective function reaching a steady value rather than strictly zero. Note that this iterative approach involves re-computing the SVD of the data at every iteration whose computational cost can be mitigated by performing rank-1 updates to SVD~\cite{levey2000sequential, brand2006fast}. We now show how to solve \eqref{e:CPOD_min}, which is a linearly constrained least-squares problem~\cite{boyd2004convex}. We introduce the penalized cost-function $J$ and re-write the constrained optimization problem \eqref{e:CPOD_min} as

\begin{equation}
   \mbf{a}^*, \bs{\lambda}^* = \underset{\mbf{a}, \mbf{\lambda}}{arg\text{min}} \quad J(\mbf{a}, \bs{\lambda}) = \f{1}{2} (\mbf{\Phi}_k  \mbf{a} - \tilde{\mbf{u}})^\top (\mbf{\Phi}_k ^\top \mbf{a} - \tilde{\mbf{u}}) + \bs{\lambda}^\top (\mbf{H}^\top \mbf{\Phi}_k \mbf{a} - \mbf{z})
    \label{e:CPOD_cost}
\end{equation}

where $\bs{\lambda} \in \mbb{R}^{m\times1}$ is the vector of Lagrange parameters. The partial derivatives of the cost function in \eqref{e:CPOD_cost} with respect to $\mbf{a}$ and $\bs{\lambda}$ are then set equal to zero

\begin{equation}
\begin{split}
    \frac{\partial J}{\partial \mbf{a}} &= \mbf{\Phi}_k^\top (\mbf{\Phi}_k \mbf{a} - \tilde{\mbf{u}})  + \mbf{\Phi}_k^\top \mbf{H}\bs{\lambda} = 0 \\  
    \frac{\partial J}{\partial \bs{\lambda}} &= \mbf{H}^\top \mbf{\Phi}_k \mbf{a} - \mbf{z} = 0
\end{split}
\label{e:kkt}
\end{equation}

where the first line of \eqref{e:kkt} consists of $k$ equations and the second consists of $m$ equations. The full set of equations in \eqref{e:kkt} represent the \emph{Karush-Kuhn-Tucker} (KKT) conditions~\cite{nocedal2006numerical, boyd2004convex} for constrained optimization which can be written in matrix form as 

\begin{equation}
    \begin{bmatrix}
        \mbf{I}_k & \mbf{\Phi}_k^\top \mbf{H} \\
        \mbf{H}^\top \mbf{\Phi}_k & \mbf{0}
    \end{bmatrix} 
    \begin{bmatrix}
        \mbf{a} \\
        \bs{\lambda}
    \end{bmatrix} =
    \begin{bmatrix}
        \mbf{\Phi}_k^\top \tilde{\mbf{u}} \\
        \mbf{z}
    \end{bmatrix}
    \label{e:kkt_matrix}
\end{equation}

where the $\mbf{I}_k$ is a $k \times k$ identity matrix that results from $\mbf{\Phi}_k^\top \mbf{\Phi}_k = \mbf{I}_k$ and $\mbf{0}$ is a $m \times m$ matrix of zeros. It can be shown that the coefficient matrix in \eqref{e:kkt_matrix} is invertible provided $\mbf{H}$ is full rank (see Appendix~\ref{app:lclsq}) and hence the least-squares problem has a unique solution. The solution of \eqref{e:kkt_matrix} involves computation of a matrix inverse during every iteration but is computationally cheap since the matrix has the reduced $(k+m) \times (k+m)$ size instead of the full $n \times n$ size where, $k+m << n$. 

The CPOD method is summarized in Algorithm~\ref{a:Alg_2} where it treats the available data to be devoid of any uncertainties and hence the $\mbf{u}_i$ are equivalent to the $\bs{\mu}$ used the in the Bayesian method. Furthermore, the CPOD algorithm relies on a rich basis set obtained from POD and hence the snapshot matrix ($\mbf{U}$) combines all the available data (CFD and wind-tunnel) in step-1 of Algorithm~\ref{a:Alg_2}. 

\begin{algorithm}[H]
\SetAlgoLined
\KwResult{Fused vector $\tilde{\mbf{u}}$}
\KwData{Snapshot matrix $\mbf{U} = [\mbf{u}_1, \hdots, \mbf{u}_q]$,\\ 
Fields at target flight condition $\mbf{u}_{CFD}$, $\mbf{u}_{WT}$,\\
QoIs at target flight condition $\mbf{z}$, \\ Stopping criterion $c=5,~\epsilon_c=10^{-6}$} 
\begin{enumerate}
    \item Extract POD modes $\mbf{U} = \mbf{\Phi} \mbf{D} \mbf{\Psi}^\top$\, Determine rank $k$;
    \subitem $\mbf{\Phi}_k = \mbf{\Phi}(:,1:k)$
    \item Guess $\tilde{\mbf{u}}$ as $\theta \times \mbf{u}_{CFD} + (1-\theta) \times \mbf{u}_{WT} $ where $\theta \subseteq [0,1]$\;
    \item \While{$\Delta_c \geq \epsilon_c$}{
 \begin{itemize}
     \item Compute $\mbf{a}^*$ via solving \eqref{e:CPOD_min}
     \item Update $\tilde{\mbf{u}} \leftarrow \mbf{\Phi}_k \mbf{a}^*$
     \item Update $\mbf{\Phi}_k$: $[\mbf{U}, \tilde{\mbf{u}}] \underset{\text{thin-svd}}{=} \mbf{\Phi} \mbf{D} \mbf{\Psi}^\top$
     \item Compute cost function, \eqref{e:CPOD_cost}
     \item Update $\Delta_c$
 \end{itemize}
 }
\end{enumerate}

 \caption{CPOD - Proper Orthogonal Decomposition with Constraints}
 \label{a:Alg_2}
\end{algorithm}

\subsection{Discussion}
We now compare and contrast the performance of the CPOD against the Bayesian approach. As we shall demonstrate the CPOD relies on \emph{learning} the flow features from the entire dataset. Therefore when it is deemed necessary to enrich the POD basis set, additional data is generated via CFD since experimental data are not available on demand. Additionally, since the cost of generating CFD data for the RAE test case is significantly cheaper than the CRM model, we restrict comparison of the methods based on the RAE test case only. Experimental data corresponding to 11 flight conditions are available (see Table~\ref{t:RAE_cases}) for the RAE test case; the number of snapshots, $q$ is therefore the CFD data generated in addition to these 11 cases. We pick two contrasting cases (2 \& 11) to demonstrate the comparison in favor of keeping the discussion precise. 

\subsubsection{Effect of Snapshot Size $q$}
\begin{figure}[htb!]
    \centering
    \begin{subfigure}{0.5\linewidth}
    \centering
        \includegraphics[width = 3.50 in]{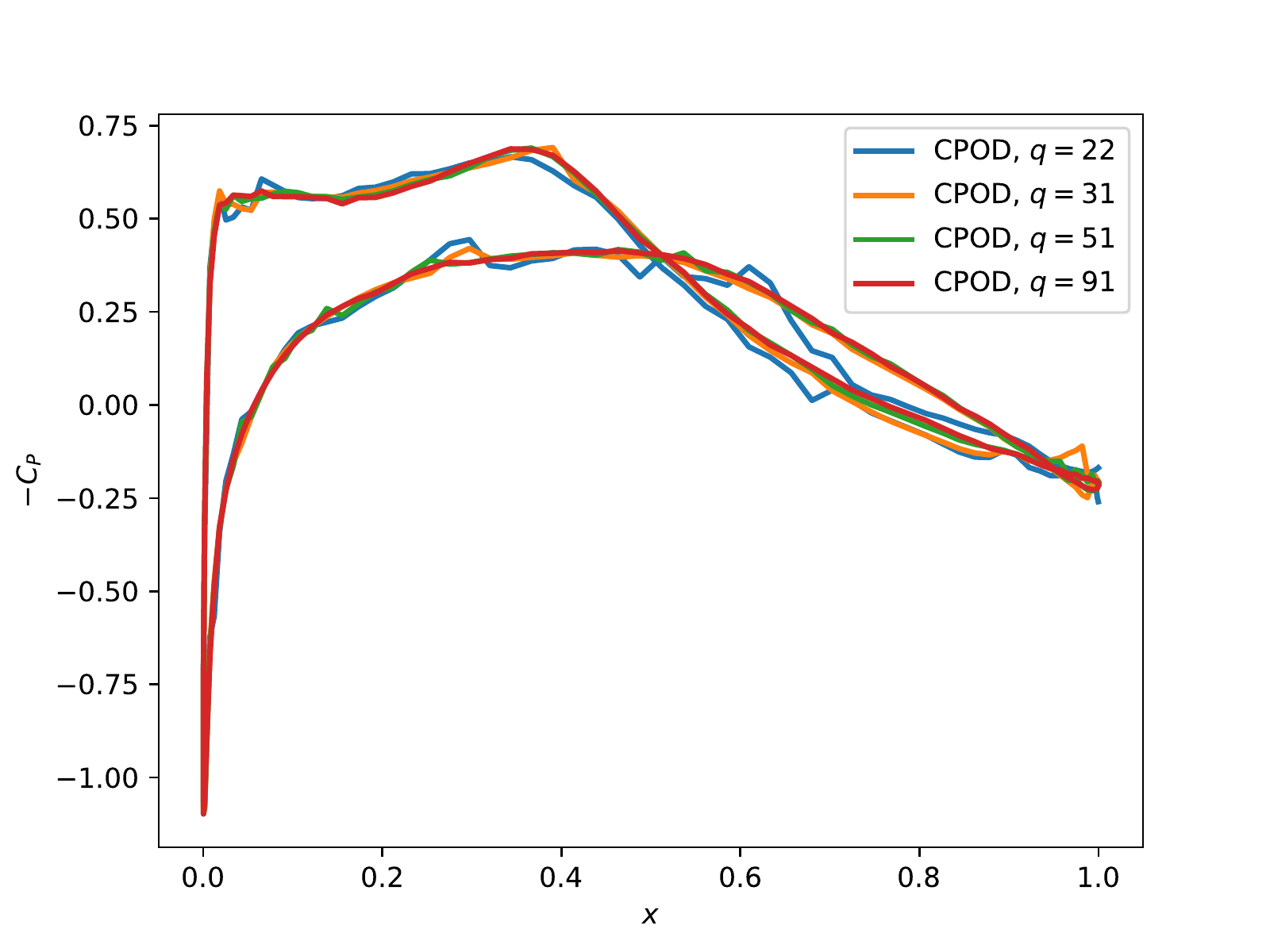}
        \caption{Case-2: $Mach=0.676$, $Re=5.7 \times10^{6}$, $\alpha=-2.18~deg.$}
        \label{sf:cpod_effect_of_M_a}
    \end{subfigure}%
    \begin{subfigure}{0.5\linewidth}
    \centering
        \includegraphics[width = 3.50 in]{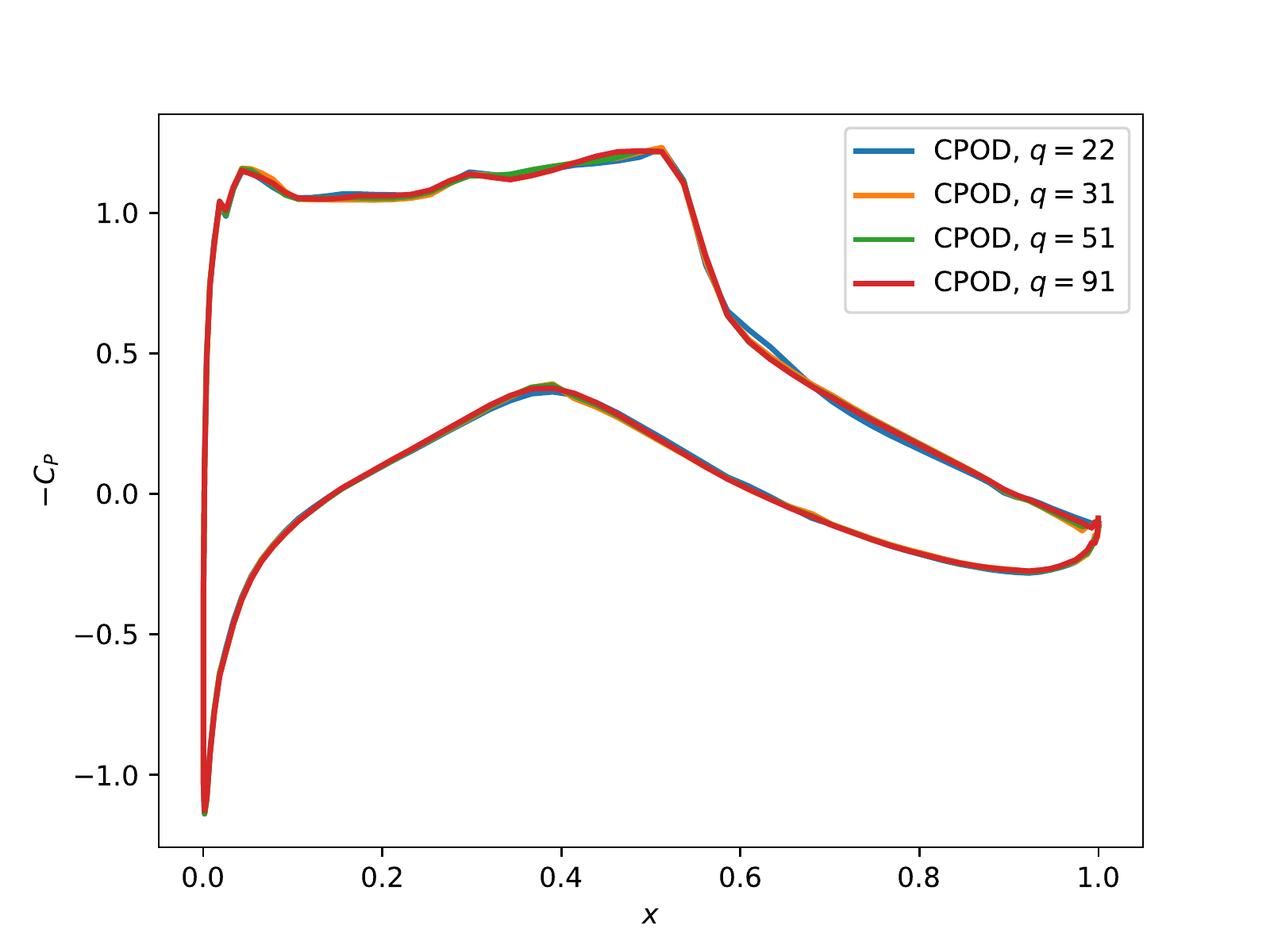}
        \caption{Case-11: $Mach=0.74$, $Re=2.7\times 10^{6}$, $\alpha=3.19~deg.$}
        \label{sf:cpod_effect_of_M_b}
    \end{subfigure}   
    \caption{Effect of snapshot size $q$ on the smoothness of CPOD predictions.}
    \label{f:cpod_effect_of_M}    
\end{figure}

Since the CPOD relies on searching for the unknown $\tilde{\mbf{u}}$ from a subspace, the richness of the snapshot matrix plays a major role in the performance of the method unsurprisingly; this is demonstrated in Figure~\ref{f:cpod_effect_of_M}. Here, a maximin Latin Hypercube designs~\cite{santner2003design} of size 20, 40 and 80 are generated in the $(Mach, Re, \alpha)$ space to augment the experimental data leading to $q=31, 51$ and $91$ snapshots respectively. Note that the $q=22$ case contains 11 CFD cases corresponding to exactly the same flight conditions as in Table~\ref{t:RAE_cases}. 

With too few snapshots ($q=22$), the CPOD determines a $\tilde{\mbf{u}}$ that satisfies the forward model within the specified tolerance but results in a $C_P$ curve that looks noisy and physically unrealistic (Figure~\ref{sf:cpod_effect_of_M_a}). This also demonstrates the ill-conditioning of the problem mentioned in section~\ref{s:Intro}, i.e. there are more than one $C_P$ curve that satisfies the forward model to match the measured QoIs. Recall that the Bayesian approach discussed earlier overcomes this issue with the prior regularization. With more snapshots, the CPOD subspace is enriched leading to smoother $C_P$ curves. The sensitivity to the snapshot size manifests more prominently only when the discrepancy between the forward model predictions and the measured QoIs is significant, as in Case 2 shown in Figure~\ref{sf:cpod_effect_of_M_a}. For this specific case, $q=91$ is necessary to ensure sufficiently smooth $C_P$ curves. As a counter example, Figure~\ref{sf:cpod_effect_of_M_b} shows that fewer snapshots are sufficient for this specific case. However, based on the worst-case scenario (Case-2), $q$ is fixed at $91$ for the rest of the results.

An aspect we care about as part of the data fusion framework is uncertainty quantification. Whereas the MAP results are a function of the input uncertainties in the data, the CPOD results do not account for any of the uncertainties and treats the dataset $\lbrace \mbf{u}_i \rbrace$ as ouputs of deterministic experiments as mentioned before. However, in the next subsection we show how approximate confidence bounds can be constructed for the CPOD predictions.

\subsubsection{Confidence bounds for CPOD}
One of the obvious limitations of the CPOD is its inability to account for uncertainty in the datasets. However, one can use a frequentist approach to obtain confidence bounds as demonstrated here. Assuming that the unknown fused $C_P$ is the true mean of a multi-variate normally distributed population with unknown covariance function, the confidence intervals are approximated via the Student's $t-$ distribution~\cite{ross2014first} described as follows. Given $T$ realizations of the $\tilde{\mbf{u}}$ that results from repeating Algorithm~\ref{a:Alg_2} with independent draws of $\theta$ from a uniform distribution $\mc{U}(0,1)$, a $t-$distribution of $\nu=T-1$ degrees of freedom is defined as the distribution of the location of the sample mean relative to the true mean, divided by the sample standard deviation. This way, the $t-$distribution can used to construct confidence bounds for the true mean. The sample mean ($\hat{\bs{\mu}}$) and covariance ($\hat{\mbf{\Sigma}}$) are given by

\begin{equation}
    \hat{\bs{\mu}} = \frac{1}{T} \sum_{i=1}^T \tilde{\mbf{u}}_i
\end{equation}

\begin{equation}
    \begin{aligned}
        \hat{\mbf{\Sigma}} &= \frac{1}{T-1} \sum_{i=1}^T [\tilde{\mbf{u}}_i - \hat{\bs{\mu}}][\tilde{\mbf{u}}_i - \hat{\bs{\mu}}]^\top \\
    \end{aligned}
\end{equation}
 
\begin{figure}[htb!]
    \centering
    \begin{subfigure}[b]{\textwidth}
    \centering
        \includegraphics[width=0.475\linewidth]{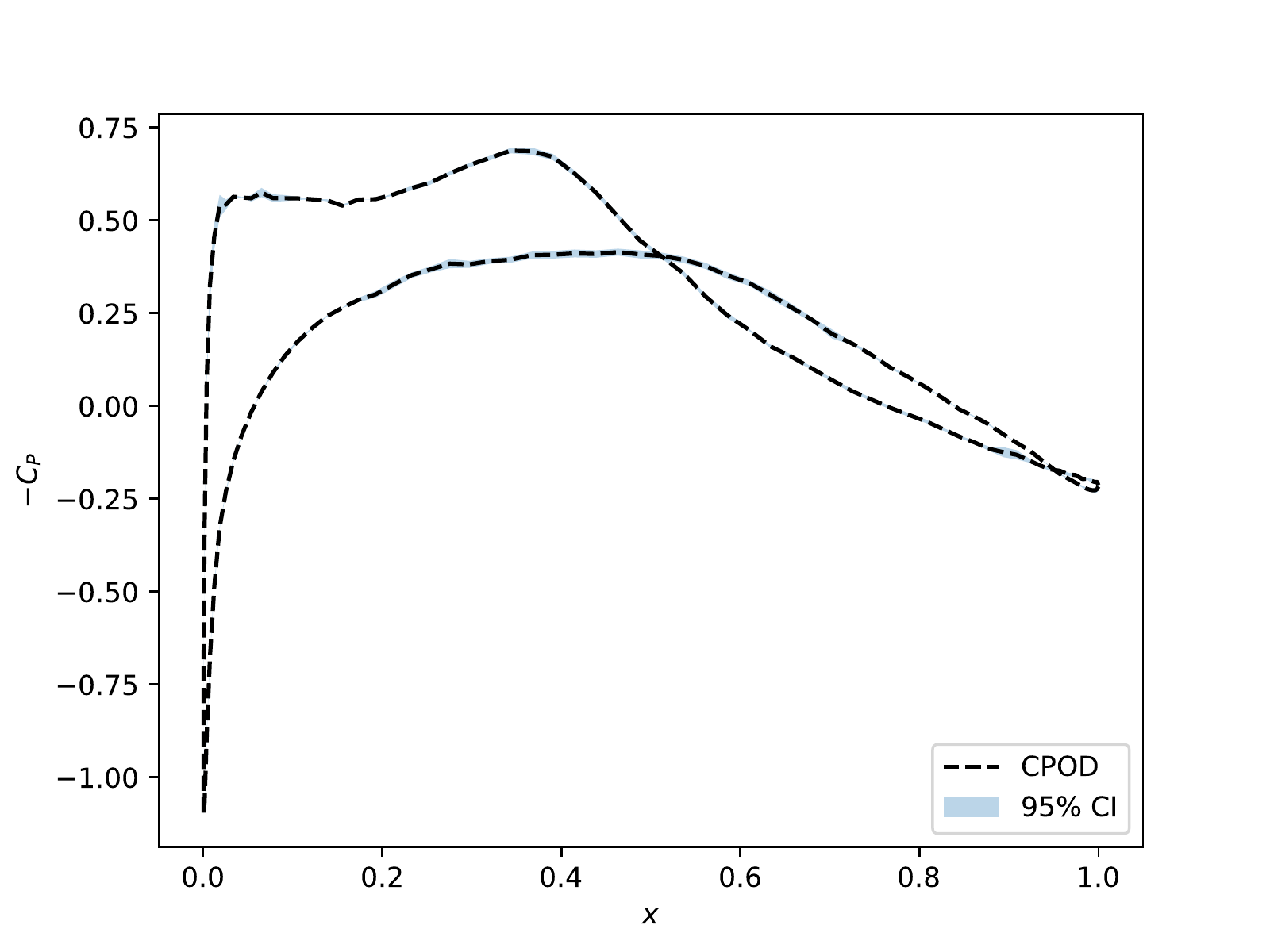}
        \hfill
        \includegraphics[width=0.475\linewidth]{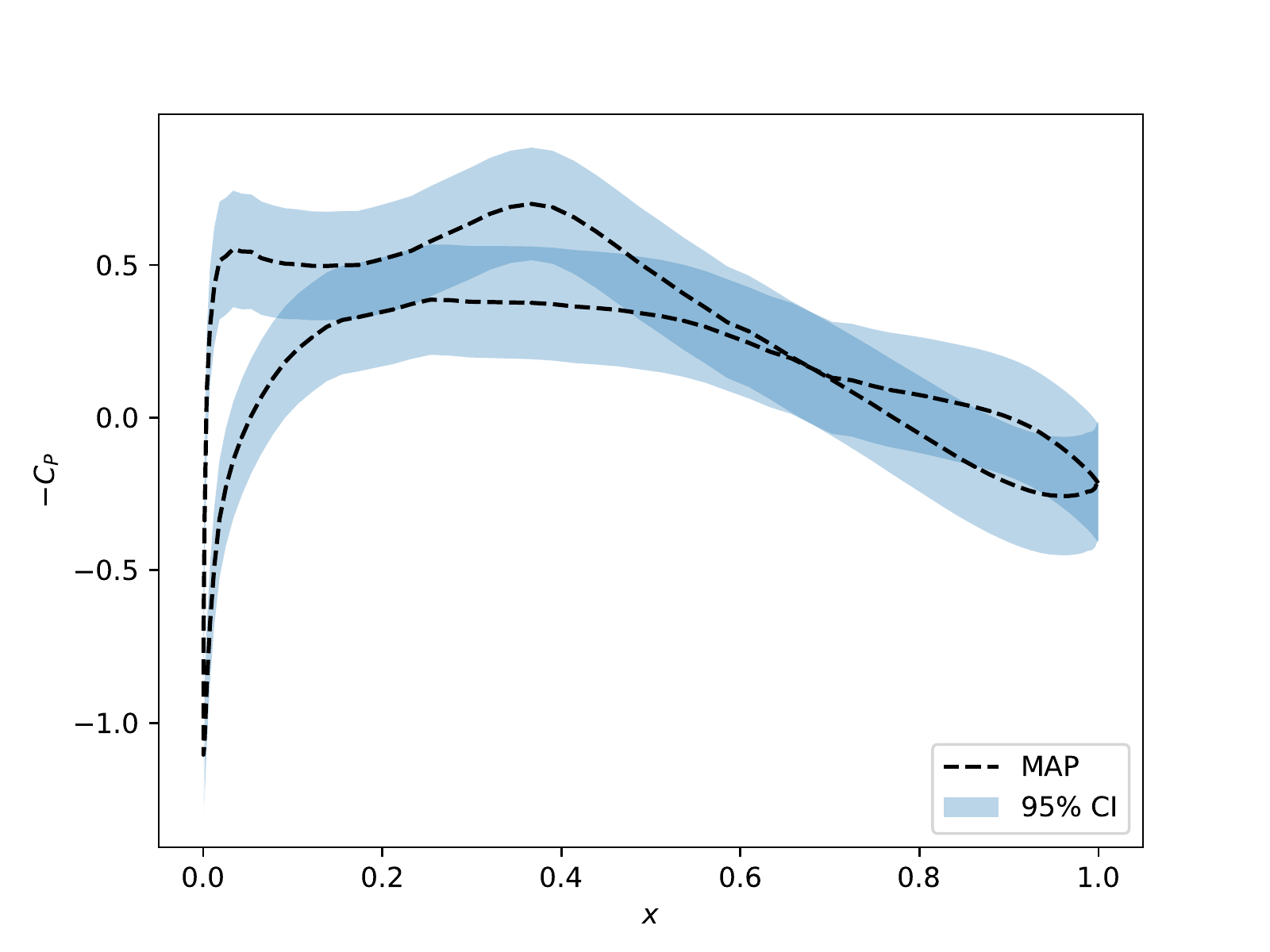}
        \caption{Case-2: $Mach=0.676$, $Re=5.7~10^{6}$, $\alpha=-2.18~deg.$}
        \label{f:cpod_confbounds_a}
    \end{subfigure}
    \vskip\baselineskip
    \begin{subfigure}[b]{\textwidth}
    \centering
        \includegraphics[width=0.475\linewidth]{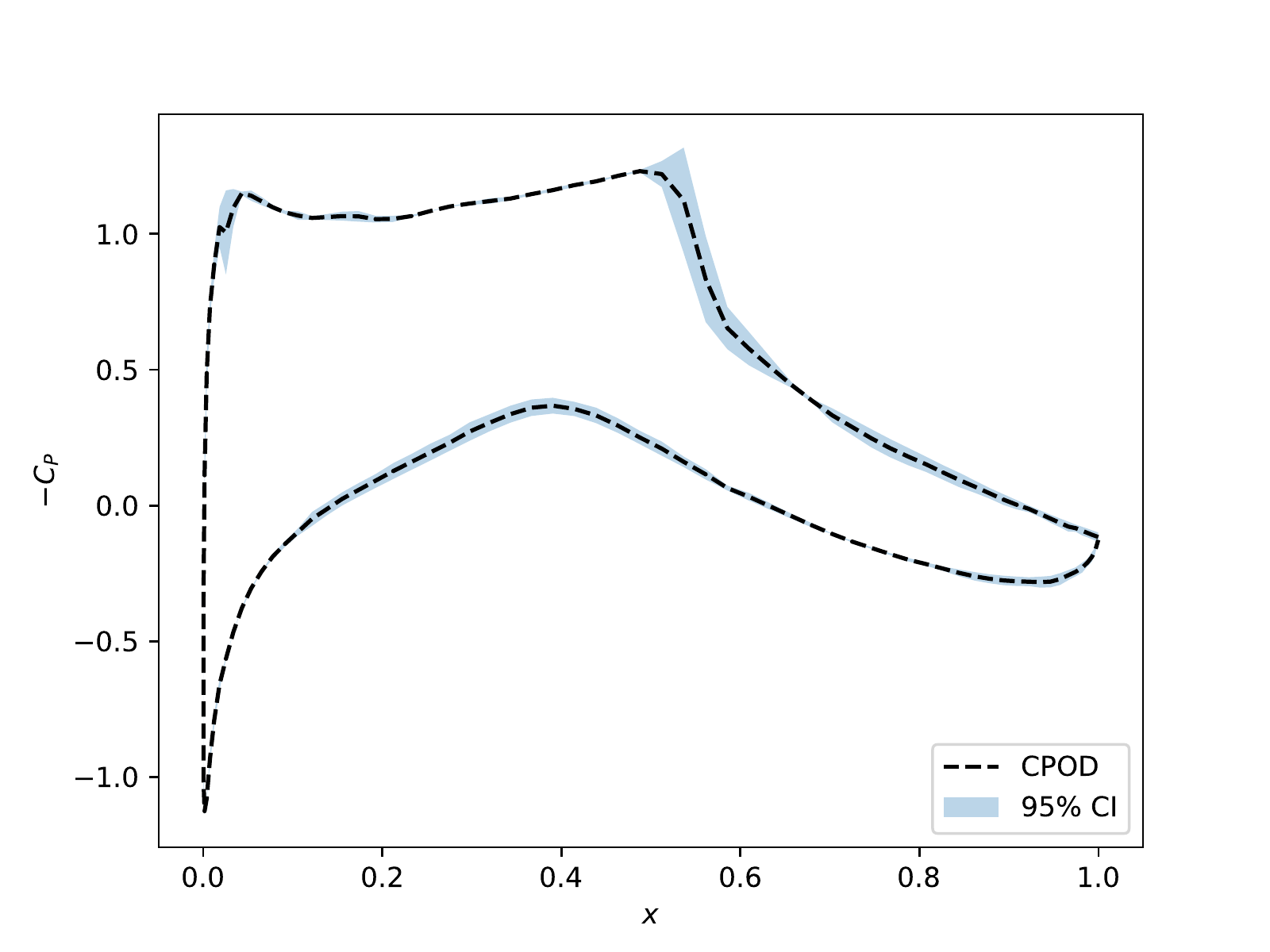}
        \hfill
        \includegraphics[width=0.475\linewidth]{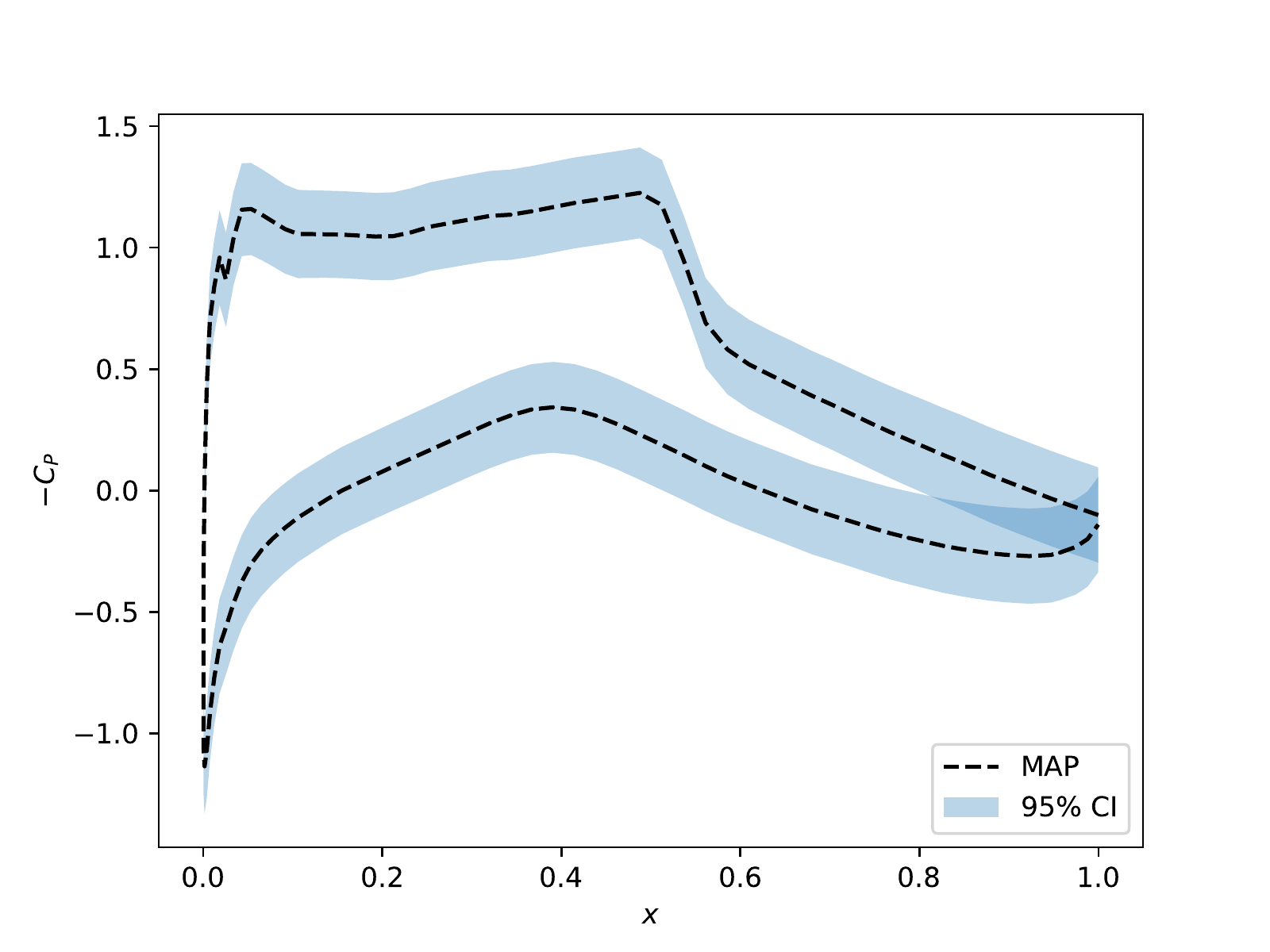}
        \caption{Case-11: $Mach=0.74$, $Re=2.7~10^{6}$, $\alpha=3.19~deg.$}
        \label{f:cpod_confbounds_b}
    \end{subfigure}   
    \caption{ Confidence bound comparison}
    \label{f:cpod_conf_bounds}
\end{figure}

The $1-\beta$ confidence intervals can then be provided as $\hat{\bs{\mu}} \pm t_{1-\beta, \nu}~ \texttt{diag}(\hat{\mbf{\Sigma}})/\sqrt{T}$ where $\texttt{diag}(\hat{\mbf{\Sigma}})$ are the diagonal elements of $\hat{\Sigma}$ and $t_{1-\beta, \nu}$ is the $t-$value corresponding to a $(1-\beta)\times 100$\% confidence. The 95\% confidence bounds are plotted for the two RAE cases in Figure~\ref{f:cpod_conf_bounds} with $T=1000$. The bounds are more pronounced in for the CPOD results in \ref{f:cpod_confbounds_b} compared to \ref{f:cpod_confbounds_a} particularly around the shock. This is because the CPOD relies mainly on minimizing the QoI data misfit whereas the Bayesian method accounts for uncertainties in the data. It is emphasized that the confidence bounds in CPOD are an artifact of the randomness in the initial guess provided (via the parameter $\theta$) and does not account for the the input uncertainties, unlike the Bayesian method. In that sense, the CPOD confidence bounds might not be highly useful in uncertainty quantification although they visualize the effect of the mis-specification of the initial guess to the method. In all the plots in this section showing CPOD predictions, $T=1000$ unless otherwise mentioned.

The CPOD and Bayesian methods present two approaches towards fusing data from multiple fidelity experiments. The common thread between the two approaches is that they both they find the best fit between the predictions of the forward model and the measurements of the QoI's in the least-squares sense. However, they differ in a fundamental way - whereas the CPOD is a data-driven approach that essentially depends on the flow-features learned from the data (in the form of POD modes), the Bayesian approach relies on subjective prior specifications. Another important aspect that distinguish the approaches is that while the Bayesian approach accounts and propagates uncertainties in the available data, the CPOD does not. Despite the differences, the authors have observed that with adequate snapshots the CPOD predictions are very \emph{similar} to the MAP predictions of the Bayesian method.

\subsubsection{CPOD vs MAP}
\begin{figure}[htb!]
    \centering
    \begin{subfigure}{0.5\linewidth}
    \centering
        \includegraphics[width = 3.50 in]{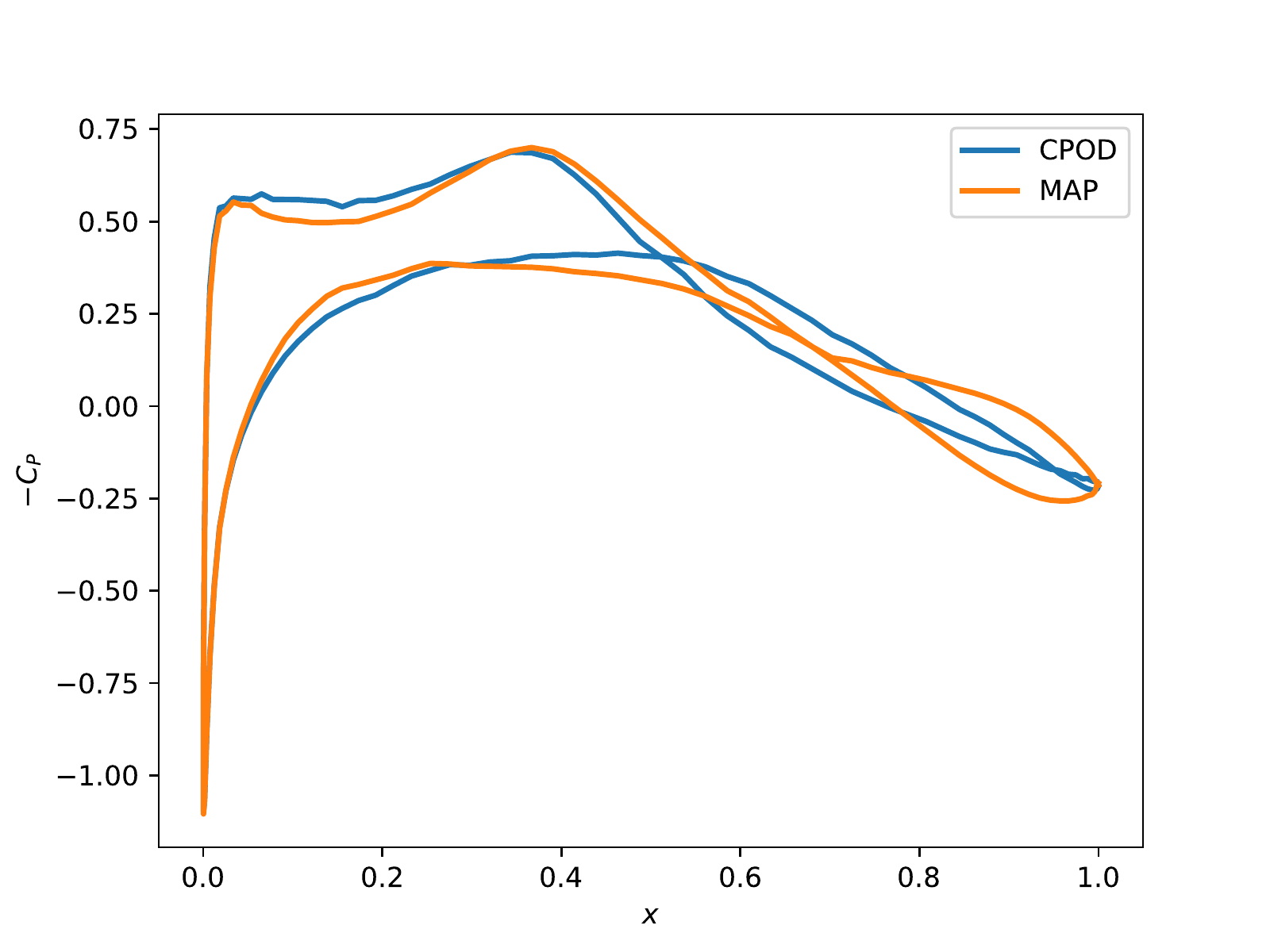}
        \caption{Case-2: $Mach=0.676$, $Re=5.7~10^{6}$, $\alpha=-2.18~deg.$}
        \label{sf:cpod_vs_map_a}
    \end{subfigure}%
    \begin{subfigure}{0.5\linewidth}
    \centering
        \includegraphics[width = 3.50 in]{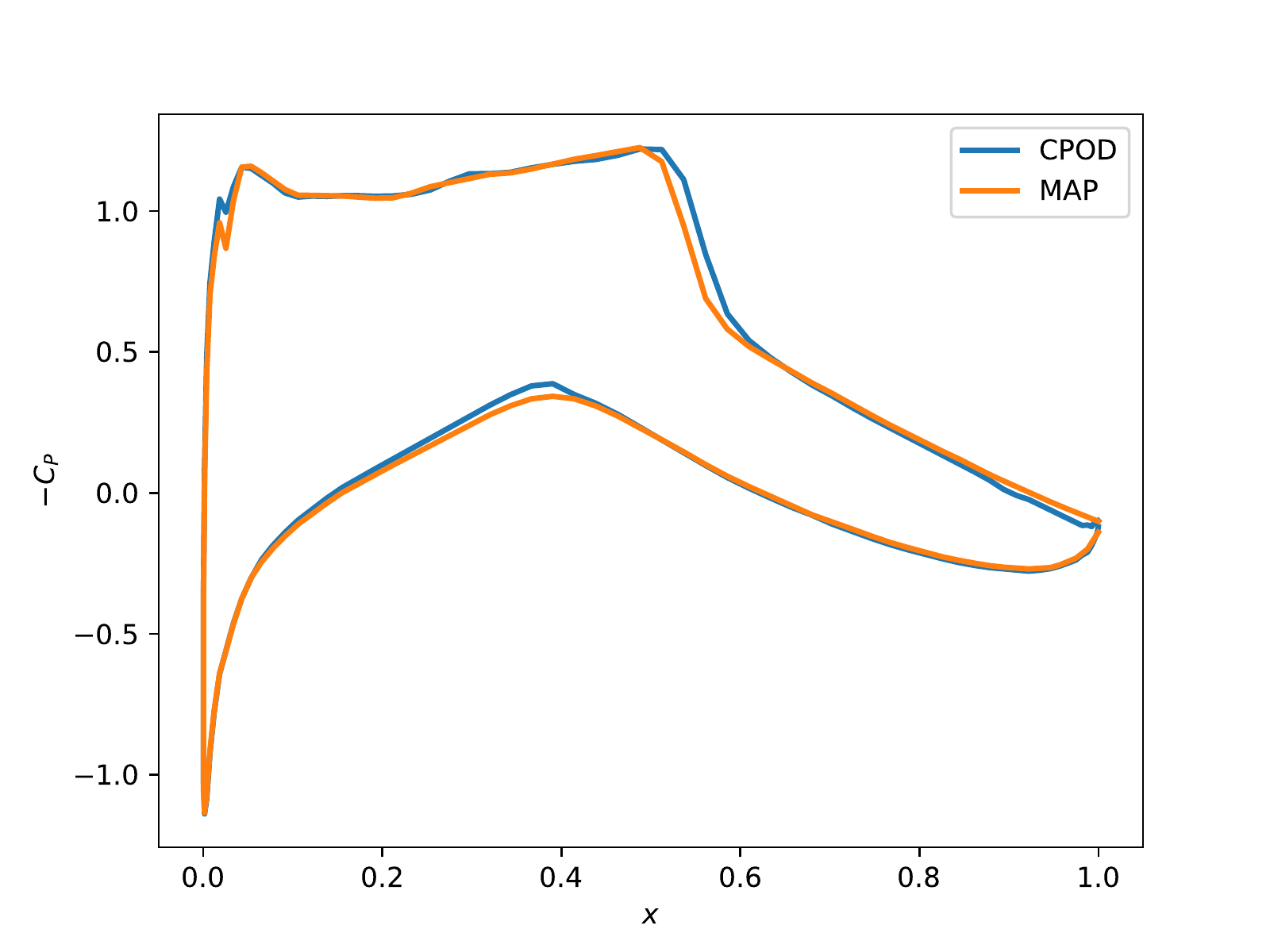}
        \caption{Case-11: $Mach=0.74$, $Re=2.7~10^{6}$, $\alpha=3.19~deg.$}
        \label{sf:cpod_vs_map_b}
    \end{subfigure}
    \caption{Comparison of the CPOD and Bayesian methods for data fusion}
    \label{f:CPOD_vs_MAP}
\end{figure}

The predictions from the Bayesian (MAP) and the CPOD (with $q=91$) methods are compared for the RAE cases 2 and 10 in Figure~\ref{f:CPOD_vs_MAP}. Surprisingly, the MAP and the CPOD predictions are very similar despite the fact that the CPOD depends on the entire dataset ($q=91$ in this case) to learn the flow features in the data and provide a physically realistic result whereas the MAP required only one snapshot each from the two fidelity experiments. The difference is more pronounced in cases where the discrepancy between the QoI's predicted from the forward model and the measurements is more pronounced, as shown in Figure~\ref{sf:cpod_vs_map_a}. Although the CPOD has learned flow-features from data, in this case the snapshots are dominated by CFD data which inherits the bias in the data as well. Therefore it leads to a slightly different prediction compared to the MAP which depends only on the model misfit and prior specification (which depends only on $\mbf{u}_{CFD}$ and $\mbf{u}_{WT}$). Despite the differences, one can appreciate the similarity between the predictions at a very high level. 

\subsubsection{Computational Aspects}
The CPOD relies on an appropriate initial guess very similar to the requirement of the prior in the Bayesian method. As mentioned in section~\ref{s:Intro}, both methods can lead to physically unrealistic results while still satisfying the forward problem within a specified tolerance. This work relies on the assumption that the unknown (fused) $C_P$ is somewhere in the neighborhood of the available $C_P$'s from the various fidelity experiments. And within such a sufficiently small search space, it is observed that the CPOD and the MAP predictions are quite similar.

\begin{figure}
    \centering
    \begin{subfigure}{0.5\textwidth}
    \centering
        \includegraphics[width=1\linewidth]{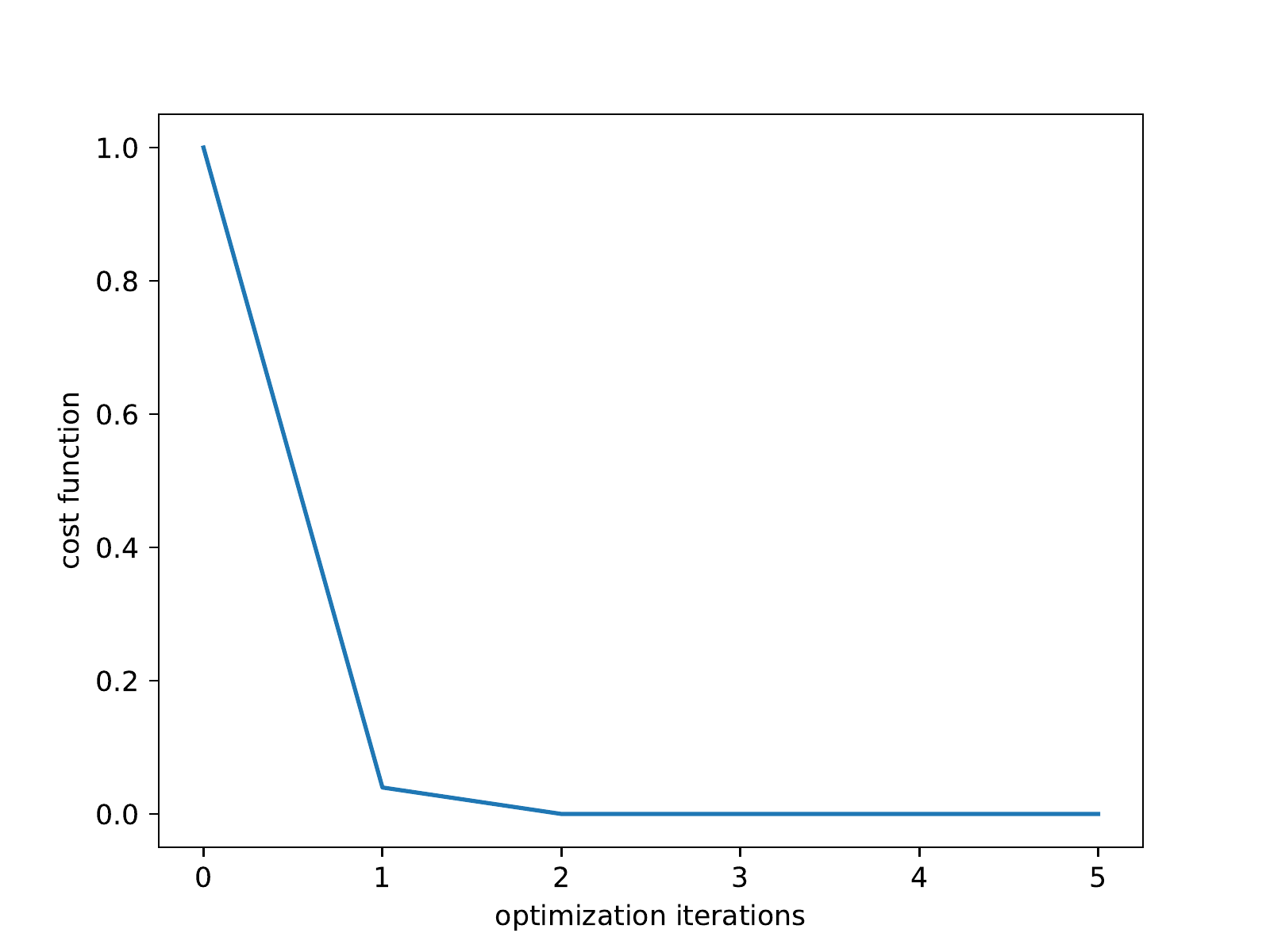}
        \caption{Convergence of cost function}
        \label{sf:jhist}
    \end{subfigure}%
    \begin{subfigure}{0.5\textwidth}
    \centering
        \includegraphics[width=1\linewidth]{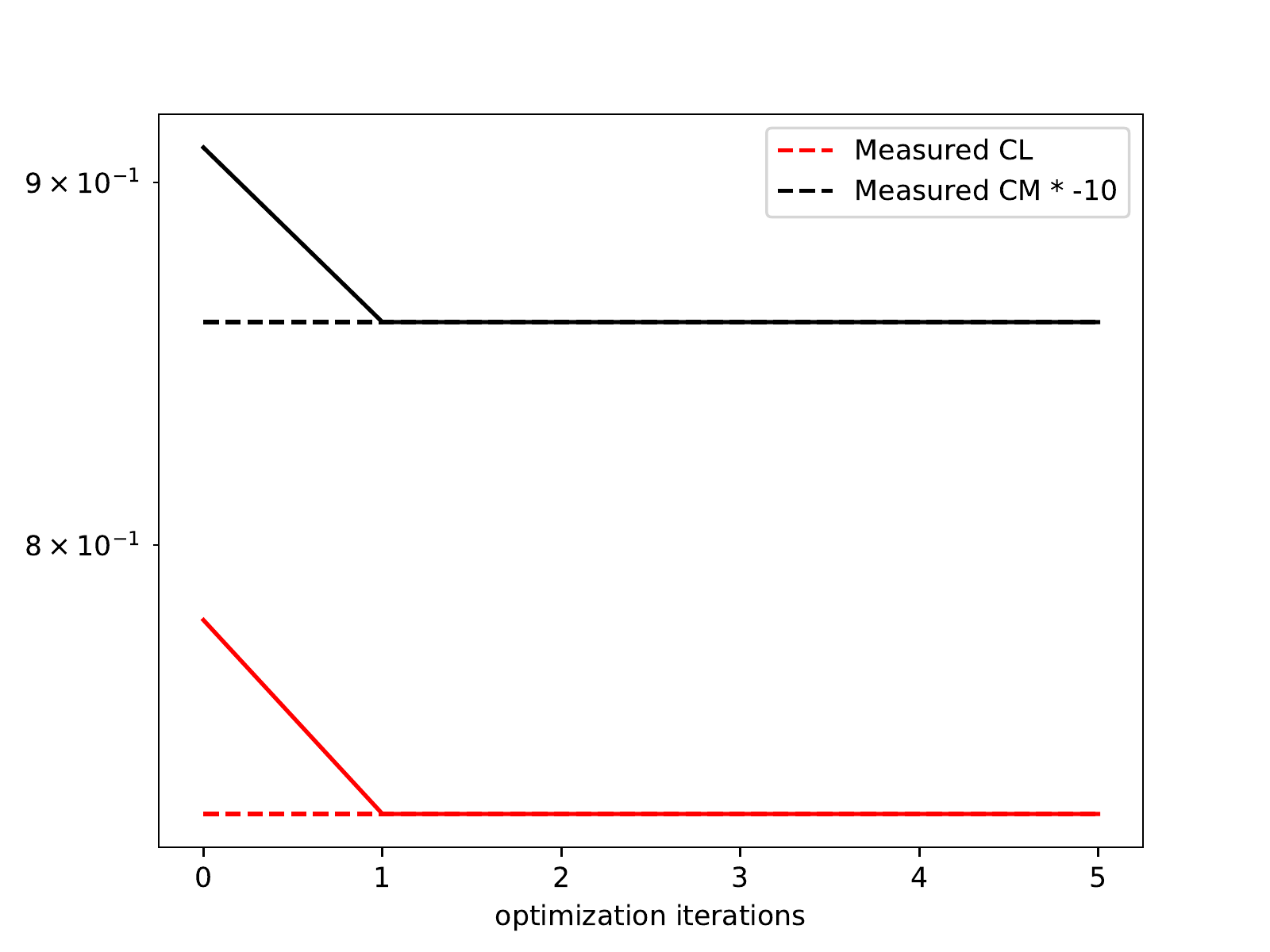}
        \caption{Convergence of QoIs. Solid lines are QoI histories}
        \label{sf:zhist}
    \end{subfigure}    
    \caption{Optimizer history for the CPOD}
    \label{f:cpod_hist}
\end{figure}

One of the significant differences between the two methods (in addition to those mentioned in the previous subsection) is the computational time complexity. It would appear that the CPOD incurs significantly more computational cost since it operates on a snapshot set in contrast to the Bayesian approach. Additionally, the SVD (and recursive SVD-update) steps scale with the data size. However, the Bayesian approach involves the inversion of matrices of size $n\times n$ which is expensive as $n$ gets bigger. On the other hand, the CPOD method is iterative although it was observed to converge monotonically within 3-5 iterations for all the cases tested in this work; an example of which is shown in Figure~\ref{f:cpod_hist}. It is however worth recalling that the reduced computations in the Bayesian approach are a consequence of the choice of conjugate priors that lead to analytically closed-form expression for the MAP and the confidence bounds. This will not be the case in a more general setup with generic priors and the estimation of the hyperparameters included in the method. A very rough comparison of the computational time-complexity for both methods are provided in Table~\ref{t:computational_cost} in terms of the order of magnitude ($\mc{O}$) of floating point operations. Note that in the comparison, the Bayesian method assumes Gaussian prior and likelihoods and known hyperparameters whereas the CPOD represents \emph{one} repetition at a fixed $\theta$. Although from a time-complexity point of view, the Bayesian method seems expensive, from a wall-clock time point of view either methods take $\mc{O}(secs)$ in a standard workstation (Intel i7, 1.9GHz, 4 Core) at the time of writing this article for both the 2D and 3D cases. This demonstrates the feasibility of the present implementation with real large-scale data.

\begin{table}[]
    \centering
        \caption{Comparison of computational costs}
    \begin{tabular}{ccc}
        \hline
        Computation & CPOD & Bayesian  \\
        \hline
         SVD & $nq^2$  & -\\
         linear system solve & $(k+m)^3$ & -\\
         Cost function computation & $n^2$ & -\\
         Estimation of $\theta$ & - & $n+m^3$\\
         Compute $\mbf{y}_{MAP}, \mbf{\Gamma}$ & - & $n^3 + n^2$\\ 
         \hline
    \end{tabular}
    \label{t:computational_cost}
\end{table}

\section{Conclusion}
A novel method to fuse noisy, incomplete and biased aerodynamic field information from wind-tunnel measurements and high-fidelity mathematical model predictions is presented. The method applies Bayesian inference to solve for the fused $C_P$ distribution given predictions from computer experiments and wind-tunnel measurements. The approach depends on providing a prior belief on the unknown $C_P$ distribution as a linear combination of the field data. Then a likelihood model is defined that minimizes the least-squares misfit between the output QoIs (computed via the forward model \eqref{e:fwd_model}) and the measurements. Combining the prior belief with the likelihood via Bayes' rule, the fused $C_P$ distribution is estimated as a probability distribution. The parameters of the methodology are the uncertainties associated with the datasets as well as the measurements in addition to a correlation length-scale that determines the smoothness of the prior. One of the main assumptions made in this work is the specification of the prior via the estimation of the parameter $\theta$. Future directions in this work shall add more rigor to the method by acknowledging our ignorance on $\theta$ and estimating it via a Hierarchical Bayes method. The method is successfully demonstrated on the transonic flow past a wing section (RAE2822 airfoil) as well as an entire aircraft (CRM).
Overall, the proposed method performs a statistical adjustment to the available field data based on very measurements of QoIs treated as ground truth. The adjusted data can then be used to construct data-driven models towards realizing the DT goal. 

As an alternative method to solve the same problem, the CPOD method is introduced which, like the Bayesian method minimizes the misfit between the predicted and measured QoIs, however the search space is constrained by the POD subspace. Comparison of the CPOD and Bayesian methods revealed that they result in surprisingly similar results despite significant differences in their approach. A known limitation of the CPOD is that it currently does not account for the uncertainty in the dataset; future directions in this work aims to devise methods that account for the noise in the SVD step. Another observed limitation of the CPOD is that in cases where experimental measurements of the fields is limited, additional snapshots are likely to be generated from CFD. This could lead to biased results since the POD basis could inherit the bias present in the CFD data.

Another focus of future work is the hybridization of the Bayesian and CPOD method - i.e. specification of informative priors derived from the POD. While this is a relatively straightforward extension, the authors are investigating the remedies to the known limitations of the proposed methodologies as a first step.

\section*{Acknowledgments}
The material is based upon work supported by Airbus in the frame of the Airbus / Georgia Tech Center for MBSE-enabled Overall Aircraft Design and by the U.S. Department of Energy, Office of Science, under contract number DE-AC02-06CH11357. The insets used in Figures~\ref{f:overall_method} and \ref{f:cpod_method} showing wind-tunnel measurements are adapted from \url{www.nasa.gov}.

\begin{mdframed}
\textbf{Government License:} The submitted manuscript has been created by UChicago Argonne, LLC, Operator of Argonne National Laboratory ("Argonne”). Argonne, a U.S. Department of Energy Office of Science laboratory, is operated under Contract No. DE-AC02-06CH11357. The U.S. Government retains for itself, and others acting on its behalf, a paid-up nonexclusive, irrevocable worldwide license in said article to reproduce, prepare derivative works, distribute copies to the public, and perform publicly and display publicly, by or on behalf of the Government. The Department of Energy will provide public access to these results of federally sponsored research in accordance with the DOE Public Access Plan (http://energy.gov/downloads/doe-public-access-plan).
\end{mdframed}

\bibliography{bibliography}

\section*{Appendix}
\appendix

\subsection{Proof of existence of solution for \eqref{e:kkt_matrix}}
\label{app:lclsq}

Here we show that a unique solution exists for the Equation~\ref{e:kkt_matrix}. 

\begin{theorem}
The KKT matrix $\begin{bmatrix}
        \mbf{I}_k & \mbf{\Phi}_k^\top \mbf{H} \\
        \mbf{H}^\top \mbf{\Phi}_k & \mbf{0}
\end{bmatrix}$
is invertible if $\mbf{H}$ is full rank. Consider the following matrix inversion lemma by Press et al~\cite{press1992numerical}. If $\mbf{A} = \begin{bmatrix}
\mbf{P} & \mbf{Q} \\
\mbf{R} & \mbf{S}
\end{bmatrix}$, \text{then} 
$\mbf{A}^{-1} = \begin{bmatrix}
\tilde{\mbf{P}} & \tilde{\mbf{Q}} \\
\tilde{\mbf{R}} & \tilde{\mbf{S}}
\end{bmatrix}$

where,
\begin{equation*}
    \begin{split}
       \tilde{\mbf{P}} &= \mbf{P}^{-1} + \mbf{P}^{-1}\mbf{Q}\mbf{M}\mbf{R}\mbf{P}^{-1}\\
       \tilde{\mbf{Q}} &= -\mbf{P}^{-1}\mbf{Q}\mbf{M}\\
       \tilde{\mbf{R}} &= -\mbf{M}\mbf{R}\mbf{P}^{-1}\\
       \tilde{\mbf{S}} &= \mbf{M} \\
       \text{and} \\
       \mbf{M} &= (\mbf{S} - \mbf{R}\mbf{P}^{-1}\mbf{Q})^{-1}
    \end{split}
\end{equation*}
\end{theorem}

\textbf{Proof.} For the coefficient matrix in \eqref{e:kkt_matrix} to be invertible, it suffices to show that $\mbf{M}$ exists, since $\mbf{P}=\mbf{I}_m$ and hence $\mbf{P}^{-1}$ exists. The $\mbf{M}$ can be written as

\begin{equation}
    \mbf{M} = [\mbf{0} - (\mbf{H}^\top \mbf{\Phi}_k) \times (\mbf{\Phi}_k ^\top \mbf{H}) ]^{-1}
\end{equation}

The $m$ columns of $\mbf{H}$ contain the integral operators to compute the forces and moments from the $C_P$ distributions. More specifically, the first column computes $C_L$ and the second column computes $C_M$. If the columns of $\mbf{H}$ were linearly dependent, then $C_L$ and $C_M$ would be scalar multiples of each other which is not the case. Therefore it is concluded that $\mbf{H}$ has full column rank of $m$. Since the orthogonal transformation preserves the rank, rank$\left( (\mbf{H}^\top \mbf{\Phi}_k) \times (\mbf{\Phi}_k ^\top \mbf{H})\right)$ is equal to rank$\left( \mbf{H}^\top \mbf{H} \right)$ which is full-rank. Therefore $\mbf{M}$ exists and hence the inverse to the coefficient matrix in \eqref{e:kkt_matrix} exists.

\subsection{Pre-Processing}
\label{app:pre-process}
\begin{figure}[htb]
    \centering
    \begin{subfigure}{0.5\textwidth}
        \centering
        \includegraphics[scale=0.12]{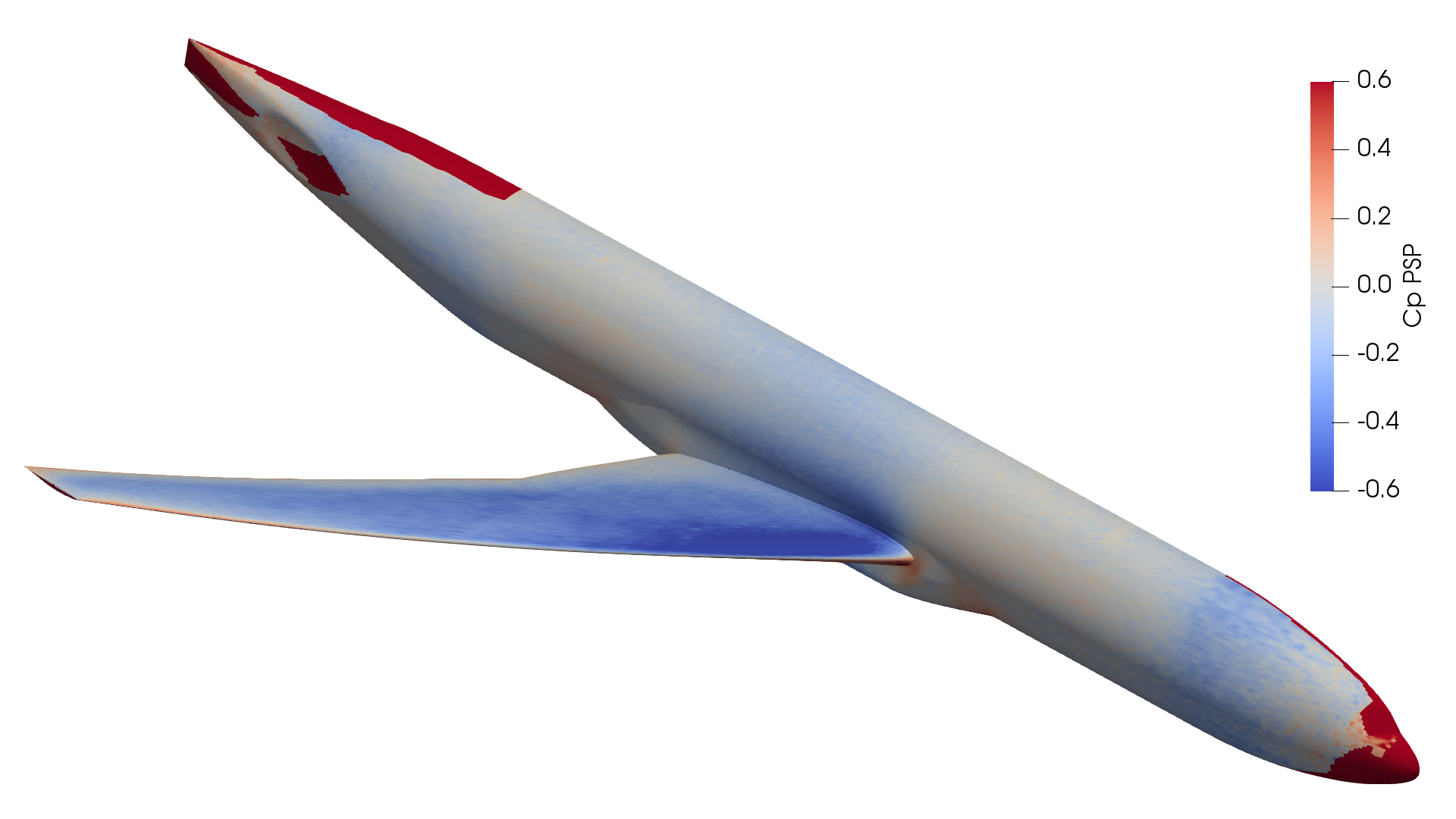}
        \caption{Raw Data}
        \label{f:CRM_iso_raw}
    \end{subfigure}~
    \begin{subfigure}{0.5\textwidth}
        \centering
        \includegraphics[scale=0.12]{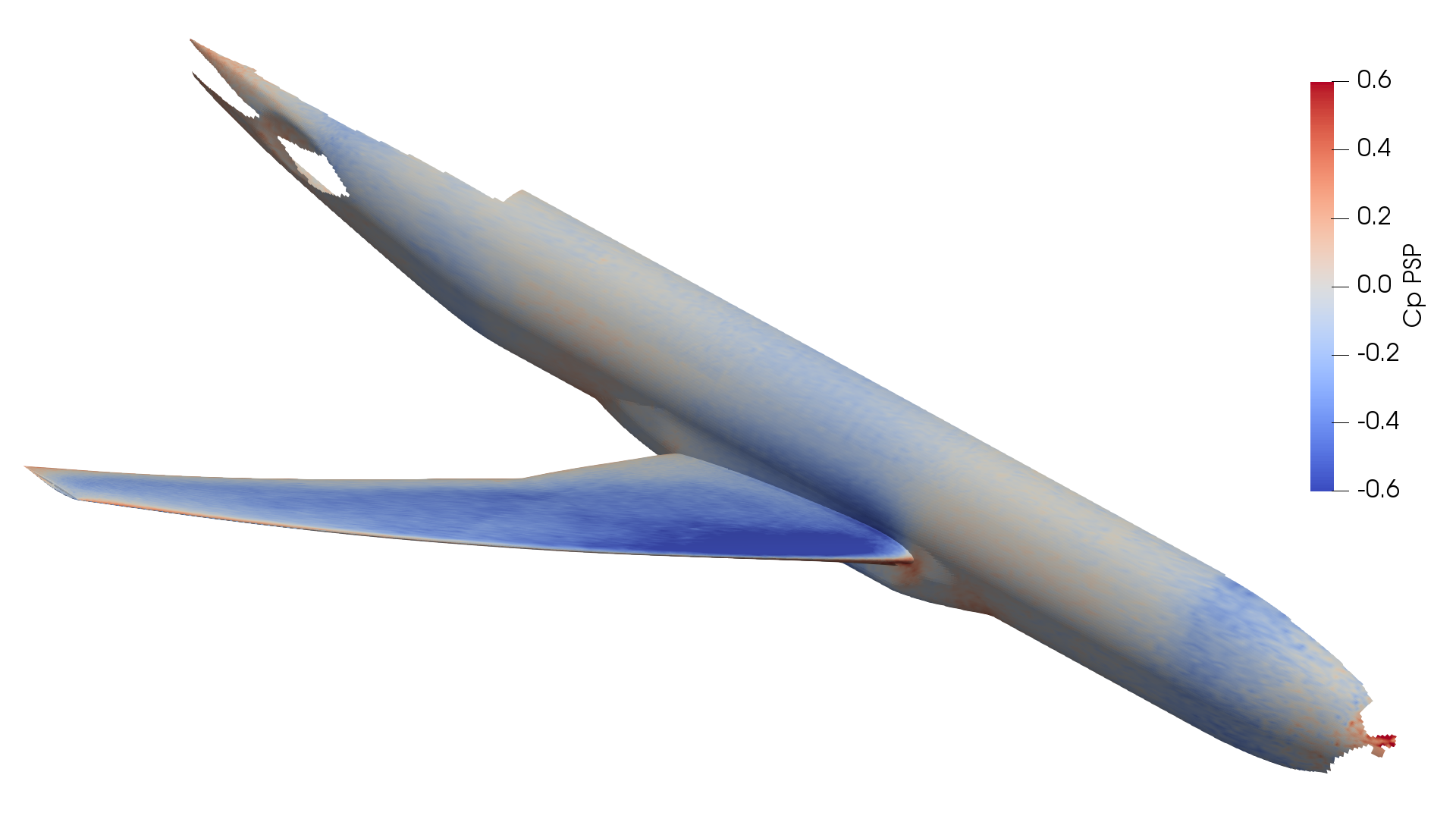}
        \caption{Invalid Data Removed}
        \label{f:CRM_iso_invalid_no_fill}
    \end{subfigure}
    \caption{PSP measurements containing invalid data points}
    \label{f:CRM_invalid_iso}
\end{figure}

\begin{figure}[htb]
    \centering
    \begin{subfigure}{0.5\textwidth}
        \centering
        \includegraphics[scale=0.12]{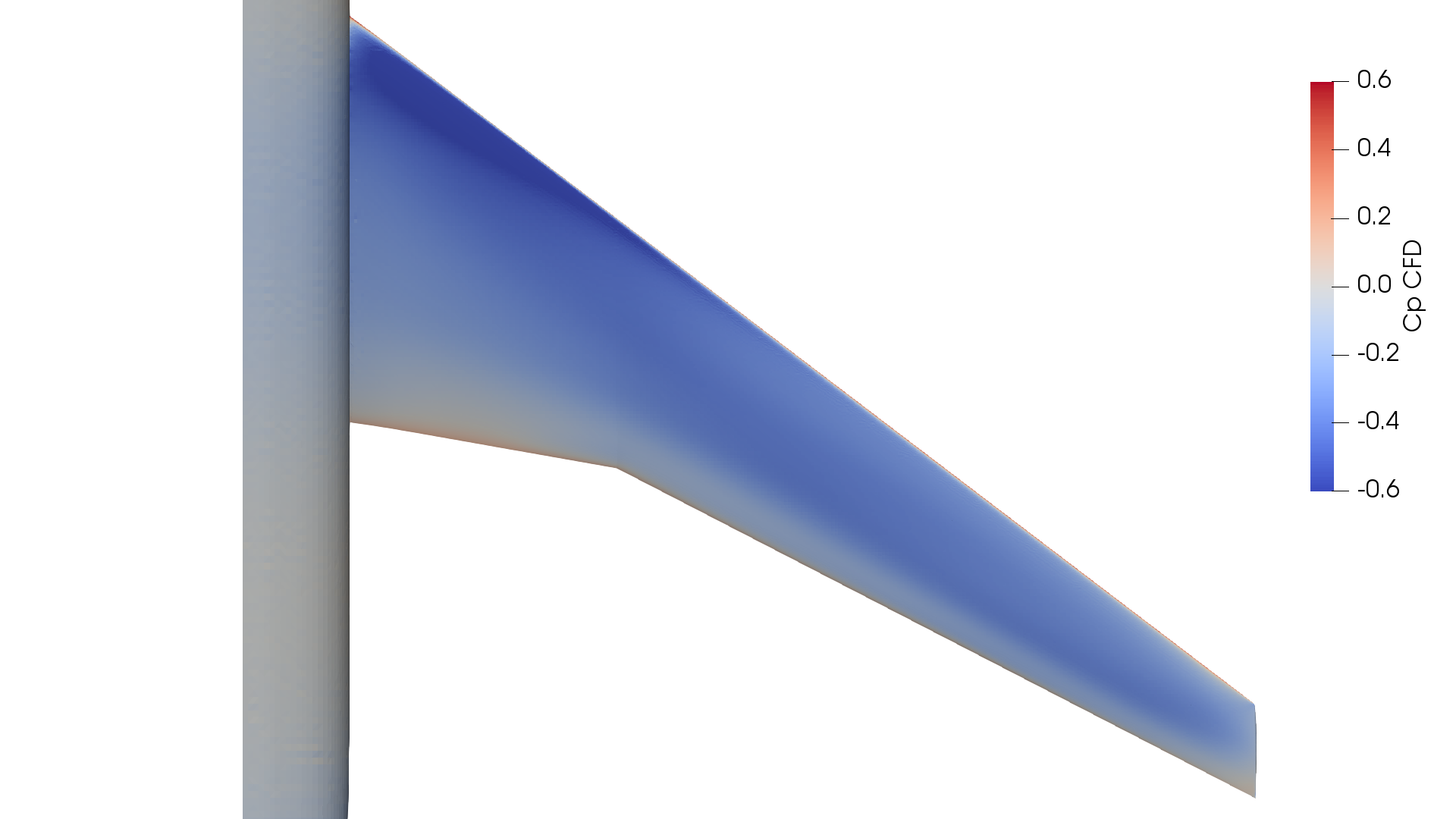}
        \caption{Original data}
    \end{subfigure}~
    \begin{subfigure}{0.5\textwidth}
        \centering
        \includegraphics[scale=0.12]{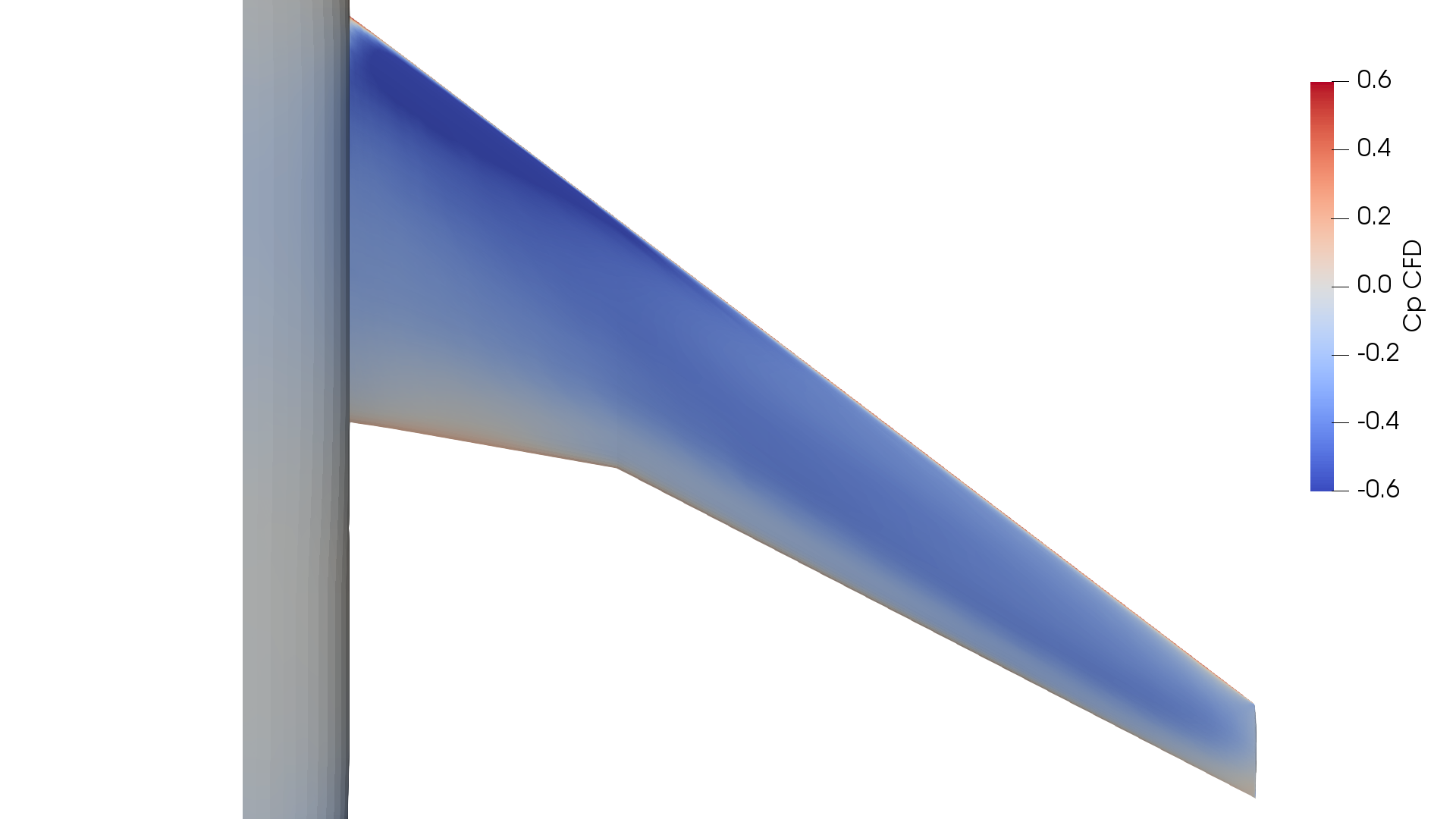}
        \caption{Coarsen data}
    \end{subfigure}
    \caption{CFD pressure distribution on upper wing}
    \label{f:CRM_CFD_comp}
\end{figure}

\begin{figure}[htb]
    \centering
    \begin{subfigure}{0.5\textwidth}
        \centering
        \includegraphics[scale=0.12]{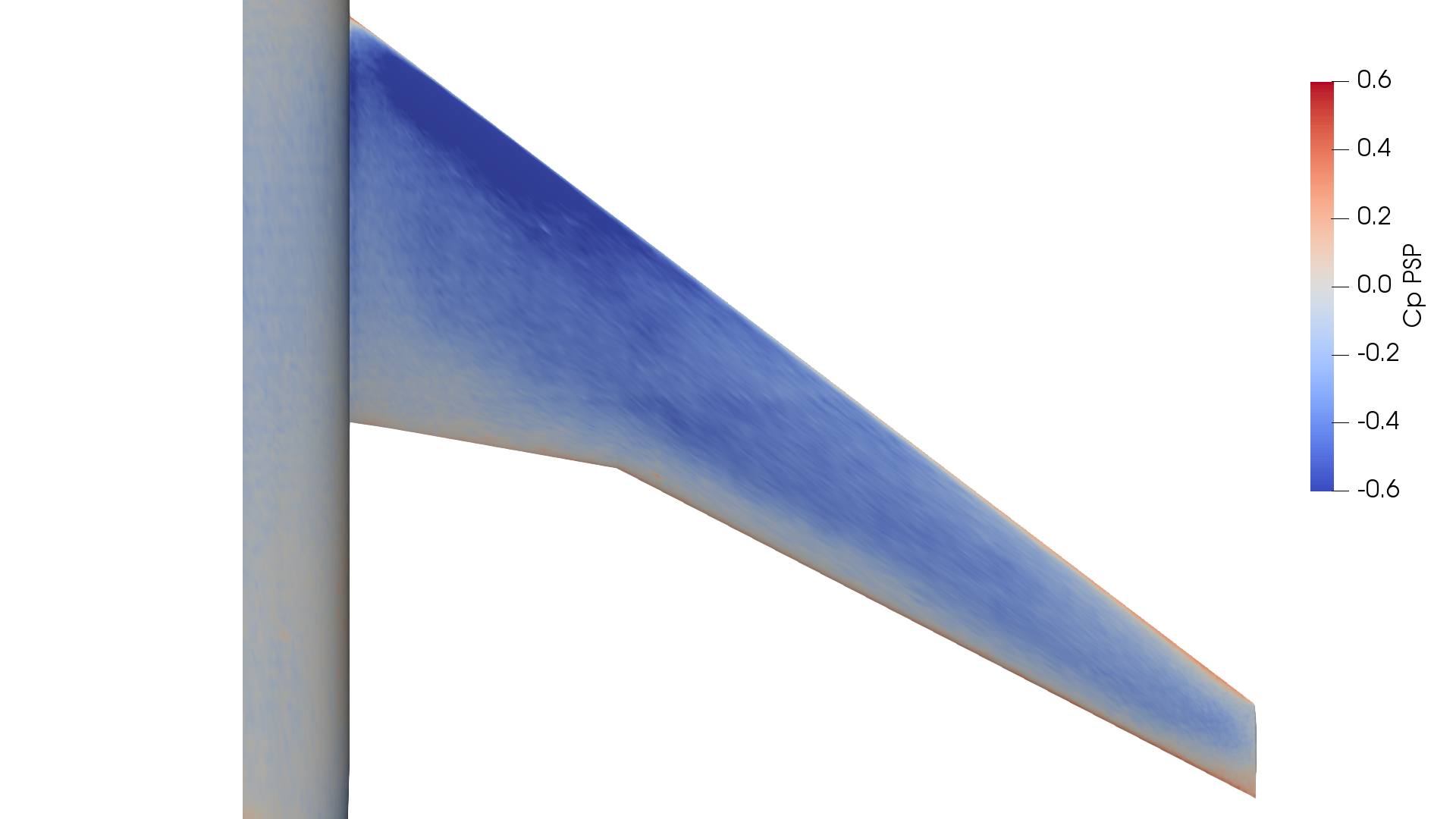}
        \caption{Original data}
    \end{subfigure}~
    \begin{subfigure}{0.5\textwidth}
        \centering
        \includegraphics[scale=0.12]{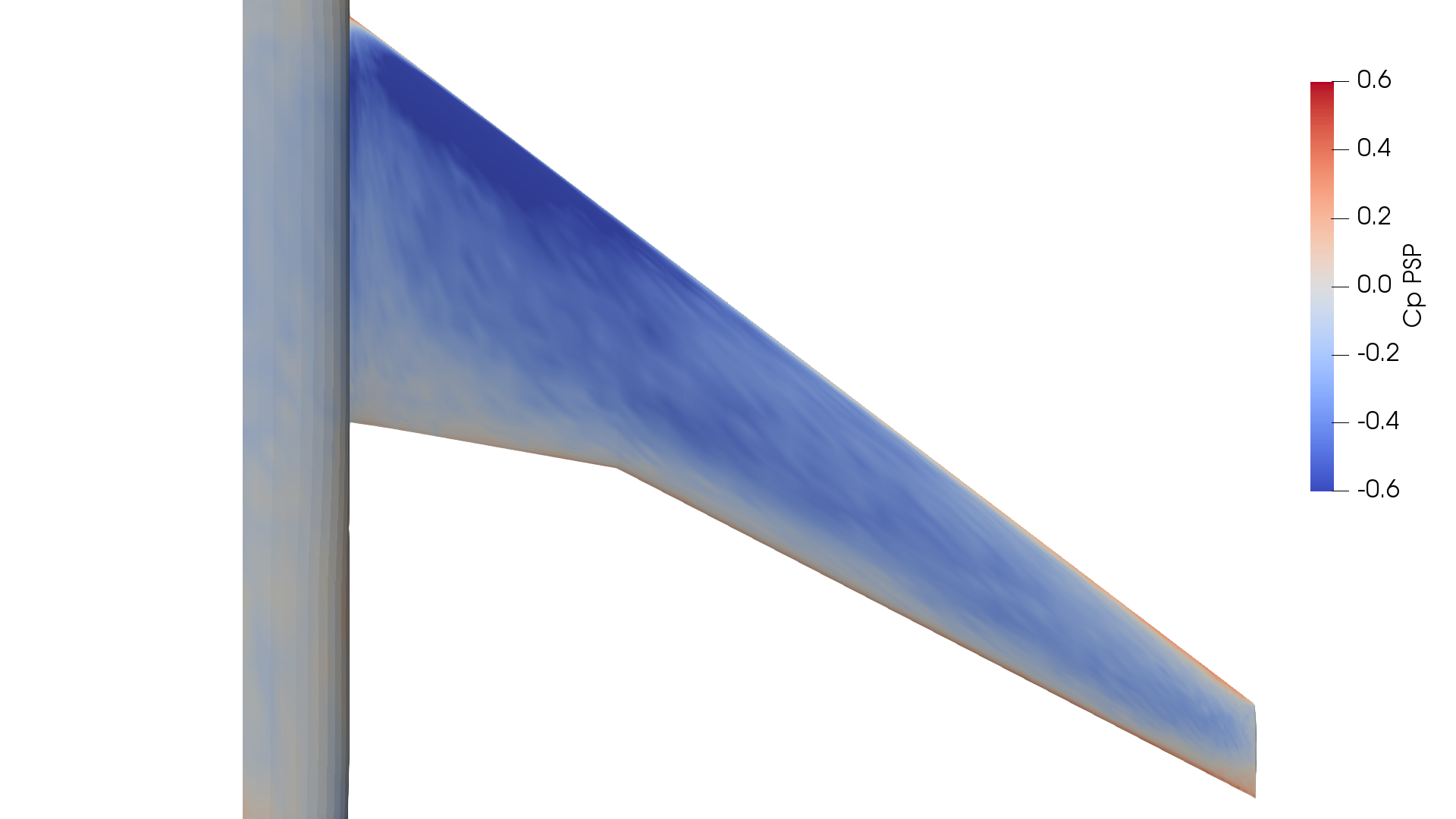}
        \caption{Coarsen data}
    \end{subfigure}
    \caption{PSP pressure distribution on upper wing}
    \label{f:CRM_PSP_comp}
\end{figure}

The pressure distribution from the PSP measurement contains missing data and in the datasets considered in the present study, they are found predominantly in regions near nose and tail of the fuselage; see Figure~\ref{f:CRM_invalid_iso}. These missing regions are then interpolated from the surrounding surface data using the voronoi kernel interpolation available in open-source visualization tool Paraview ~\cite{ahrens2005paraview}, although other techniques for interpolation may be used. Once the missing data are imputed, we interpolate them onto a common grid to perform the data fusion. The common grid is created by coarsening the PSP measurement grid to keep the computational costs tractable. The final number of cells and points of the grids are shown in the Table~\ref{tb:grid_summary}. The comparison of the computed coefficients of lift and moment for the original and coarse grids are shown in Table~\ref{tb:grid_error} where the summary statistic of the percent error across all the test cases is provided; a maximum error of 1.72\% error confirms that the final grid is not too coarse. The Figure~\ref{f:CRM_CFD_comp} and Figure~\ref{f:CRM_PSP_comp} also show the contour plots of pressure distribution comparing the original grid and the coarse grid. Overall we observe that the grid coarsening has a negligible impact on the quantities of interest while using only $\sim 10\%$ of the information on the original grids which leads to improved computational efficiency.  

\begin{table}[htb]
\centering
\parbox{.45\textwidth}{
\centering
    \begin{tabular}{c|cc}
            & Num. of Cells & Num. of Points \\ \hline
        CFD & 595642 & 533443 \\
        PSP & 84288 & 85711 \\
        Coarse & 8688 & 9103 \\
    \end{tabular}
    \caption{Summary of grids geometry}
    \label{tb:grid_summary}
}
\parbox{.45\textwidth}{
\centering
    \begin{tabular}{cc|ccc}
                             &    & Mean & Max  & Min  \\ \hline
        \multirow{2}{*}{CFD} & CL & 0.53 & 0.74 & 0.37 \\
                             & CM & 0.39 & 0.61 & 0.09 \\ \hline
        \multirow{2}{*}{PSP} & CL & 1.07 & 1.72 & 0.02 \\
                             & CM & 0.45 & 1.30 & 0.03 
    \end{tabular}
    \caption{Absolute percent difference between original and coarse grids}
    \label{tb:grid_error}
    }
\end{table}

\end{document}